\documentclass[fleqn,usenatbib]{mnras}
\usepackage{graphicx}
\usepackage{multirow}
\usepackage{amsmath}
\usepackage{soul}

\newcommand{\kms}{~km\,s$^{-1}$}

\newcommand{\MHI}{M$_{\rm HI}$}

\newcommand{\Msun}{M$_{\odot}$}

\newcommand{\MHm}{M$_{\rm H_2}$}

\newcommand{\Mdust}{M$_{\rm dust}$}

\newcommand{\Mg}{M$_{\rm gas}$}
\newcommand{\Mb}{M$_{\rm bar}$}
\newcommand{\Ms}{M$_{\star}$}

\newcommand{\MMoy}{M$_{\odot}$\,yr$^{-1}$}

\newcommand{\Yeff}{Y$_{\rm eff}$}
\newcommand{\sMZ}{$\Sigma_{\rm *} -$ $\rm Z$}
\newcommand{\sMSFR}{$\Sigma_{\rm *}-$ $ \Sigma_{\rm SFR}$}
\newcommand{\sMgSFR}{$\Sigma_{\rm {gas}}-$ $ \Sigma_{\rm SFR}$}
\newcommand{\alphaCO}{$\upalpha_{\rm{CO}}$}
\newcommand{\sMs}{$\Sigma_{\rm *}$}
\newcommand{\sSFR}{$\Sigma_{\rm SFR}$}
\newcommand{\sMdust}{$\Sigma_{\rm{dust}}$}
\newcommand{\sMgas}{$\Sigma_{\rm {gas}}$}
\newcommand{\sMbar}{$\Sigma_{\rm {bar}}$}
\newcommand{\sHmol}{$\Sigma_{\rm H_{2}}$}
\newcommand{\sHI}{$\Sigma_{\rm HI}$}
\newcommand{\sMHISFR}{$\Sigma_{\rm {HI}}-$ $ \Sigma_{\rm SFR}$}
\newcommand{\sMmolSFR}{$\Sigma_{\rm H_{2}}-$ $ \Sigma_{\rm SFR}$}

\usepackage{savesym}
\usepackage{amsmath}
\savesymbol{iint}
\usepackage{txfonts}
\usepackage{booktabs}
\usepackage{xcolor}
\restoresymbol{TXF}{iint}
\bibpunct{(}{)}{;}{a}{}{,}
\usepackage{newtxtext,newtxmath}
\usepackage[T1]{fontenc}
\usepackage{ae,aecompl}

\usepackage{graphicx}	
\usepackage{amsmath}	
\usepackage{newtxtext,newtxmath}

\title[Metal-THINGS: scaling relationships NGC 1569]{Metal-THINGS: a panchromatic analysis of the local scaling relationships of the dwarf irregular galaxy NGC 1569}

\author[Gardu\~no, L. E.]{Gardu\~no, L. E.$^{1}$\thanks{E-mail: luis@inaoep.mx}, Zaragoza-Cardiel, J.$^{1,2}$, Lara-L\'opez, M. A.$^{3}$, Zinchenko, I. A.$^{4,5}$, \newauthor Zerbo, M.C.$^{6,7}$, De Rossi, M. E.$^{6,7}$, Fritz, Jacopo$^{8}$, Dib, S.$^{9}$, Pilyugin, L. S.$^{10,5}$, \newauthor S\'anchez-Cruces, M.$^{11}$, Heesen, V.$^{12}$, O'Sullivan, S. P.$^{3}$, L\'{o}pez-Cruz, O.$^{1}$, Valerdi, M.$^{1,2}$, \newauthor Rosado, M.$^{11}$       
\\
$^{1}$Instituto Nacional de Astrof\'isica, \'Optica y Electr\'onica (INAOE), Luis Enrique Erro No.1, Tonantzintla, Pue., C.P. 72840, M\'exico\\
$^{2}$Consejo Nacional de Humanidades, Ciencias y Tecnolog\'{i}as, Av. Insurgentes Sur 1582, 03940, Ciudad de M\'{e}xico, Mexico\\
$^{3}$Departamento de F\'{ı}sica de la Tierra y Astrof\'{ı}sica, Instituto de F\'{ı}sica de Part\'{ı}culas y del Cosmos, IPARCOS. \\
Universidad Complutense   de Madrid (UCM), E-28040, Madrid, Spain\\
$^{4}$Faculty of Physics, Ludwig-Maximilians-Universit{\"a}t, Scheinerstr. 1, 81679 Munich, Germany\\
$^{5}$Main Astronomical Observatory, National Academy of Sciences of Ukraine, 27 Akademika Zabolotnoho St, 03680, Kiev, Ukraine\\
$^{6}$Universidad de Buenos Aires, Facultad de Ciencias Exactas y Naturales y Ciclo B\'{a}sico Com\'{u}n. Buenos Aires, Argentina\\
$^{7}$CONICET-Universidad de Buenos Aires, Instituto de Astronom\'{i}a y F\'{i}sica del Espacio (IAFE). Buenos Aires, Argentina\\
$^{8}$Instituto de Radioastronomia y Astrofisica, UNAM, Campus Morelia, A.P. 3-72, C.P. 58089, Morelia, Mexico\\
$^{9}$Max Planck Institute for Astronomy, K\"{o}nigstuhl 17, 69117, Heidelberg, Germany\\
$^{10}$Institute of Theoretical Physics and Astronomy, Vilnius University, Sauletekio av. 3, 10257, Vilnius, Lithuania\\
$^{11}$Instituto de Astronom\'{i}a, Universidad Nacional Autonoma de M\'{e}xico, Apartado Postal 70-264, CP 04510 M\'{e}xico, CDMX, M\'{e}xico\\
$^{12}$Hamburg University, Hamburger Sternwarte, Gojenbergsweg 112, 21029 Hamburg, Germany\\
}

\date{Accepted 2023 August 31. Received 2023 August 28; in original form 2023 May 9}

\pubyear{2023}

\begin{document}
\label{firstpage}
\pagerange{\pageref{firstpage}--\pageref{lastpage}}
\maketitle

\begin{abstract}
We investigate several panchromatic scaling relations (SRs) for the dwarf irregular galaxy NGC 1569 using IFU data from the Metal-THINGS Survey.
Among the spatially resolved properties analyzed, we explore SRs between the stellar mass, SFR, molecular gas, total gas, baryonic mass, gas metallicity, gas fraction, SFE and effective oxygen yields. Such multiwavelength SRs are analyzed at a spatial resolution of 180 pc, by combining our IFU observations with data from the surveys THINGS, CARMA, and archival data from DustPedia. Although we recover several known relations, our slopes are different to previously reported ones. Our star formation main sequence, Kennicutt-Schmidt (KS) and molecular KS relations show higher SFRs, lower scatter, and higher correlations, with steeper (1.21), and flatter slopes (0.96, 0.58) respectively. The shape of the SRs including metallicity, stellar mass, and gas fraction are flat, with an average value of 12+log(O/H) $\sim$ 8.12 dex. 
The baryonic mass vs effective oxygen yields, and the stellar, gas and baryonic mass vs SFE show higher dispersions and lower correlations. 
Since we use the dust mass as a tracer of gas mass, we derive the Dust-to-Gas Ratio and the CO luminosity-to-molecular gas mass conversion factors, showing differences of 0.16 and 0.95 dex for the total and molecular gas surface density, respectively, in comparison to previously reported values. We use a self regulated feedback model to conclude that stellar feedback plays an important role  generating outflows in NGC 1569.
\end{abstract}

\begin{keywords}
galaxies: abundances, galaxies: dwarf, galaxies: irregular, galaxies: starburst, galaxies: star formation, galaxies: statistics
\end{keywords}


\section{Introduction}\label{Intro}




The evolutionary path of galaxies is driven by distinct processes, at different time scales. The history of star formation in galaxies, gas accretion, mergers or gas inflows/outflows -among others- can be studied through the current galaxy properties, such as star formation rate (SFR), gas metallicity (Z), stellar mass (\Ms), gas mass (\Mg), baryonic mass (\Mb), star formation efficiency (SFE) and effective yields (\Yeff). Since all of these properties contain information about the past and current evolution of galaxies, scaling relations (SRs) between these properties are an important tool to understand the most important mechanisms that drive galaxy evolution.
 
During the last few decades, several global SRs were explored using fiber spectroscopic surveys, showing critical physical properties of galaxies in the local universe. On one hand, using the Sloan Digital Sky Survey \citep[SDSS,][]{SDSS2000} and the Galaxy and Mass Assembly Survey \citep[GAMA,][]{GAMA2011} data, the relation between stellar mass vs. SFR \citep[\Ms-SFR,][]{Brinchmann2004,Lara-Lopez2013b}, and stellar mass vs. metallicity \citep[\Ms-Z,][]{Tremonti2004,Lara-Lopez2013a} were established for thousands of galaxies. On the other hand, additional SRs were established such as the baryonic mass-effective yield \citep[\Mb-\Yeff,][]{Tremonti2004,LaraLopez2019}, the gas fraction-metallicity \citep[$\upmu$-Z,][]{Pilyugin2004,LaraLopez2019}, the gas fraction-effective yield \citep[$\upmu$-\Yeff,][]{Dalcanton2007} or the relation between baryonic mass, stellar mass and gas mass with the star formation efficiency \citep[\Mb-SFE, \Ms-SFE, \Mg-SFE,][]{ALFALFA2011,LaraLopez2019}. All the SRs mentioned above help to understand a part of the galaxy evolution process by analyzing the gas inside galaxies and how different feedback processes affect their evolutionary path.

Since stars form out of collapsing gas clouds, a correlation is expected between the surface densities of star formation and gas. Indeed, this is described by the Kennicutt-Schmidt (KS) relation, an empirical scaling relation between the gas surface density and the SFR surface density given by $\Sigma_{\rm SFR}$ = $\rm a~{\Sigma_{\rm gas}}^{\rm n}$ \citep{Kennicutt1989} which was first proposed in the pioneering work of \cite{Schmidt1959}. A classic example of the global KS relation is presented in \cite{Kennicutt1998}, who also demonstrate the importance of radio and infrared observations in order to get robust estimations of gas mass and SFRs, respectively.  
 
With the advent of new instrumentation such as the Integral Field Units (IFU), new possibilities of study have opened up; what was first done on a global scale, now can be done on a spatially resolved scale. Recently, IFU surveys have established  local SRs, such as the stellar mass surface density vs. metallicity relation \citep[\sMZ,][]{Rosales2012,Barrera2016,Baker2023}, the  stellar mass surface density vs. SFR surface density relation \citep[\sMSFR,][]{Sanchez2021, Pessa2022, Baker2023} and the spatially resolved gas fraction vs. metallicity relation \citep[$\upmu$ - Z,][]{Barrera2018}. Although the spatially resolved oxygen yields have not been studied by many authors, \cite{Vilchez2019} explored resolved effective yield profiles for two galaxies (NGC 5457 and NGC 628). In all of the above cases, it is important to note that local SRs mimic the global ones, which suggests that  processes understood at local scales  could be the key to understand global galaxy evolution processes.

The KS relation has also been analyzed locally and it preserves a similar shape to the global relation \citep{Kennicutt2007,Blanc2009,Casasola2022,Pessa2022}. 
Other studies have extended the relation by considering different star formation regions in a vast range of galactic environments, from the outer disks of dwarf galaxies to spiral disks, merging galaxies and individual molecular clouds \citep{Dib2011,Dib2017,Shi2018,Pessa2021}.

By analyzing the spatially resolved KS relation, some authors conclude that the gas mass (usually molecular gas) could play the most important role in the SFR regulation process, suggesting that this could be a more fundamental relation instead of the \sMSFR\ relation \citep{Morselli2020,Ellison2021,Pessa2021,Pessa2022}. However it is still a matter of debate. Indeed, other authors \citep[{\it{e.g.}}, ][]{Dib2017} explain the importance of the \sMSFR\ relation since the gravity of stars, over scales of a few hundred parsec in galactic disks, is as important (and very often dominates) as that of gas. As \cite{Dib2017} (and references there in) mentioned, this implies that existing stars can play a fundamental role in generating large scale gravitational instabilities which in turn lead to star formation.

The correlation between the \Ms, Z and SFR led to the simultaneous discovery of a 3-dimensional structure (whose shape is still debatable),  refereed to the Fundamental Plane  \citep[][]{LaraLopez2010a}, and the Fundamental Metallicity Relation \citep[FMR,][]{Mannucci2010}, highlighting the importance of the already known galactic gas inflows and outflows. Once the observational capabilities of radio wavelength reached the point at which large surveys could be pursued, the consequent analysis was to explore the relation between the gas mass (atomic, molecular or both) and the metallicity \citep[{\it{e.g.}}, ][]{Lara-Lopez2013a}. Some studies confirmed the correlation between such properties and even showed the possibility of being more fundamental by driving the FMR \citep{Bothwell2013a}. Deeper studies set that the dependence with the gas mass is probably stronger than with SFR, and thus the underlying FMR is between stellar mass, metallicity and gas mass \citep{Bothwell2016a}. This latter is supported by \cite{Bothwell2016b}, who affirm that the metallicity dependence on SFR is a derivation of the dependence on the molecular gas content given by the KS relation. However, this is also a matter of debate. As other author mentioned, there is a dual dependence of the SFR/SFE on metallicity. One is due to the metal content in the star forming gas which governs its ability to cool, and the second is via metallicity dependent feedback \citep[stellar winds,][]{Dib2011,Dib2011b}. Both of the effects play against each other. A lower metallicity implies less cooling and thus less molecular gas. A lower metallicity also means weaker stellar winds and so collapsing clouds encounter less gas expulsion, which leads to more SFR. Which of these effects wins or what is their relative importance is yet to be quantified in detail.
 
The total gas mass is obtained by the mass in atomic gas, \MHI, which is derived from HI observations, in combination with the molecular gas, \MHm, whose determination relies on measurement of the CO molecule emission, and on the CO luminosity-to-molecular gas mass conversion factor, $\rm \upalpha_{CO}$. For simplicity, $\rm \upalpha_{CO}$ factor was considered constant for all type of galaxies  \citep[{\it{e.g.}}, ][]{Kennicutt1998}. However, there is evidence proving that $\rm \upalpha_{CO}$ can vary with gas density and metallicity \citep{Rubio1993,Lequeux1994,Bolatto2013}. In particular, low metallicity galaxies have very faint CO emission, which makes its detection difficult and hinders a reliable calibration for the conversion factors. As pointed out also in \cite{Bolatto2013}, gas rich galaxies with active star formation and low metallicity usually have a very faint CO emission. A very particular case is seen for low mass dwarf irregular galaxies, that show almost null CO emission, and have regularly low metallicities \citep{Draine2007}. 

Both the metallicity and the dust grains have been highly studied since they provide important information for understanding the chemical evolution of galaxies. Indeed, it is thought that metals in the ISM are encapsulated inside of dust grains. Under the proper physical processes, {\it{e.g.}}, supernovae shocks, the dust grains can be destroyed, carrying metals back to the ISM. As mentioned in \cite{Remy2014}, the amount of metals that are locked up in the dust grains can be quantify by the Dust-to-Gas Ratio (DGR). Thus, it is expected that the DGR depends strongly on metallicity and changes from one galaxy to another \citep[{\it{e.g.}}, because the ISM density also changes,][]{Clark2023}, especially in the low metallicity regime \citep{Remy2014}. Some studies, such as \citet{Leroy2011} and \citet{Sandstrom2013}, developed methodologies to compute not only the DGR but the $\upalpha_{\rm CO}$, taking into consideration the dependence of each other with the metallicity, obtaining reliable results. 

In the local Universe, dwarf galaxies look like scaled down versions of high redshift massive galaxies due to their similarity in star formation, metallicity, physical size, high mass fractions or morphology ({\it{e.g.}}, thickness and clumpiness) \citep{Elmegreen2012,Elmegreen2013,Motino2021}. The variety of dwarf galaxies is extensive and there are different types (elliptical, spheroidal, irregular, blue compact, ultra faint, ultra compact). It is thought that the dwarf irregular galaxies are the most common in the universe, regularly found in isolation, with the particularity that they are still forming stars \citep{Ellis1997,Spaans1997, Tolstoy2000, Gallart2015, Simon2019}. Nowadays, there is an effort trying to find a connection among dwarf galaxies, establishing that irregular dwarfs evolve to Blue Compact Dwarf (BCD) through several starbursts, that enrich the ISM until their gas is exhausted to finally fade to gas free dwarf spheroidals \citep{Kong2019}. Dwarf irregular galaxies are important since they could reach the extreme low regime of mass and metallicity. Since possibly they were the first structures that were formed \citep{Belokurov2006, Mateo1998}, they provide unique clues to explore the early galaxy formation and evolution.

We present an analysis of the irregular dwarf galaxy NGC 1569 using local scaling relationships. This galaxy is a particularly interesting object since it is a starburst with a low stellar mass and low metallicity. A summary of the relevant physical properties can be seen in Table \ref{NGC1569_properties}. We use IFU spectroscopy observations of the Metal-THINGS Survey \citep{LaraLopez2021,LaraLopez2023} in combination with other multiwavelength data. Such combination of data allows us to study, in a spatially resolved way, SRs involving both the stellar and the gas component: $\Sigma_{\rm *}-$ $ \Sigma_{\rm SFR}-$ Z, $\Sigma_{\rm gas}-$ $ \Sigma_{\rm SFR}-$ Z. Given that NGC 1569 has a very limited CO emission that results in a poor estimation of \MHm\ and \Mg, we compute the dust mass (\Mdust) by fitting of the SED using multiwavelength data, and then we estimate the DGR according to methods already reported in the literature to obtain a reliable \Mg.


\begin{table}
\centering
\begin{tabular}{|ccc|}
\hline
\multicolumn{3}{|c|}{NGC 1569} \\ \hline
\multicolumn{1}{|c|}{Distance} & \multicolumn{1}{c|}{3.25 Mpc} & \cite{Tully2013} \\ \hline
\multicolumn{1}{|c|}{\multirow{2}{*}{Stellar mass}} & \multicolumn{1}{c|}{ 8.61 log(\Msun)} & \cite{Leroy2019} \\ \cline{2-3} 
\multicolumn{1}{|c|}{} & \multicolumn{1}{c|}{8.6 log(\Msun)} & This work \\ \hline
\multicolumn{1}{|c|}{\multirow{2}{*}{Dust mass}} & \multicolumn{1}{c|}{ 4.88 log(\Msun)} & \cite{Young1989} \\ \cline{2-3} 
\multicolumn{1}{|c|}{} & \multicolumn{1}{c|}{5.3 log(\Msun)} & This work \\ \hline
\multicolumn{1}{|c|}{\multirow{2}{*}{Gas mass}} & \multicolumn{1}{c|}{8.7 log(\Msun)} & \cite{Young1989} \\ \cline{2-3} 
\multicolumn{1}{|c|}{} & \multicolumn{1}{c|}{8.33 log(\Msun)} & This work \\ \hline
\multicolumn{1}{|c|}{\multirow{3}{*}{\begin{tabular}[c]{@{}c@{}}Metallicity\\ 12+log(O/H)\end{tabular}}} & \multicolumn{1}{c|}{8.16-8.19} & \cite{Kobulnicky1997} \\ \cline{2-3} 
\multicolumn{1}{|c|}{} & \multicolumn{1}{c|}{8.12} & This work \\ \hline
\multicolumn{1}{|c|}{\multirow{4}{*}{SFR}} & \multicolumn{1}{c|}{0.95 \MMoy} & \cite{Leroy2019} \\ \cline{2-3} 
\multicolumn{1}{|c|}{} & \multicolumn{1}{c|}{0.91 \MMoy} & DustPedia \\ \cline{2-3}
\multicolumn{1}{|c|}{} & \multicolumn{1}{c|}{0.48 \MMoy} & This work \\ \cline{2-3} 
\multicolumn{1}{|c|}{} & \multicolumn{1}{c|}{0.27 \MMoy} & Outputs CIGALE \\ \hline
\multicolumn{1}{|c|}{Scale} & \multicolumn{1}{c|}{0.015 kpc/arcsec} &  This work\\ \hline
\end{tabular}
\caption{Global important properties of NGC 1569. The adopted distance is using TRGBs. We compare some values computed by our own data with values in the literature. A SFR estimation is taken from  the reported data of DustPedia. The other SFR estimation is taken from the outputs modules of CIGALE in our own runs. The scale is computed according to the distance.}
\label{NGC1569_properties}
\end{table}

Since the galaxies from the Metal-THINGS survey are nearby (D < 15 Mpc), the physical spatial resolution of the survey is up to two orders of magnitude better than other surveys such as CALIFA \citep{Sanchez2012} and MANGA \citep{Bundy2015} at the same wavelength coverage, which implies that galaxy properties are analyzed at scales of parsecs. The Field of View (FOV) of the George Mitchel Spectrograph allows us to cover almost the entire galaxy, in contrast with previous studies in which only the central structure was analyzed \citep{Westmoquette2007a,Westmoquette2007b,Mayya2020}. Another important key to note is that neither MANGA nor CALIFA have studied the metallicity and SFR in detail for the regime of low mass galaxies because they are biased towards higher mass ones (log(M/M$_\odot$) > 9 ). By analyzing galaxies such as NGC 1569, we are filling the gap to the low mass regime.

The structure of this work is as follows. In \S \ref{SampleSelection}, we describe our data acquisition and in \S \ref{physicalproperties}, the estimation of the main galaxy properties. In \S \ref{SR1_section}, we derive the spatially resolved SRs for the galaxy properties: \sMs, \sMgas, \sSFR\ and Z. We mainly explore the local relations for oxygen yields (\Yeff) and SFE in \S \ref{SR2_section}. Finally in \S \ref{discuss_section} and \ref{conclusion_section}, we present a general discussion of this work and a summary of our conclusions, respectively.



\section{Observations and data reduction}\label{SampleSelection}

\subsection{Metal-THINGS}\label{metthings}
In this paper, we use data from the Metal-THINGS survey \citep{LaraLopez2021, LaraLopez2023}, which is obtaining IFU spectroscopy of  34 nearby galaxies mapped in HI with the Very Large Array (VLA) \citep[The THINGS survey,][]{Walter2008}.  

Metal-THINGS is an ongoing survey currently observing at the McDonald Observatory with the 2.7m Harlan Schmidt telescope using the George Mitchel Spectrograph (GMS), formerly known as  VIRUS-P \citep{Hill2008}. GMS is an IFU with 246 fibers, each one with a diameter of 4.2 arcsecs, disposed on a square array of 100 x 102 arcsecs. The planning and development of the  observing program as well as the technical details and general procedures are  described in Lara-López et al. 2023 (in preparation). For the case of NGC 1569, our observations and methodology are the same as to those described in \cite{LaraLopez2021}.

\begin{figure}
    \centering
    \includegraphics[trim={0.7cm 0 1.5cm 1.3cm},clip,width=1\linewidth]{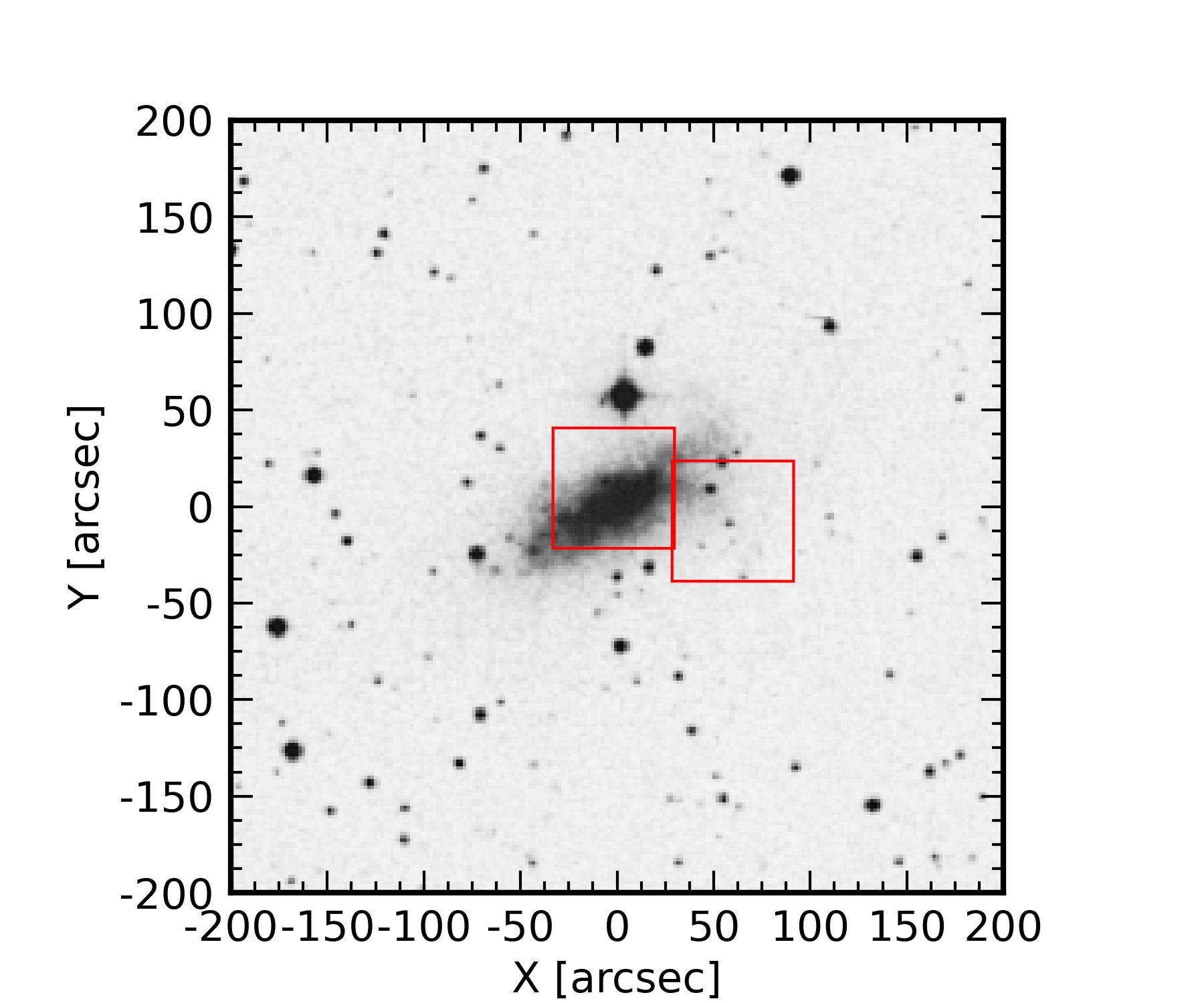}
    \caption{NGC 1569 image showing the field of view of the GMS in red boxes, 2 pointings were observed. The archival image in B band was taken from the Palomar Observatory Sky Survey, NGS-POSS.}
    \label{Pointings}
\end{figure}

We observed two pointings of NGC 1569 (Fig. \ref{Pointings}), the first one (left) was observed in January 2018 and the second one (right) in October 2020, both using the red setup, with a spectral coverage from 4400 to 6800 {\AA} and spectral resolution of 5.3 {\AA} (low resolution grating VP1). Since GMS has a 1/3 filling factor, each pointing is observed in three dither positions to ensure a high surface coverage of the galaxy. The motivation for taking the second pointing is to investigate the faint tail in this galaxy (in the upper left corner of the pointing 2 can be seen the tail structure that goes towards the center). 

During the observations, we integrated 15 minutes per dither, followed by 5 minutes of sky exposure. This process is repeated three times until a total integration time of 45 minutes is reached per dither (2.25 hours of total time for the three dithers). The observing nights for NGC 1569 were clear and had an average seeing of 2 arcsecs. 

As usual, every night exposures of bias, calibration lamps (Neon + Argon), and sky flats were taken, as well as a calibration star using a six dither position in order to get a full flux coverage for flux calibration.

For the data reduction we use P3D \footnote{\url{http://p3d.sourceforge.io}}, for bias subtraction, flat frame correction, and wavelength calibration. Next, our own routines in Python are used to make the sky subtraction and dither combination. Finally, we use the common tasks of IRAF \footnote{\url{http://iraf.nao.ac.jp/}} \citep{Tody1986} to make the flux calibration. First, we use the \emph{Standard} task to extract the spectrum of the standard star observed the same night of the observations with the same observational conditions. Second, we use the \emph{Sensfunc} task to compute both, the sensitivity and extinction functions, and finally, for the flux calibration, we use the \emph{Calibrate} task to apply the sensitivity curve and extinction to the spectra.


We fit the stellar continuum of all flux-calibrated spectra using STARLIGHT \citep{Cid2005, Mateus2006, Asari2007}, by setting 45 simple stellar populations (SSP) models from the evolutionary synthesis models of \citet{BC2003} with ages from 1 Myr up to 13 Gyr and metallicities Z = 0.005, 0.02 and 0.05. Thus, the stellar mass is extracted as one of the STARLIGHT outputs. After subtracting the fitted stellar continuum from the spectra, the emission lines are measured using Gaussian line-profile fittings, according to the methodology of \citet{Zinchenko2016,Zinchenko2019a}. Such fitting allow us to analyze the spectral regions that contain important emission lines such as H$\upbeta$; [O${\rm{III}}$] $\uplambda$$\uplambda$4959, 5007; [O${\rm{I}}$] $\uplambda$$\uplambda$6300, 6364; H$\upalpha$, [N${\rm{II}}$] $\uplambda$$\uplambda$6548, 6583; [S${\rm{II}}$] $\uplambda$$\uplambda$6716, 6731.

\begin{figure}
\includegraphics[trim={0 0cm 0 0cm},clip,width=1\linewidth]{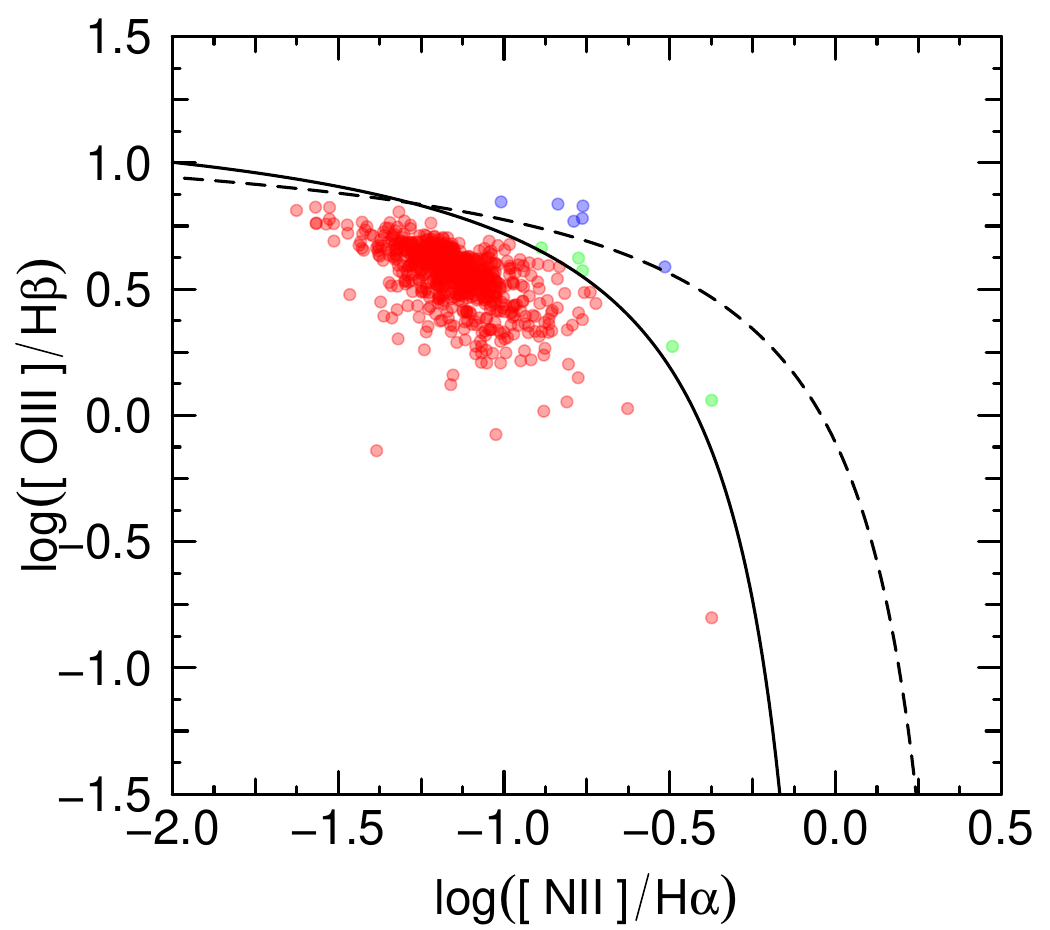}
\caption{BPT diagram for fibers with  a S/N $>$ 3 in the four emission lines used. The solid line corresponds to the \citet{Kauffmann2003} limit, the  dashed line to the \citet{Kewley2001} limit. Red, green and blue colors correspond to  SF, compound and AGN fibers, respectively.}
\label{BPT}
\end{figure}

In Fig. \ref{BPT}, we show the BPT diagnostics \citep{BPT1981} of our fibers using the classification of \citet{Kauffmann2003} and \citet{Kewley2001}, imposing a cut of S/N $>$ 3 for the H$\upalpha$, H$\upbeta$, [O${\rm{III}}$] $\uplambda$5007 and [N${\rm{II}}$] $\uplambda$6583 emission lines. As can be seen, we obtain a total of 750 (98.5\%) SF, 5 (0.6\%) Composite and 6 (0.8\%) Active Galactic Nuclei (AGN) - like fibers.

As a part of our analysis -described later- we use data at other wavelengths provided by the THINGS Survey \citep{Walter2008}, the CARMA survey \citep{Rahman2012} and DustPedia \citep{Davies2017}. DustPedia is a collaboration that provides access to multiwavelength imagery (and photometry) for nearby galaxies and model physical parameters for such galaxies. Thus, we get images of NGC 1569 in different bands, from UV (152.8, 227.1 nm, GALEX) to the mid-far infrared (3.6, 4.5, 5.8, 8, 24 $\upmu$m Spitzer; 70, 100, 160 $\upmu$m Hershel-PACS; 250 $\upmu$m Hershel-SPIRE). 

\section{Estimation of physical properties}\label{physicalproperties}

We create a tridimensional datacube by associating each fiber to its relative position on the sky and convolving them by a gaussian of FWHM = 4.2" (the diameter of the fiber) at each wavelength. Thus, we get spaxel maps of our different emission lines. Then, we convolve and reproject such maps to a common spatial resolution (see \S \ref{dustsec} below). According to the pixel size, we get a pixel area which allows to compute surface densities of some properties. Since we have an associated stellar mass at each fiber, this procedure is also applied to all the individual fibers, getting as a result a stellar mass surface density map (left panel of Fig. \ref{maps1}). We identify several stars in the field of view and use the star positions from the Two Micron All-Sky Survey \citep[2MASS,][]{2MASS} to obtain the astrometry.

The total number of spaxels for NGC 1569 before of corrections, binning and cleaning tasks is $\sim$400 (pixel size is $\sim$6", see \S \ref{dustsec}). Then, after applying a 2x2 binning and a similar S/N cut for the previously mentioned emission lines, we get a total of $\sim$100 spaxels with a pixel size of $\sim$12" that corresponds to $\sim$180 pc.   

The measured emission line spaxels are corrected for interstellar reddening using the Balmer Decrement (H\rm$\upalpha$/H\rm$\upbeta$) with the theoretical value of 2.86, using the extinction curve of \citet{Cardelli1989}.

\subsection{Stellar mass, SFR and metallicity estimation}

The SFR was computed following \cite{Kennicutt2009} using the L$_{\rm H\upalpha}$.

\begin{equation}
\rm{SFR}\left[M_\odot\;yr^{-1}\right]={5.5\times10^{-42}}\times{ L_{\rm{H\upalpha}}}
\end{equation}

\begin{figure*}
    \centering
    \includegraphics[trim={0 1.1cm 0 0cm},clip,width=0.33\linewidth]{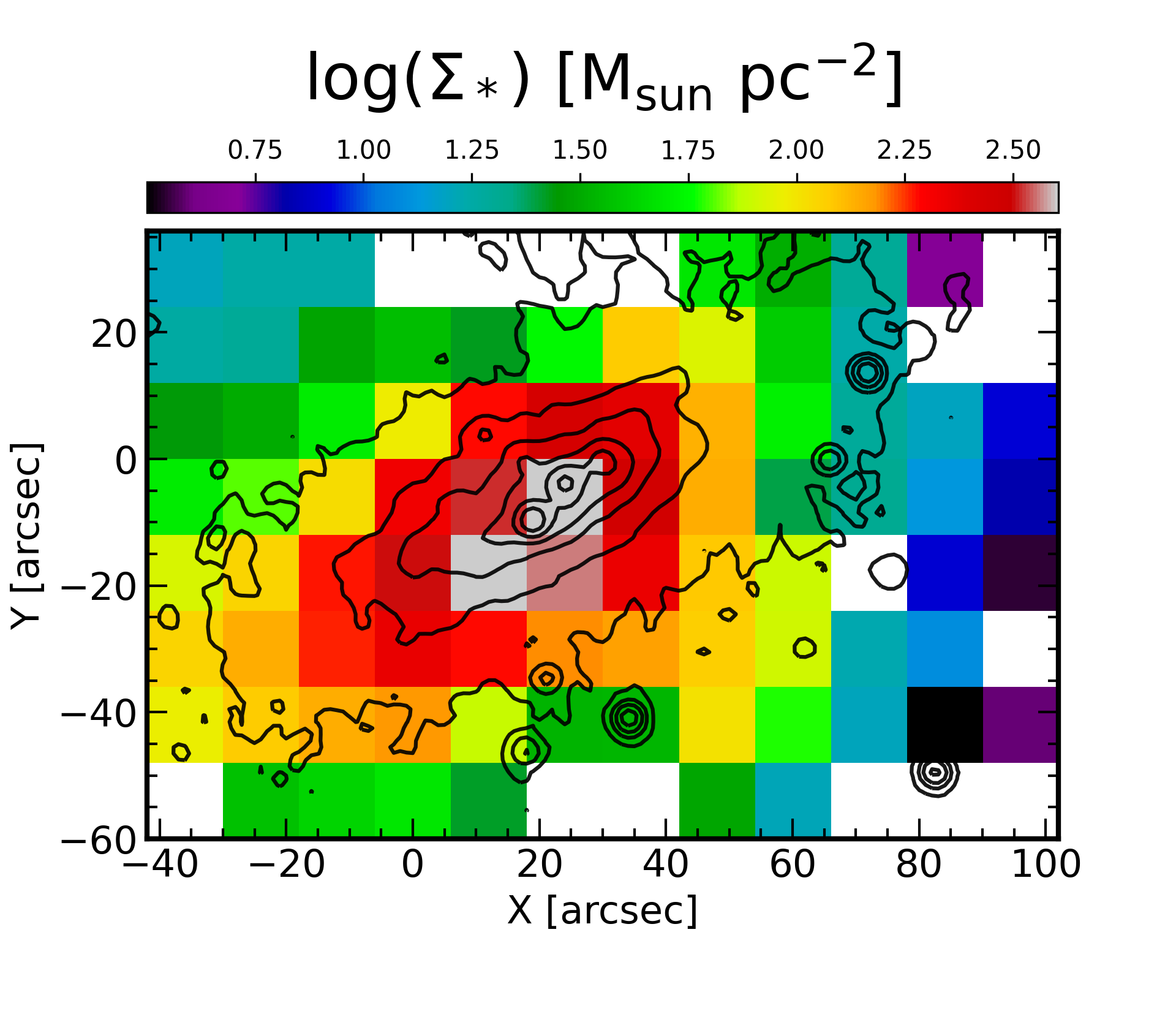}
    \includegraphics[trim={0 1.1cm 0 0cm},clip,width=0.33\linewidth]{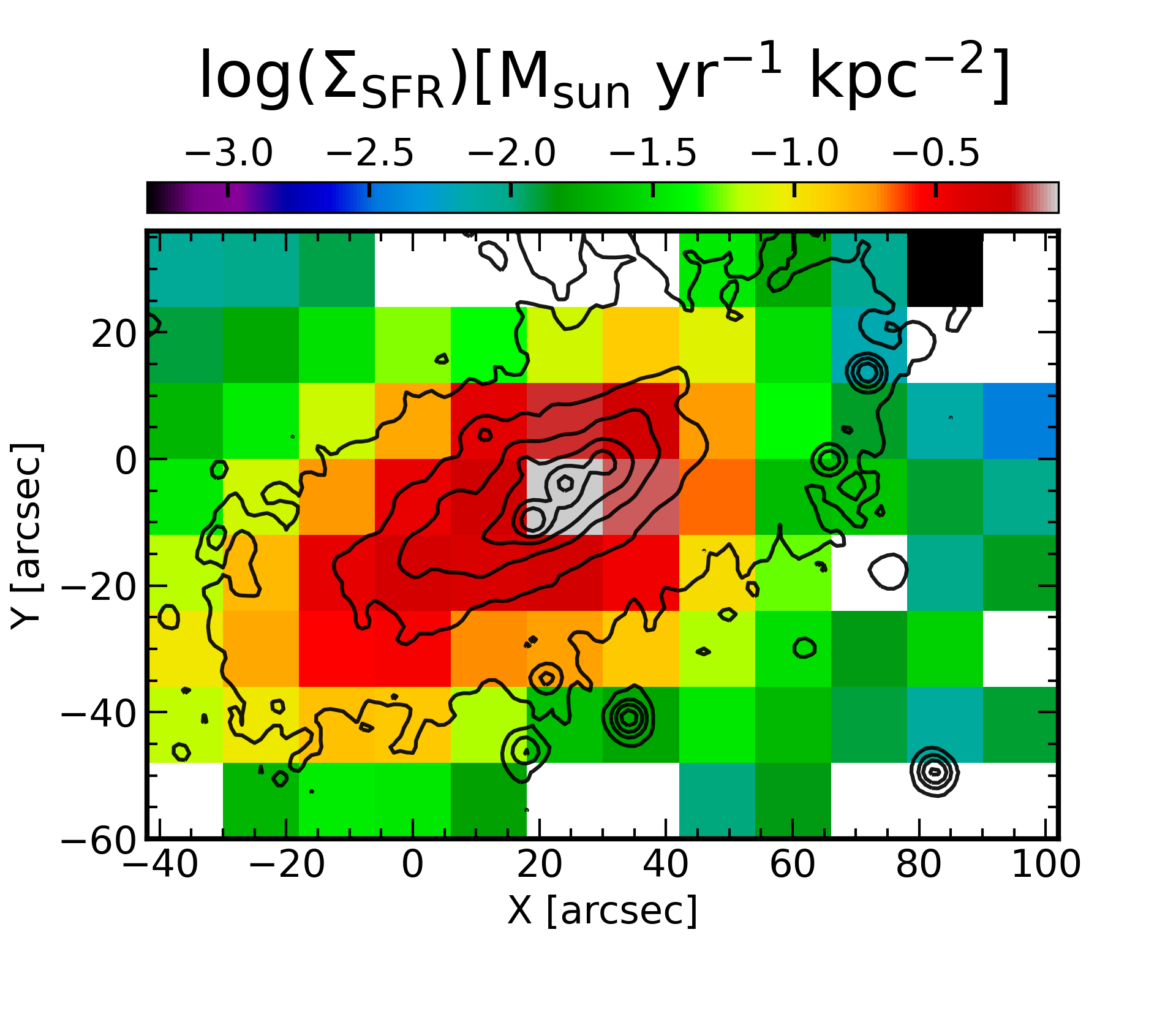}
    \includegraphics[trim={0 1.1cm 0 0cm},clip,width=0.33\linewidth]{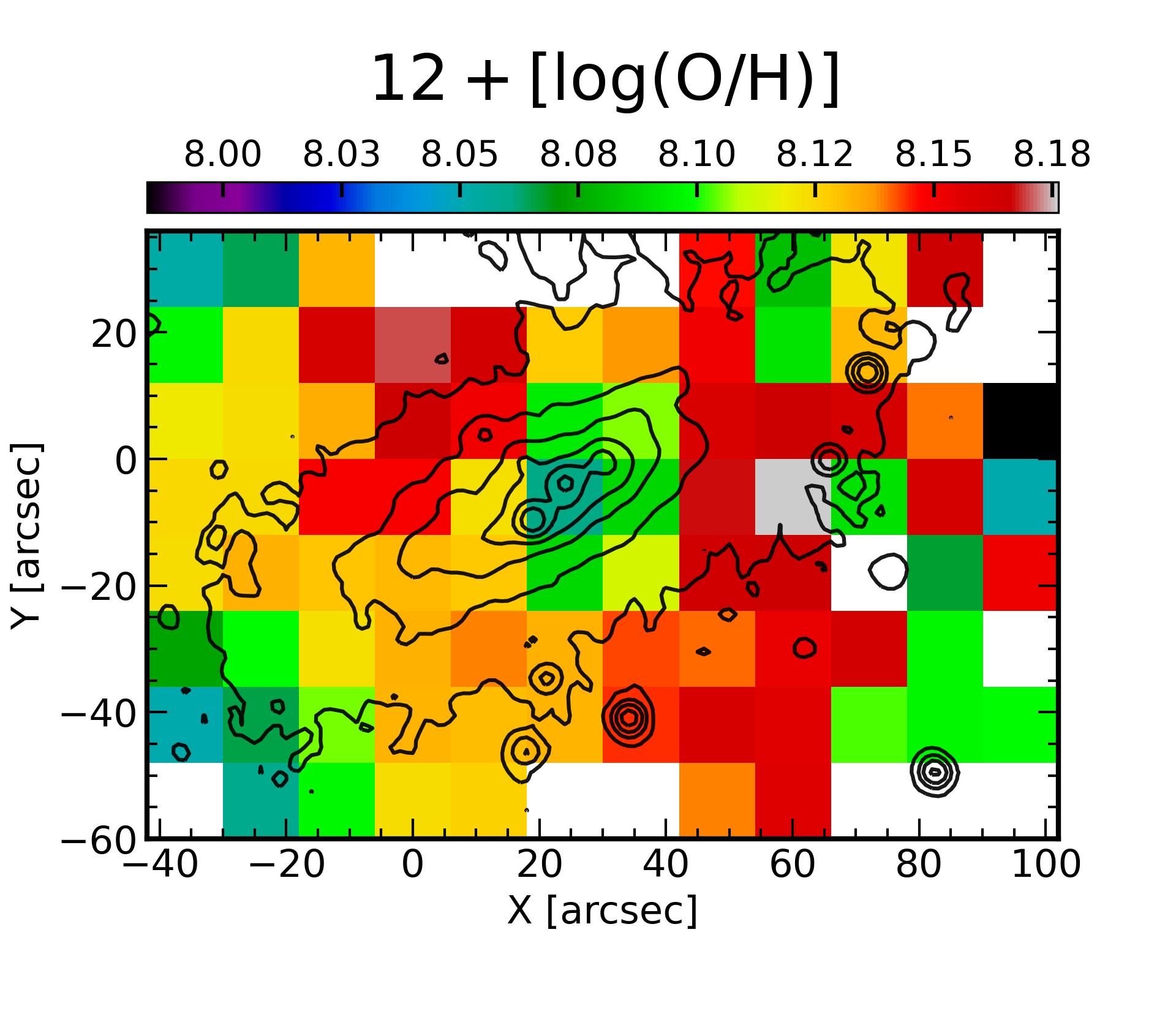}
    \caption{The stellar mass surface density, \sMs\ (left), the SFR surface density, \sSFR\ (middle) and the gas metallicity (right) maps. The black contours correspond to the galaxy structure in the J band from 2MASS. We show the 2 observed pointings in this map; after the S/N and BPT selection criteria, just a few  spaxels were recovered from the second pointing. Each spaxel in the three maps corresponds to a scale of $\sim$12" ($\sim$180 pc).}\label{maps1}
\end{figure*}

The above SFR formula considers a Kroupa IMF \citep{Kroupa2001}, which is very similar to a Chabrier IMF \citep{Chabrier2003}, with which our stellar masses are computed. Our star formation rate surface density map is shown in the middle panel of Fig. \ref{maps1}.

The gas metallicities are estimated using the S-calibration described in \citet{Pilyugin2016} which also use the emission line spectra. It is noteworthy that the computed oxygen abundances are  gas-phase abundances. They are based on the definition of the strong line ratios as $\rm N_2$ = $\rm {[N\,II]\ \uplambda 6548+ \uplambda 6584} /{\rm H}\upbeta$, $\rm S_2$ = $\rm [S\,II]\ \uplambda 6717+ \uplambda 6731 /{\rm H}\upbeta$ and $\rm R_3$ = ${\rm [O\,III]\ \uplambda 4959+ \uplambda 5007} /{\rm H}\upbeta$. Then, a relation is given according to the value of log $\rm N_{2}$. The so-called upper branch ($\log \rm N_{2} \ge -0.6$) is defined by the relation:

\begin{eqnarray}
       \begin{array}{lll}
     {\rm 12+log(O/H)}_{S,U}  & = &   8.424 + 0.030 \, \log (R_{3}/S_{2}) + 0.751 \, \log N_{2}   \\  
                          & + &  (-0.349 + 0.182 \, \log (R_{3}/S_{2}) + 0.508 \log N_{2})   \\ 
                          & \times & \log S_{2}   \\ 
     \end{array}
\end{eqnarray}

While the so-called lower branch ($\log N_{2} < -0.6$) by the relation:

\begin{eqnarray}
       \begin{array}{lll}
     {\rm 12+log (O/H)}_{S,L}  & = &   8.072 + 0.789 \, \log (R_{3}/S_{2}) + 0.726 \, \log N_{2}   \\  
                          & + &  (1.069 - 0.170 \, \log (R_{3}/S_{2}) + 0.022 \log N_{2})    \\ 
                          & \times & \log S_{2}   \\ 
     \end{array}
\end{eqnarray}

Throughout all the paper we use the term metallicity when referring to the gas-phase oxygen abundances. Our metallicity map can be seen in the right panel of Fig. \ref{maps1}.


\subsection{Dust mass estimates}\label{dustsec}
For the dust mass estimates, we follow \citet{Calzetti2018}, in which the global dust properties of a galaxy are used to compute the resolved dust masses, by fitting the entire Spectral Energy Distribution (SED) (to the available bands) and using the dust models from \citet{DraineLi2007}. 

To fit the complete SED, we use the Code Investigating GALaxy Emission \citep[CIGALE,][]{Burgarella2005,Nersesian2019}. CIGALE is a fitting code that reproduces the observed panchromatic photometry of galaxies, by means of stellar and dust emission models, imposing an energy balance to simultaneously account for UV/optical extinction, and IR dust emission. Among the most important input parameters are the Simple Stellar Population (SSP) set, the Star Formation History (SFH) and dust properties affecting both extinction and IR emission. The input parameters that we used were taken from DustPedia (see Table \ref{CIGALE_parameters}), and as input observations, we take the data of NGC 1569 also from DustPedia, in the mentioned bands at the end of \S \ref{SampleSelection} (from GALEX (UV) up to Hershel-SPIRE (FIR) at 250 $\mu\rm{m}$.) The optical photometry, also included in the SED, are centered at 5170, 5426, 5725 and 6080 {\AA} (obtained from our spectroscopic data, as described in \S \ref{SampleSelection}). CIGALE provides several output estimates (in which the dust mass can be found) according to the best computed SED fitting and a specific choice of the initial mass function. In this analysis, we use the IMF of \cite{Chabrier2003} and the synthesis models of \cite{BC2003}. 

We apply color corrections to all the Spitzer, Hershel-PACS and Hershel-SPIRE bands that we used, as \citet{Lianou2014} recommend. Spitzer bands are also corrected by aperture effects due to the condition of extended objects. 

Since NGC 1569 is in a sky projection to the galactic plane, it has a high extinction magnitude A$\rm_V$=1.58. Thus, we correct for galactic extinction all the GALEX and optical bands using the extinction curve of \citet{Cardelli1989}, and for the Spitzer bands, we use the curve extinction of \citet{Indebetouw2005}.

Finally, since all bands have different spatial resolutions, a re-projection and convolution to the worst resolution band (Hershel-SPIRE, 250 $\upmu$m) is applied in order to have a correct and fair comparison between all image data. The pixel size of such image data results in $\sim$6". Then, we also apply a 2x2 binning to the images to have a pixel size of $\sim$12" ($\sim$180 pc). The re-projection is performed using the kernels from \citet{Aniano2011}. 
Before the re-projection and convolution process, we mask the stars that appear in the field of view of the galaxy to avoid flux contamination. 

We run CIGALE on each pixel to get a dust mass surface density map (left upper panel of Fig. \ref{maps2}). We have estimated dust masses on each pixel twice: first, we use data up to Hershel-PACS 160 $\upmu\rm{m}$ (without any binning) in order to have several pixels and apply the method described in \S \ref{gasmassestimates} to estimate the DGR. Second, we also use data up to Hershel-SPIRE 250 $\upmu\rm{m}$ to estimate the dust masses used in the following sections with a 2x2 binning.
As mentioned, the specific input parameters that we use are shown in Table \ref{CIGALE_parameters}. Dust mass estimates using data up to Hershel-SPIRE 250\ $\rm{\upmu m}$ would be more sensitive to the bulk of the cold dust than the Hershel-SPIRE 160\ $\rm{\upmu m}$. For this galaxy the dust masses are very similar when using IR data up to the Hershel-PACS 160\ $\rm{\upmu m}$ wavelength, compared to using up to the Hershel-PACS 500\ $\rm{\mu m}$ wavelength (see Appendix \ref{Dustmass160250}).


\begin{table*}
\centering
\begin{tabular}{ccc}
\hline
\multicolumn{3}{c}{CIGALE Parameters}                                                                                                                                                                                                                                                                                                                                                                                                               \\ \hline
Module                              & Parameter                                                                                                                                                                                        & Value                                                                                                                                                                                                      \\ \hline
Delayed SFH                    & \begin{tabular}[c]{@{}c@{}}$\tau_{\rm{main}}$\\ Age$_{\rm{main}}$ \\ $\tau_{\rm{burst}}$ \\ Age$_{\rm{burst}}$ \\ $f_{\thinspace\rm{burst}}$ \\ SFR$_{\rm{A}}$\\\end{tabular}                                                                      & \begin{tabular}[c]{@{}c@{}}7000, 8000, 10000, 12000, 13000\\ 9500, 13000\\ 5.0,10.0,20.0, 50.0, 80.0, 110.0\\ 10,30,50,70,90,150,200,250,300,400\\ 0.1, 0.2, 0.3, 0.4, 0.5, 0.75\\ 1.0\\\end{tabular} \\ \hline
SSP \\ \cite{BC2003}                          & \begin{tabular}[c]{@{}c@{}}IMF\\ Stellar metallicity\end{tabular}                                                                                                                                        & \begin{tabular}[c]{@{}c@{}}\cite{Chabrier2003}\\ 0.004, 0.008\end{tabular}                                                                                                                                        \\ \hline
Dust attenuation \\ \cite{Calzetti2000} and \cite{Leitherer2002} & \begin{tabular}[c]{@{}c@{}}$E(B-V)_{\rm{young}}$\\     $E(B-V)_{\rm{old\thinspace factor}}$ \\      Central wavelength of the UV bump \\     Width (FWHM) of the UV bump\\     Amplitude of the UV bump\\     Slope delta of the power law\\     Filters \end{tabular} & \begin{tabular}[c]{@{}c@{}} 0.0075, 0.011, 0.017, 0.026, 0.038, 0.058, 0.087, 0.13, 0.20, 0.29, 0.44\\ 0.50\\ 217.5\\ 35.00.0\\ 0.0\\ 0.0\\ B$\rm B90$ \& V$\rm B90$ \& FUV\end{tabular}                     \\ \hline
Dust emission \\ \cite{Draine2014}               & \begin{tabular}[c]{@{}c@{}}$q_{\rm PAH}$ \\ $U_{\rm min}$\\ $\alpha$ \\ $\gamma$ \end{tabular}                                                                                                                             & \begin{tabular}[c]{@{}c@{}}0.47\\ 35\\  2.0\\ 0.015, 0.025, 0.035\end{tabular}                                                                                                  \\ \hline
\end{tabular}
\caption{Modules and physical parameter values used as input to CIGALE. $\tau_{\rm{main}}$, Age$_{\rm{main}}$, $\tau_{\rm{burst}}$ and Age$_{\rm{burst}}$ are expressed in 10$^{6}$yr. The SSP, dust attenuation models and dust emission model are described in the cited references.}
\label{CIGALE_parameters}
\end{table*}

\subsection{Gas mass estimates}\label{gasmassestimates}

In order to properly take into account the low gas metallicity of this galaxy, we followed the method presented in \citet{Leroy2011} and \citet{Sandstrom2013} to compute the DGR and the CO to molecular gas mass factor. This method combines the masses of the atomic (M$_{\rm{HI}}$) and molecular (M$_{\rm{H_2}}$) gas, together with the dust masses, to compute the DGR and $\upalpha_{\rm{CO}}$, where $\upalpha_{\rm{CO}}$ is the factor to convert observed CO luminosity (L$_{\rm{CO}}$) to molecular gas mass, M$_{\rm{H_2}}=\upalpha_{\rm{CO}}\ {\rm{L_{CO}}}$. Briefly, the method is based on the relation between these three masses:

\begin{equation}
\rm M_{\rm{gas}}=\frac{M_{\rm{dust}}}{\rm{DGR}}=\upmu_{\rm{gal}}(M_{\rm{HI}}+\upalpha_{\rm{CO}}\ L_{\rm{CO}}), 
\label{eq_masses}
\end{equation}

where $\upmu_{\rm{gal}}$ is the mean atomic weight, so the gas mass term includes the contribution of hydrogen and other elements, as defined in \citet{Remy2014}. We use  the Galactic mass fraction of Helium, Y$_{\odot}=0.270$  \citep{2009ARA&A..47..481A}, and the metallicity derived from the observations presented in this work, $12+\log(\rm{O/H})=8.12$, in order to estimate $\upmu_{\rm{gal}}$ and take into account the mass of the different elements.
Assuming a unique solution for similar metallicity regions to Eq. \ref{eq_masses}, one can find the best values of DGR and $\upalpha_{\rm{CO}}$ varying the value of $\upalpha_{\rm{CO}}$ while deriving the solution for DGR, using Eq. \ref{eq_masses} for all the pixels inside an specific region, and estimating the standard deviation of DGR of all those pixels inside that region. The best value of $\upalpha_{\rm{CO}}$, and then the solution, is that which minimizes the scatter of DGR.  

We obtain the HI intensity from the THINGS survey \citep{Walter2008} that we convert to HI gas mass using the equation:

\begin{equation}
\frac{\rm M_{\rm{HI}}}{\rm{M_{\odot}}}=2.36\times10^5 \left(\frac{\rm D}{\rm{Mpc}}\right)^2\times \frac{\rm F_{\rm{HI}}}{\rm{\rm{Jy\thinspace km/s}}}
\end{equation}

where F$_{\rm{HI}}$ is the flux measured from the moment-0 map, and D is the adopted distance according to Table \ref{NGC1569_properties}. 

The CO intensity is obtained using the data from the CARMA survey \citep{Rahman2012}. We derived the CO(1-0) moment-0 map from the CARMA datacube estimating the noise in the emission free channels, and looking for the peak which fulfills a signal to noise ratio larger than 3. We estimate the CO luminosity using the following equation \citep{Solomon1992}: 

\begin{equation}
\begin{split}
 \frac{\rm L_{\rm{CO}}}{\mathrm{K}\thinspace \mathrm{km}\thinspace \mathrm{s}^{-1}\thinspace \mathrm{pc}^2} & =
 3.25\thinspace \times 10^7 \left(\frac{\upnu_{\rm{rest}}}{\mathrm{GHz}}\right)^{-2}(1+\rm z)^{-1} \\
 &\times \left(\frac{\rm D}{\mathrm{Mpc}}\right)^2 \left(\frac{\rm F_{\rm{CO}}}{\mathrm{Jy}\thinspace \mathrm{km}\thinspace \mathrm{s}^{-1}}\right),
\end{split}
\end{equation}

where $\upnu_{\mathrm{rest}}$ is the rest frequency of the line ($115.27\thinspace \mathrm{GHz}$ in the case of CO(1-0)), D is the distance (seen in Table. \ref{NGC1569_properties}), z is the redshift of the galaxy (in this case it is negligible due to the proximity of the galaxy), and F$_{\mathrm{CO}}$ is the velocity-integrated flux measured from the moment-0 map.

Both, the  CO and HI flux maps were reprojected and convolved to the worst resolution when deriving the dust mass, which we choose to be that of PACS 160\ $\rm{\mu m}$ (FWHM = 11.2") in order to  have enough pixels in the CO flux map and being able to apply the method.

\begin{figure*}
    \centering
    \includegraphics[trim={0 1.1cm 0 0cm},clip,width=0.33\linewidth]{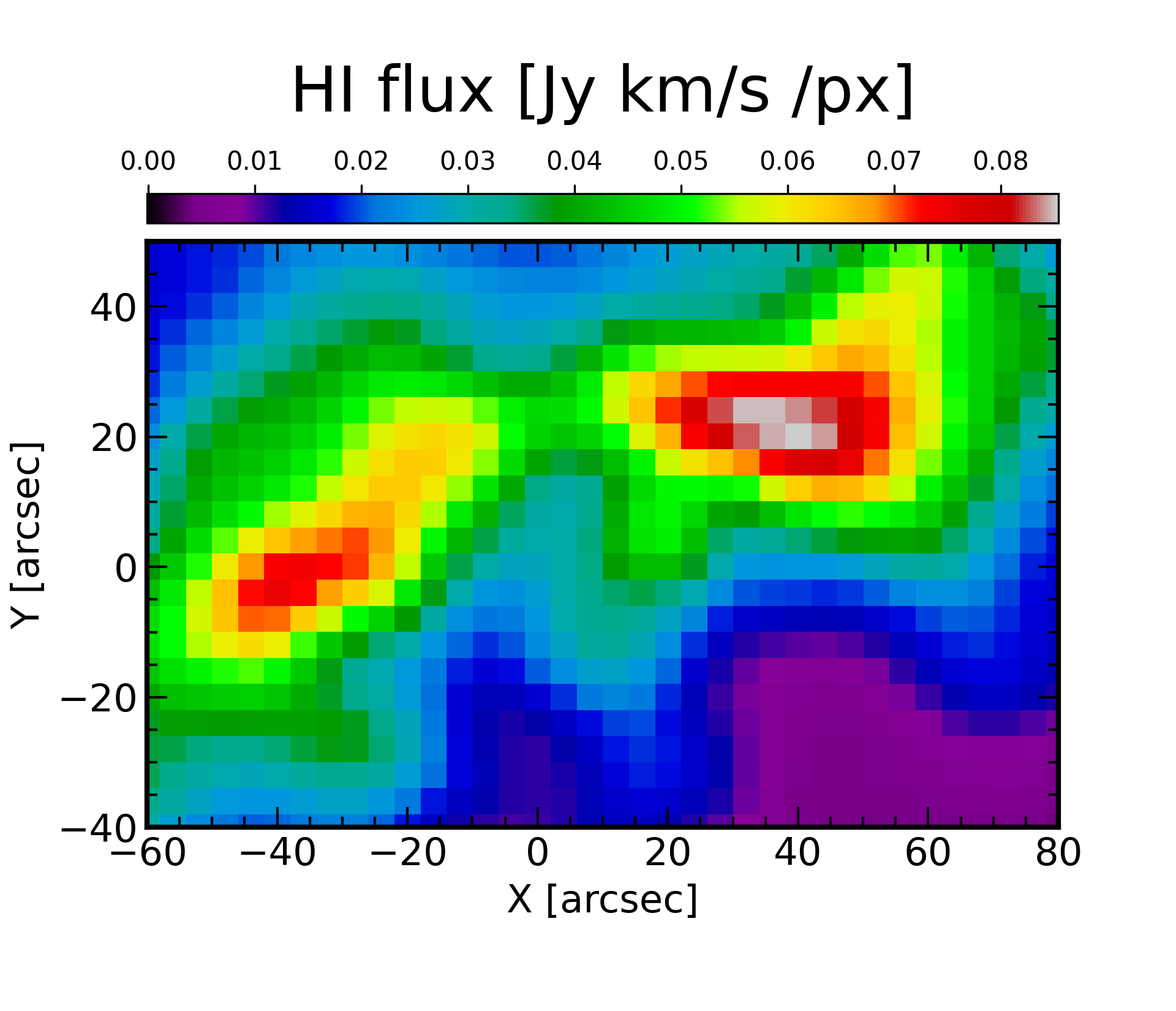}
    \includegraphics[trim={0 1.1cm 0 0cm},clip,width=0.33\linewidth]{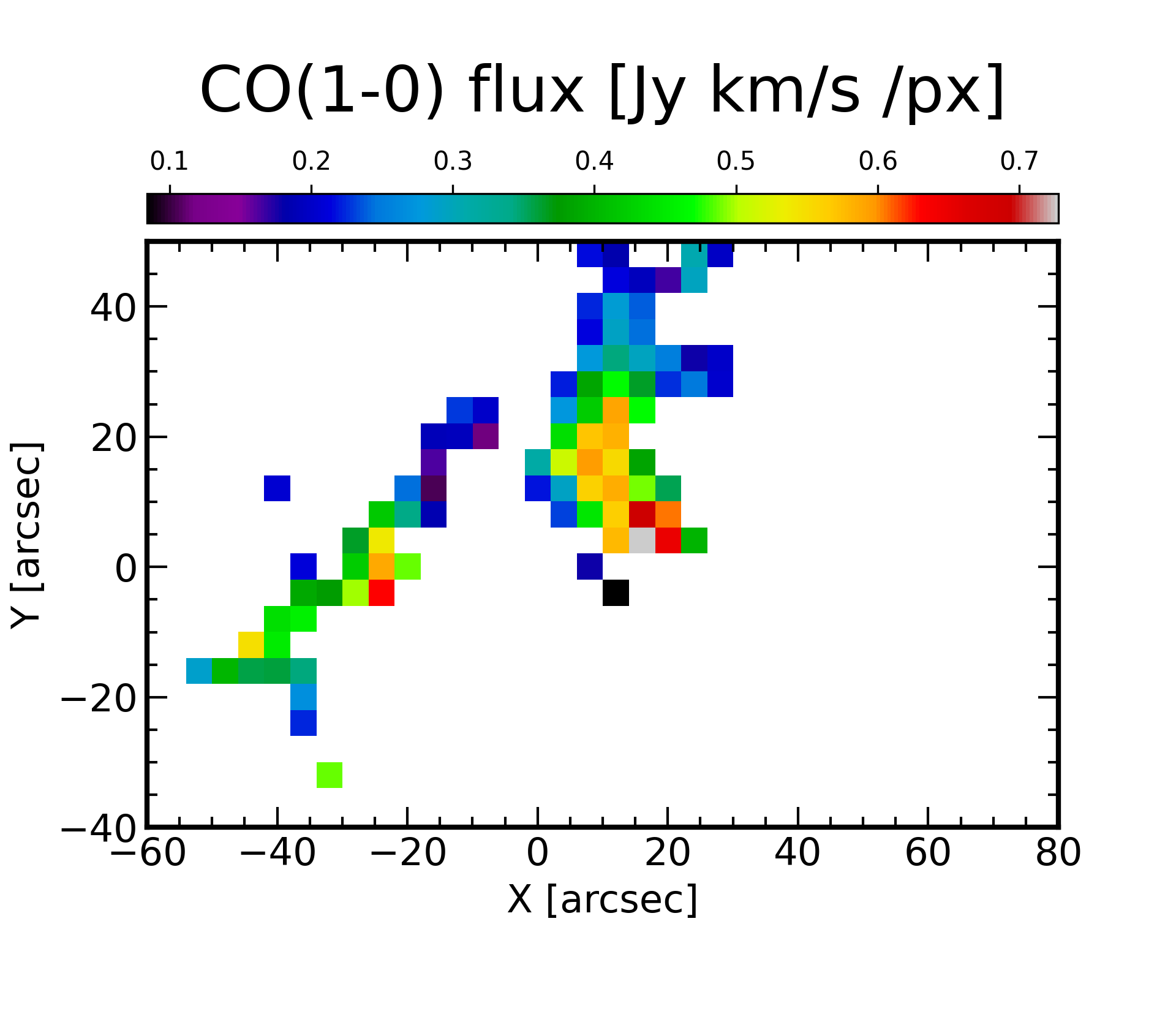}
    \includegraphics[trim={0 1.1cm 0 0cm},clip,width=0.33\linewidth]{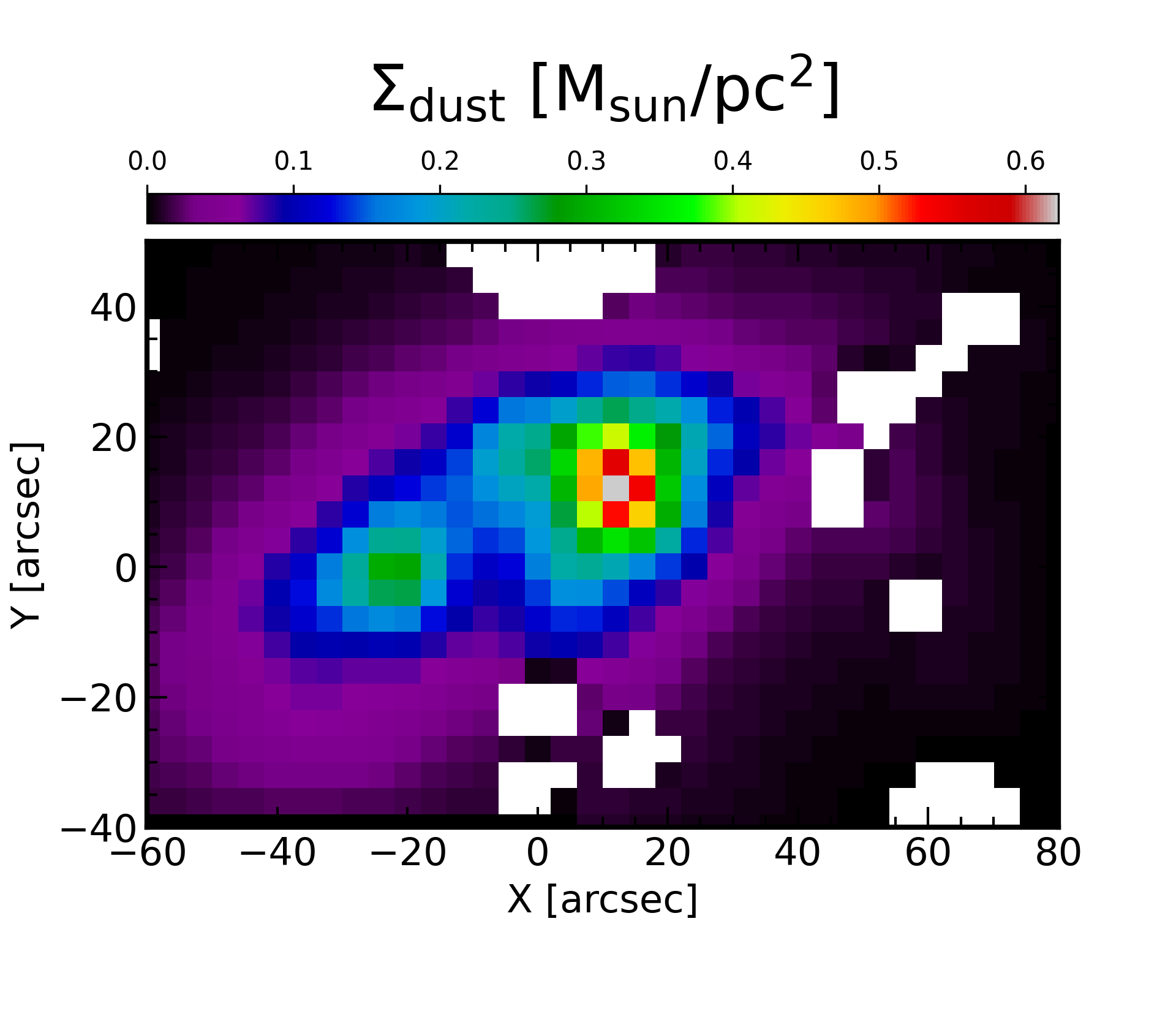}
    
    \caption{HI intensity (left), CO (1-0) intensity (middle), and dust mass surface density, $\Sigma_{\rm{dust}}$ (right), of NGC 1569. The three maps are reprojected and convolved to the worst resolution observation (FWHM$=11.2\rm{"}$, Hershel-PACS 160$\rm{\mu m}$) used in the dust mass estimation (see \S \ref{gasmassestimates}).  }\label{h_co_mdust_maps}
\end{figure*}

We show in Fig. \ref{h_co_mdust_maps} the HI flux (left), CO(1-0) flux (middle), and the dust mass surface density (right), $\Sigma_{\rm{dust}}$.
The CO map of NGC 1569 does not present much CO emission, probably due to its low metallicity. However, the CO emission is enough to compute the best DGR as a function of  $\upalpha_{\rm{CO}}$ using 84 pixel values, where we assume a unique solution of Eq. \ref{eq_masses}, since the metallicity of this galaxy is pretty constant (as we see in this work). 

\begin{figure}
    \centering
    \includegraphics[width=1\linewidth]{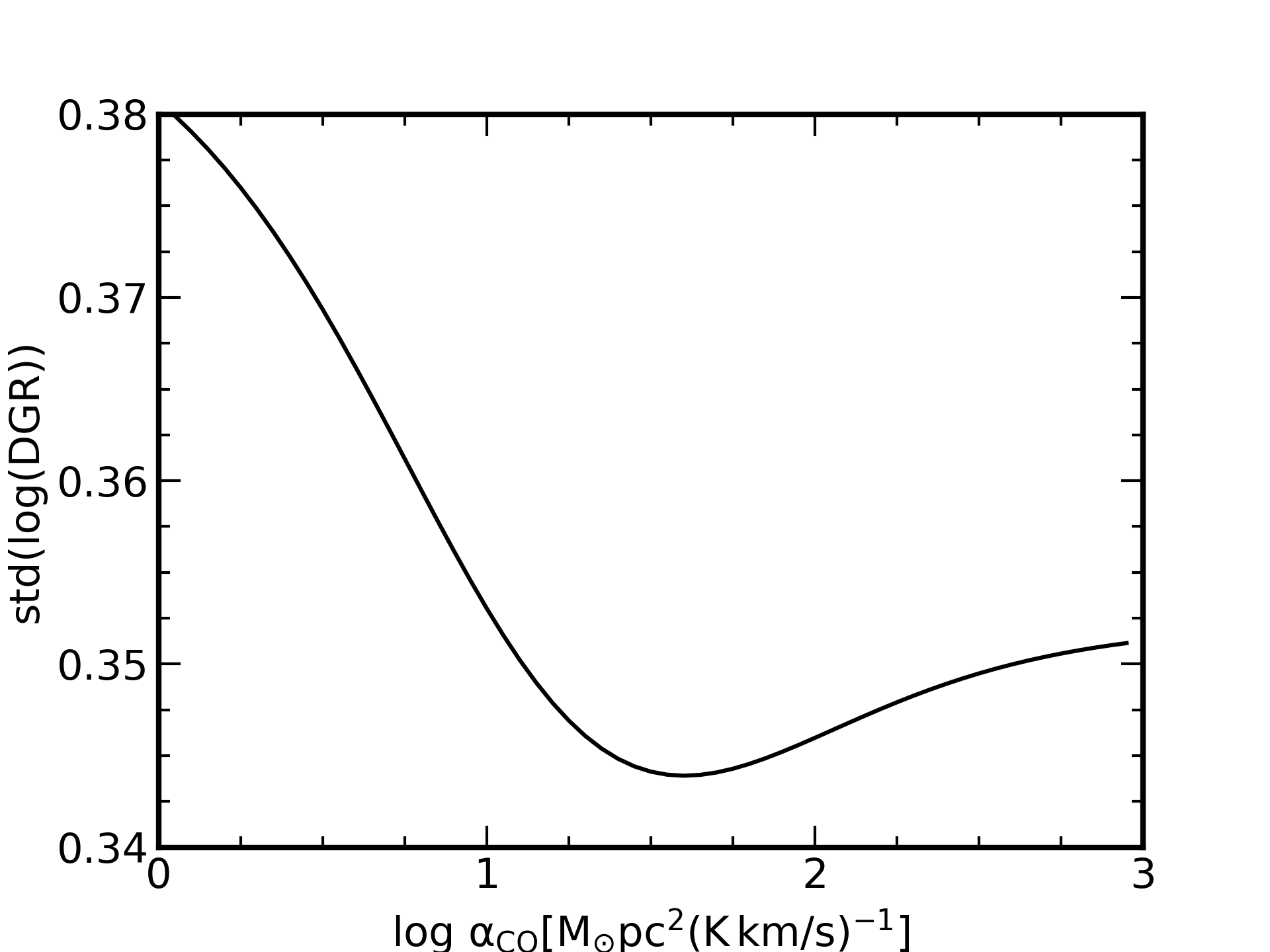}
    \caption{Standard deviation of the Dust-to-Gas Ratios (DGRs), as a function of the CO Luminosity-to-H$_2$ factor, $\upalpha_{\rm{CO}}$. The DGRs for each value of $\upalpha_{\rm{CO}}$ were estimated using Eq. \ref{eq_masses} for all the pixels where CO in NGC1569 is detected.}
    \label{stddgr}
\end{figure}

We show the result of the method in Fig. \ref{stddgr}, where we plot the standard deviation of the DGRs, versus $\upalpha_{\rm{CO}}$ for the whole set of pixels.
We obtain the best solution, {\it{i.e.}}, where the minimum value of the standard deviation of log(DGR) occurs. This solution corresponds to  $\log(\rm{DGR})=-3.08\pm0.18$  and $\log(\upalpha_{\rm{CO}})=1.6\pm0.4\thinspace\rm{M_{\odot}pc^2(K\thinspace km/s)^{-1}}$. The change of the DGR standard deviations is very small in comparison with the large variation of $\log(\upalpha_{\rm{CO}})$ in Fig. \ref{stddgr}. Therefore, the estimated uncertainty of $\upalpha_{\rm{CO}}$ shows a significant margin of error of 40\%. However, this does not affect our results since we will use the DGR value to compute gas masses presented in this work.
 
The uncertainties were obtained following  \citet{Leroy2011,Sandstrom2013}. We first estimated two different noises: i) adding random noise to the observed properties according to their errors, and ii) via bootstrapping. Each estimation is repeated 100 times independently. The errors for the two different noises were inferred with the standard deviation of the derived values, and finally we added in quadrature to obtain the final error.


The DGR and $\log(\upalpha_{\rm{CO}})$ reported values are in agreement with those expected for the metallicity of NGC 1569 (Z = 0.0044) \cite[{\it{e.g.}}, ][for \alphaCO]{Sandstrom2013,Bolatto2013} and \cite[{\it{e.g.}}, ][for DGR]{Remy2014,DeVis2019}. 

At the end, using the Eq. \ref{eq_masses} we compute the gas mass surface density map, displayed in the upper middle panel of Fig. \ref{maps2}.

\subsection{The baryonic mass, gas fraction, oxygen yields, SFE and depletion time estimates}

The physical properties that we obtained in previous sections (stellar mass, total gas mass and SFR) allow us to compute other galaxy properties following previous works \cite[{\it{e.g.}}, ][]{LaraLopez2019}. The baryonic mass is given by:

\begin{equation}
 \rm M_{\rm bar} = \rm M_{\rm \star} + \rm M_{\rm gas} 
\end{equation}

\begin{figure*}
    \centering
    \includegraphics[trim={0 0cm 0 0cm},clip,width=0.33\linewidth]{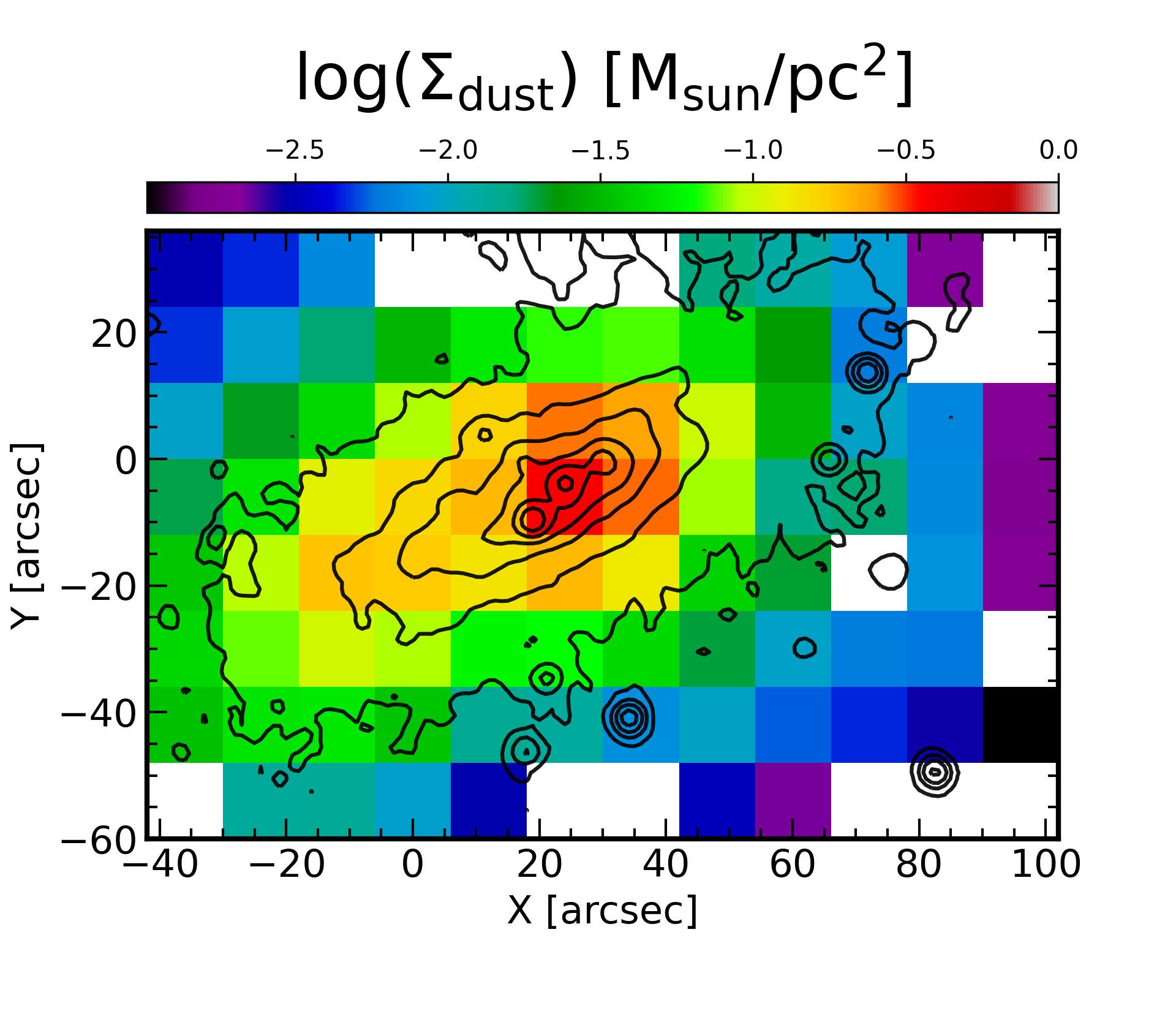}
    \includegraphics[trim={0 0cm 0 0cm},clip,width=0.33\linewidth]{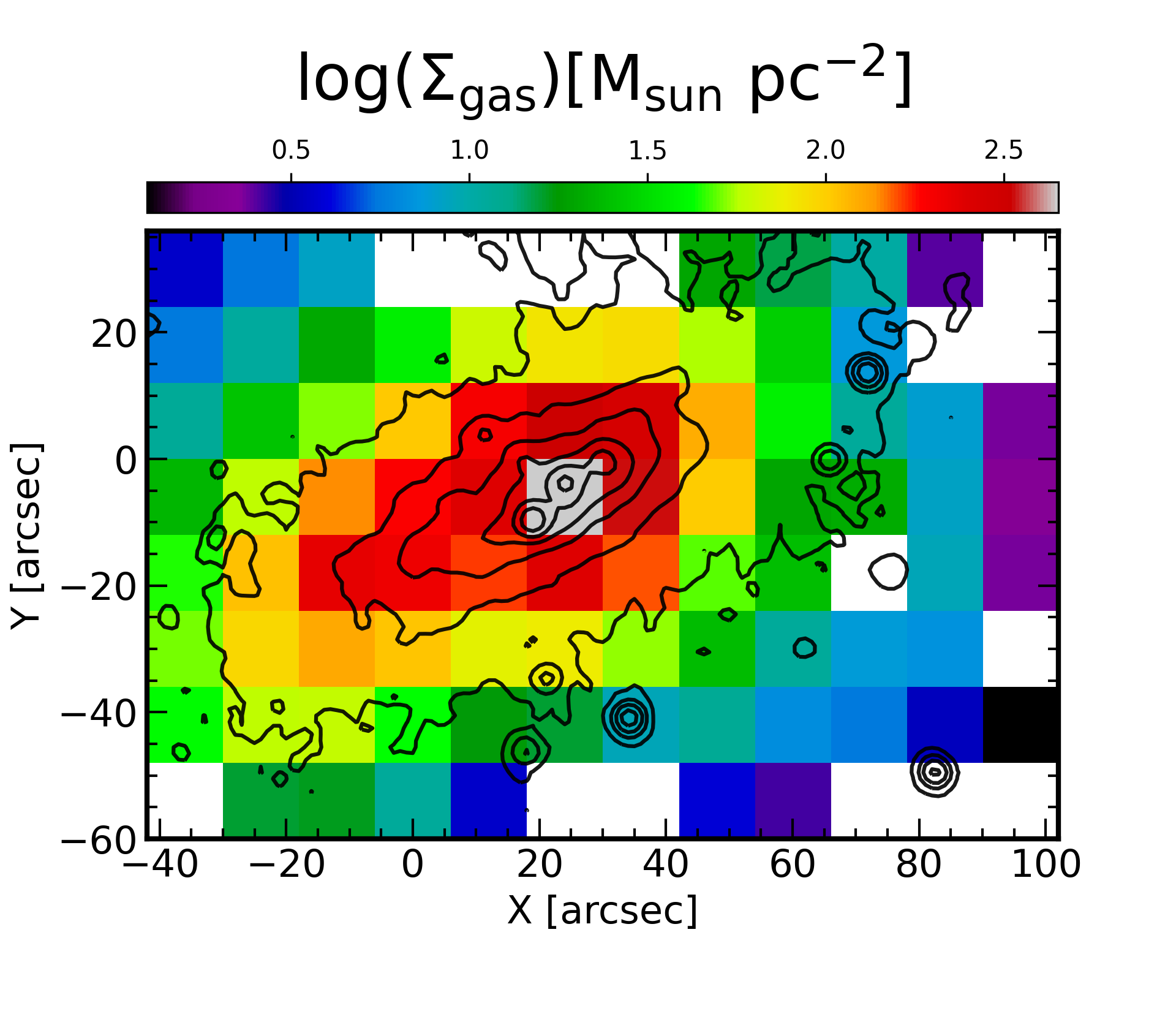}
    \includegraphics[trim={0 0cm 0 0cm},clip,width=0.33\linewidth]{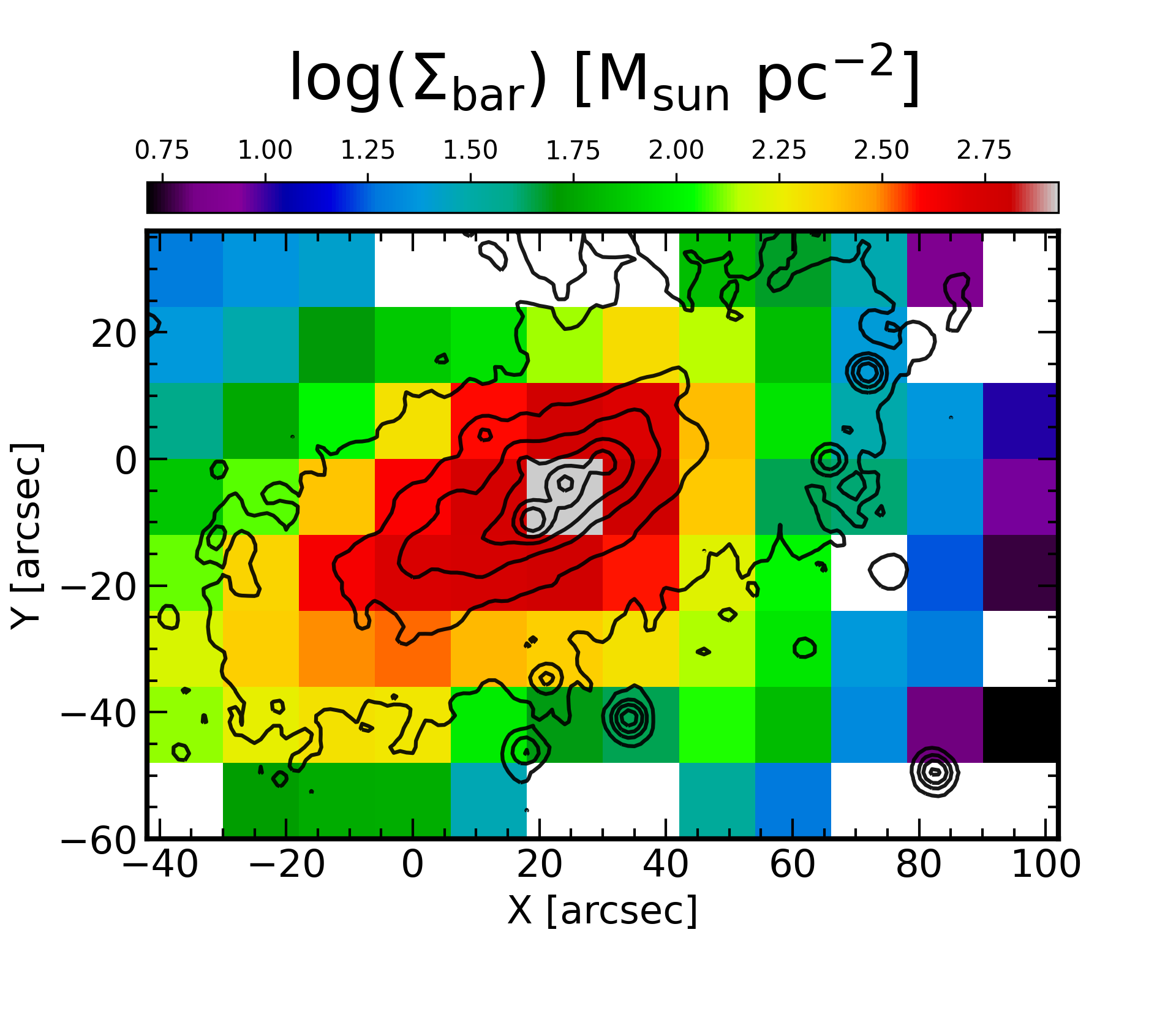}
    \includegraphics[trim={0 1.1cm 0 0cm},clip,width=0.33\linewidth]{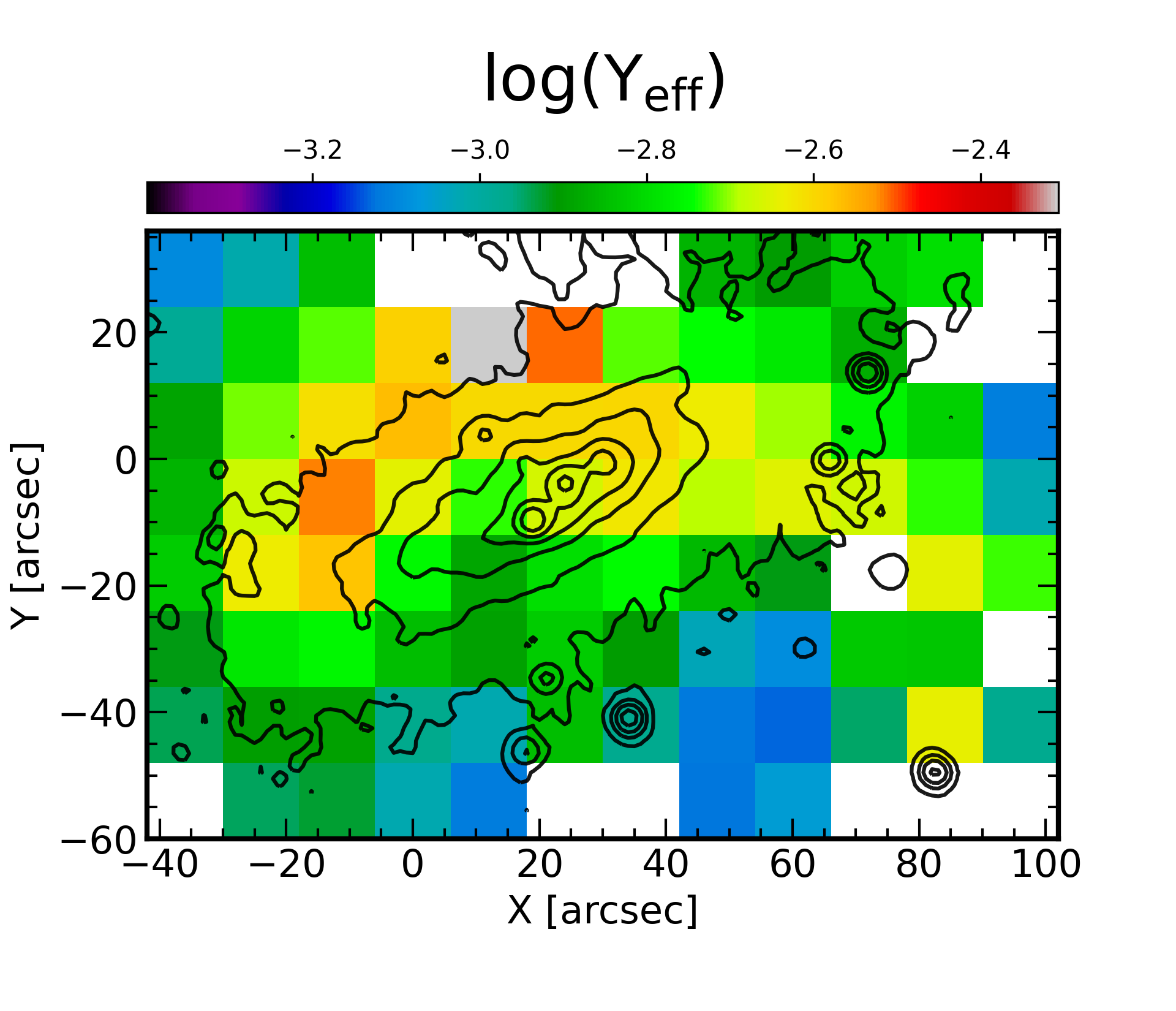}
    \includegraphics[trim={0 1.1cm 0 0cm},clip,width=0.33\linewidth]{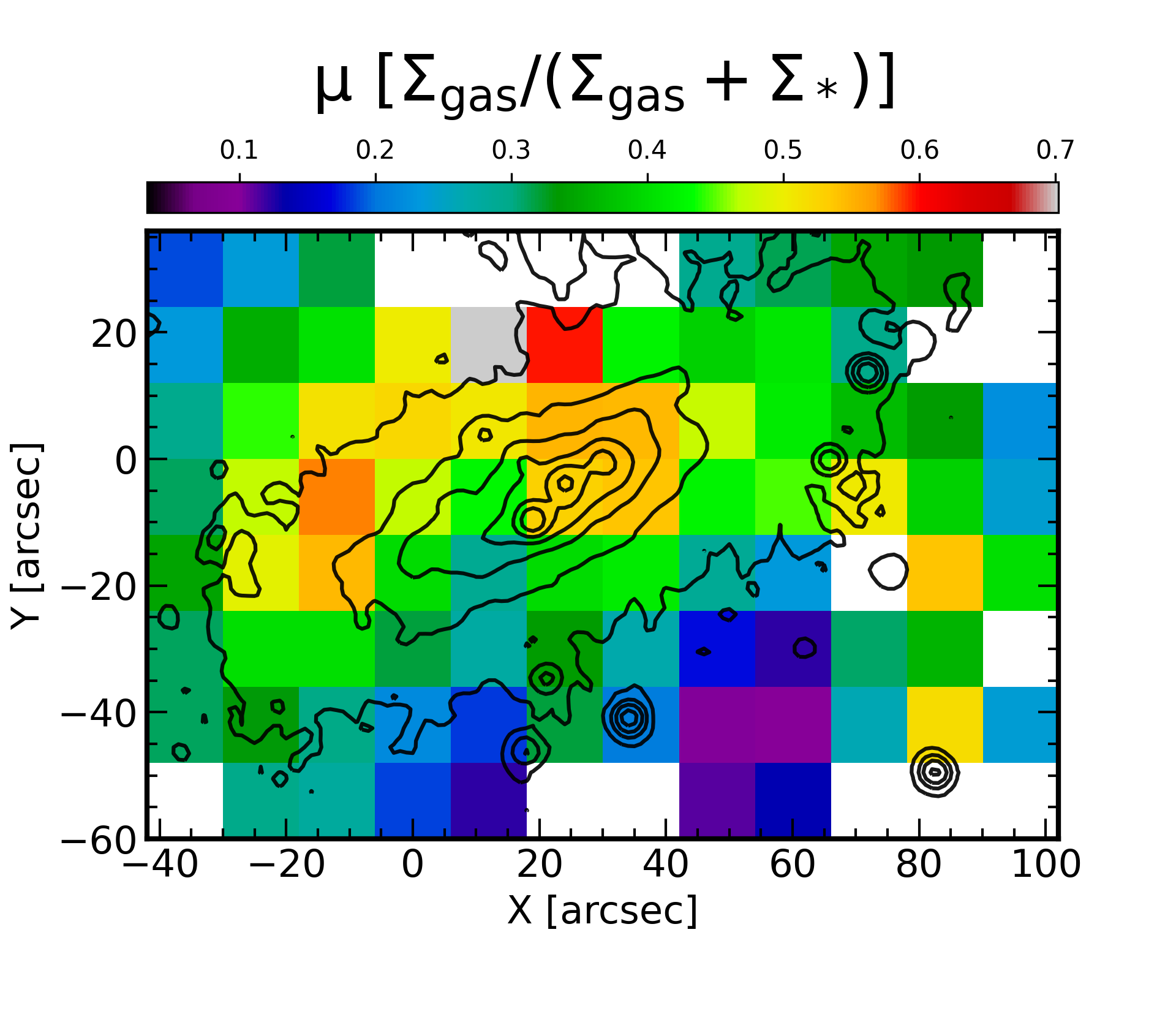}
    \includegraphics[trim={0 1.1cm 0 0cm},clip,width=0.33\linewidth]{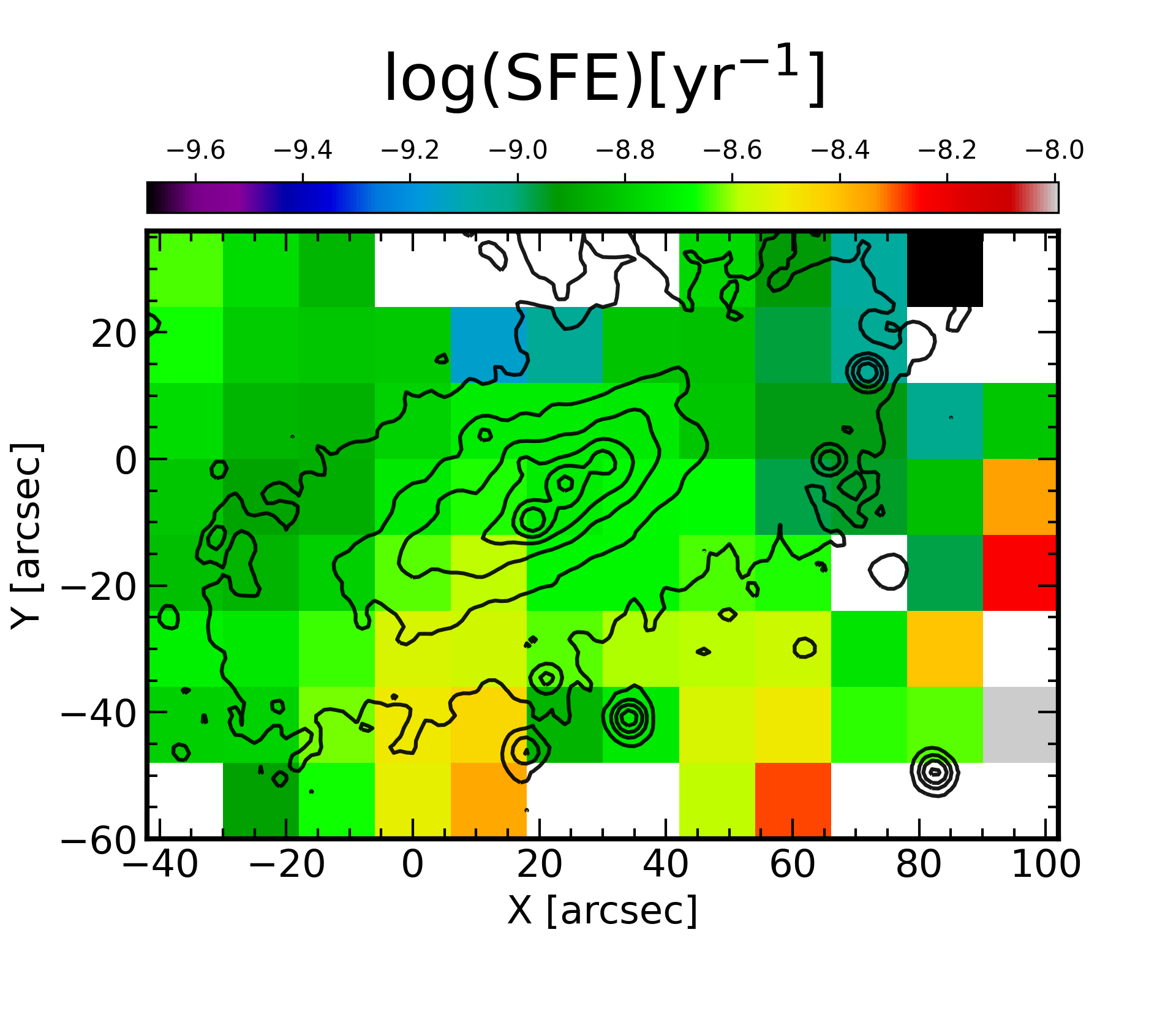}
    \caption{Upper panels: dust mass surface density \sMdust\ (left),  gas mass surface density \sMgas\ (middle) and  baryonic mass surface density \sMbar\ (right) maps. Bottom panels:  spatially resolved \Yeff\ (left),  spatially resolved gas fraction (middle) and  spatially resolved SFE (right) maps. In these maps we show the two observed pointings; after the S/N and BPT selection criteria, just a few  spaxel were recovered from the second pointing. The black contours correspond to the galaxy structure in the 2MASS J band. Each spaxel corresponds to a scale of $\sim$12" ($\sim$180 pc).} \label{maps2}
\end{figure*}

The $\rm M_{\rm bar}$ surface density map is shown in the right upper panel of Fig. \ref{maps2} 

Thus, the gas fraction is:

\begin{equation}
 \upmu =  \frac{\rm M_{\rm gas}}{\rm M_{\rm gas} + \rm M_{ \rm \star}} 
\end{equation}

For the effective oxygen yields (\Yeff), we follow the classic formalism described in \citet{Pagel1975,Searle1972}. This is:

\begin{equation}
 \rm Y_{\rm eff} =  \frac{\rm Z_{\rm gas}}{\ln(1/\upmu)} 
\end{equation}

We adopt the relation given by \cite{Garnett2002} for the value of Z$\rm _{gas}$, in which the oxygen abundance O/H is expressed in units of the number of oxygen atoms relative to hydrogen: 

\begin{equation}
 \rm Z_{\rm gas} = 12 \times \rm {(O/H)} 
\end{equation}

Finally, we define the SFE as:  

\begin{equation}
 \rm SFE = \frac{\rm SFR}{\rm M_{\rm gas}} 
\end{equation}

And the depletion time as:

\begin{equation}
 \rm t_{\rm dep} = \frac{\rm M_{\rm gas}}{\rm SFR} 
\end{equation}

Since our information is spatially resolved, we have maps for all of these properties. Thus, the effective oxygen yield map, the gas mass fraction map and the SFE map are shown in the left, middle and right bottom panel of Fig. \ref{maps2}, respectively. 

\section{Results}

\subsection{The local relationships between the stellar mass, gas mass, star formation rate and metallicity}\label{SR1_section}

In this section, we analyze the spatially resolved maps that we get for the different galaxy properties previously estimated. As mentioned in \S \ref{dustsec}, we convolved and reprojected our multiwavelength maps to the lowest resolution band (Herhsel-SPIRE 250 $\upmu\rm{m}$). The convolution and reprojection is also done to the emission lines maps, which are used to obtain the physical properties described in \S \ref{physicalproperties}. Thus, we establish a common and equivalent scale to compare all the properties with each other. A binning of 2x2 was also applied to avoid pixel correlations and to have a larger spatial scale. All the maps in Figs. \ref{maps1} and \ref{maps2} have a pixel size of 12" corresponding to a scale of 180 pc. We also compute the global galaxy properties by adding the values of the individual spaxels in the flux emission lines, and then apply the equations in \S \ref{SampleSelection}.

The spatially resolved map of stellar mass \sMs\  (left panel of Fig. \ref{maps1}) shows that the highest values of mass are located in the center of the galaxy, where the super stellar clusters (SSC) A and B are located. Around these regions the values of mass decreases in a gradual way up to the outer parts. We compute a global stellar mass of log(\Ms) = 8.6 M$_{\rm \odot}$. The SFR surface density map (middle panel of Fig. \ref{maps1}) shows similar features than the stellar mass map. Our estimation for the global SFR is log(SFR) = -0.32 \MMoy. The metallicity map does not show any pattern but it highlights some spaxels in the center of the galaxy with a lower metallicity than others in the outer part. For example, the central spaxel with a 12+log(O/H) $\sim$ 8.05 has a metallicity 0.1 dex lower than other spaxels in the outer zone with 12+log(O/H) $\sim$ 8.15. This slight change could be an evidence of outflows in the outer parts and a possible inflow just in the center of the galaxy, as we mention in the discussion of this work. In general, we have 65.3\% of spaxels with higher Z and 34.7\% lower Z, than the global value (12+log(O/H) = 8.12), respectively.

\begin{figure*}
    \centering
    \includegraphics[trim={0 0cm 0 0cm},clip,width=0.49\linewidth]{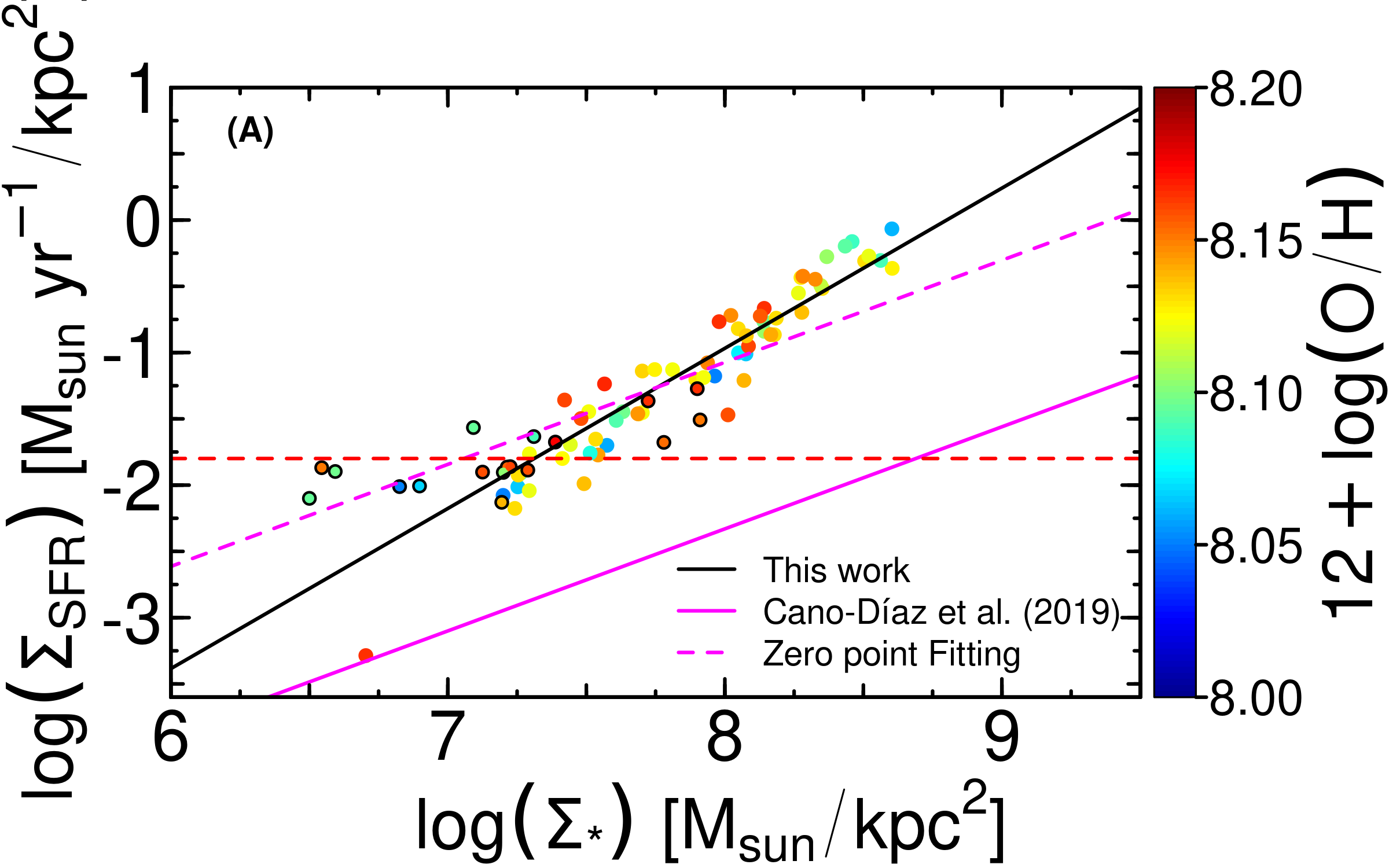}\hspace{0.15cm}
    \includegraphics[trim={0 0cm 0 0cm},clip,width=0.49\linewidth]{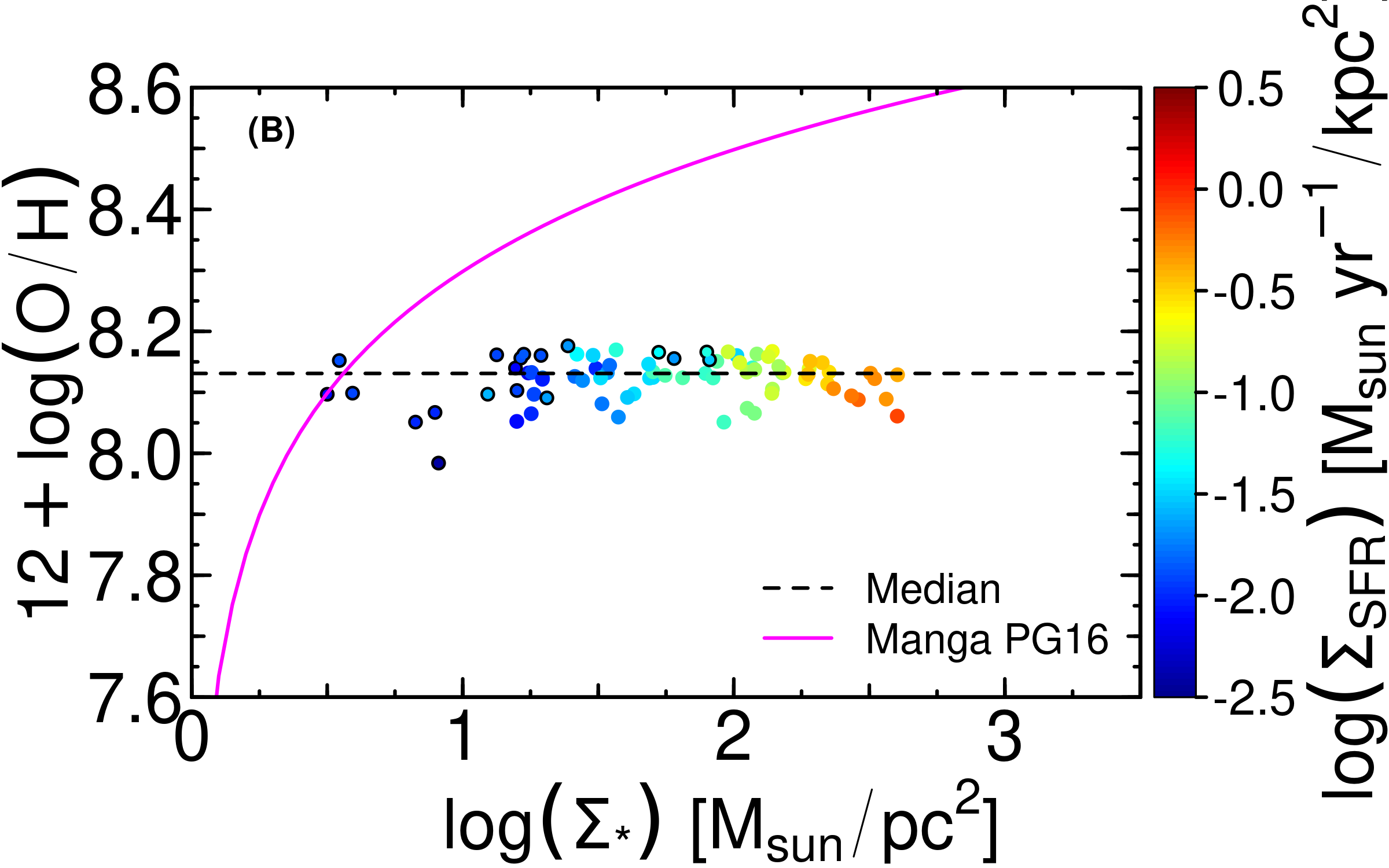}\hspace{0.15cm}
    \includegraphics[trim={0 0cm 0 0cm},clip,width=0.49\linewidth]{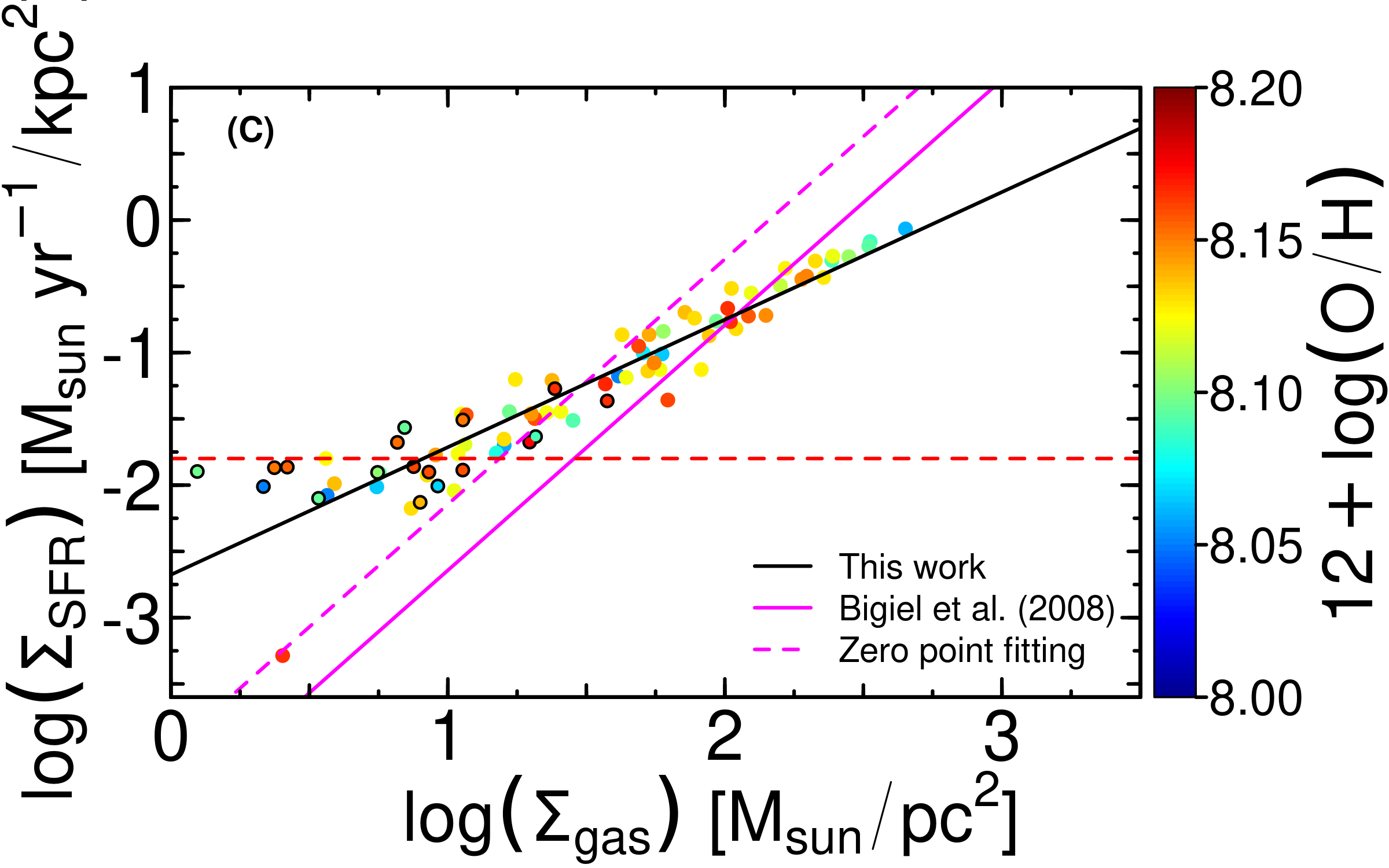}
    \caption{The \sMSFR\ (panel A), \sMZ\ (panel B) and the \sMgSFR\ (panel C) scaling relationships. The fits to the  data are shown in black solid lines. The purple  solid lines correspond to the fits of \protect\cite{Cano2019} and \protect\cite{Bigiel2008} for their respective relation. The purple solid line in the \sMZ\ relation is our own fitting for the MANGA galaxies (see Appendix \ref{Amet} for details). The horizontal dashed red line is the  limit below which there is a valid SFR. The circles with black contours correspond to the spaxels in the tail of NGC 1569. The data is color coded by the property shown in the colorbar of each panel.}\label{SR1plots}
\end{figure*}

The \sMSFR\ relation is shown in the panel A of Fig. \ref{SR1plots}, where it can be observed that we recover the well known correlation between the two properties. The solid magenta line in the figure is the power law fit reported for galaxies with SBc-Irr morphology in the MANGA survey \citep{Cano2019}. Our reported \sMSFR\ relation is color-coded by the gas metallicity, without a clear pattern or correlation. In the same figure, the dashed red line represents the limit below which the SFR value is lower than 10$^{-3}$ M$_\odot$ yr$^{-1}$. This is a typical value for which the SFR can become very uncertain due to poor sampling of the IMF. The use of synthesis stellar population (SSP) modeling to derive stellar population properties (among them the SFR), is intrinsically limited by the fact that the models are constructed using fully sampled IMFs. This means that by applying such models to an observed spectra, it is implicitly assumed that the mass distribution of stars that are producing the spectra are sufficiently complete \citep{Haydon2020}. It has been shown that this happens when the total stellar mass relative to this spectrum is of at least 10$^4$ M$_\odot$ \citep{El-Badry2017}. As the SFR prescriptions are calculated by assuming a constant SFR over $\sim$10$^7$ yr, this means that values below 10$^4$ M$_\odot$/10$^7$ yr$^{-1}$$\sim$ 10$^{-3}$ M$_\odot$ yr$^{-1}$ can possibly suffer from incomplete IMF sampling, resulting in an uncertain SFR determination.


The \sMZ\ relation can be seen in the panel B of Fig. \ref{SR1plots}. In this case we do not recover the classic polynomial shape of the global \Ms-Z relation. Instead, the metallicity spans across a range of $\sim$8.05-8.15 dex with respect to the stellar mass. This implies that spaxels have a metallicity variation of up to 0.1 dex (25\%). In general,  the metallicity is not constant throughout the galaxy but the variation is small. The plot is color-coded with the SFR and, as expected, the SF main sequence is seen ({\it{i.e.}}, \sMs\ scales with \sSFR).
 
The surface density of dust and gas mass maps (upper left and bottom panels of Fig. \ref{maps2}, respectively) share similar features with respect to the previous map of stellar mass, the highest values are in the center of the galaxy and the regions follows the structure of the galaxy. The global value of dust mass that we compute is log(\Mdust) = 5.3\Msun. Since the dust masses are multiplied by a DGR, it is expected to have the same spatial distribution in the dust and gas surface density maps. The global estimation for the gas mass that we computed is log(\Mg) = 8.33\Msun. We take the values of surface density of HI, H$\rm_2$ and the CO/H$\rm_2$ conversion factor reported by \cite{Kennicutt1998} to estimate the luminosity of CO. This value is multiplied by the $\upalpha_{\rm{CO}}$ factor that we get in \S \ref{gasmassestimates} to have our own value of H$\rm_2$ and thus we derive a new global gas mass surface density (log(\sMgas) = 1.49\Msun\ pc$^{-2}$). In comparison with the surface density gas mass estimates of \cite{Kennicutt1998}, log(\sMgas)=1.33\Msun\ pc$^{-2}$, the difference between these 2 estimations is 0.16 dex. We also derive a new value for log(\sHmol) = 1.05 \Msun\ pc$^{-2}$, in contrast with the old value log(\sHmol) = 0.10 \Msun\ pc$^{-2}$ reported by \cite{Kennicutt1998}. The large difference between these values is due to the used different \alphaCO. \cite{Kennicutt1998} assumed a constant Milky Way value $\log(\upalpha_{\rm{CO}})=0.65\pm0.4\thinspace\rm{M_{\odot}pc^2(K\thinspace km/s)^{-1}}$ for all galaxies in the sample, while we applied our own estimate in this work of $\log(\upalpha_{\rm{CO}})=1.6\pm0.4\thinspace\rm{M_{\odot}pc^2(K\thinspace km/s)^{-1}}$.

Figure \ref{SR1plots} (panel C) displays the spatially resolved KS relation color-coded by metallicity. We clearly recover the linear shape of the scaling relation, although with a different slope than previous works. It is also important to note the high dispersion towards the low mass range, specifically below the horizontal red line, which is due to the sampling problem of the IMF. When we compare our results with the previous work of \cite{Bigiel2008}, the difference between both slopes is clear. We discuss the possible origin of such differences in \S \ref{discuss_section}.    

In all the SRs shown in this section, the black contours dots represent the tail of NGC 1569 as can be appreciated in the second pointing of Fig. \ref{Pointings}. As mentioned, from the second pointing we just recover a few spaxels due to the selection criteria. Such tail is an important zone of the galaxy because it is a possible sign of interaction. However, the spaxels related to this zone, for the plots of this section, do not show a particular behaviour or strong feature to be taken into consideration.

The coefficient values of the fittings in this section are reported in Table \ref{Fittings}. 


\subsection{The local relationships between the gas fraction, baryonic mass, oxygen yields and star formation efficieny.}\label{SR2_section}

The baryonic mass surface density, gas fraction, \Yeff\ and SFE maps are shown in the lower panel of Fig. \ref{maps2}. The baryonic surface density does not show big differences with respect to the former maps. We estimate a global value for the baryonic mass of log(\Mb/\Msun) = 8.8.

In contrast with the maps in which the spaxels have high values in the center with a gradually decreasing distribution to the edges, the resolved \Yeff\ map (left lower panel of Fig. \ref{maps2}) does not show neither a pattern or a  structure. For this map, the highest values are in the center of the galaxy, and also along the horizontal plane of the map. Although it is a common practice to use
the gas-phase oxygen abundance to specify the galaxy metallicity in the construction of the different relations, in the estimation of the oxygen yield the total oxygen abundance (gas + dust) must be used. \cite{Peimbert2010} estimated the oxygen dust depletions in galactic and extragalactic HII regions finding that the fraction of oxygen atoms embedded in dust grains are a function of the oxygen abundance, being around 0.10 dex for metal-poor HII regions; thus, the total oxygen abundance for NGC 1569 is around 12 + log(O/H) = 8.22. Using the gas mass fraction $\upmu$ = 0.34, the integrated total oxygen yield is \Yeff = 0.00185 or log(\Yeff) = -2.73.

The resolved gas fraction map (middle bottom panel of Fig. \ref{maps2}) shows a relevant fact at first impression, the values go up to $\upmu=  0.5$, which implies that all the spaxels are probably dominated by stellar mass. Our estimate of the integrated gas fraction is $\upmu$ = 0.34, which reveals  its low gas fraction. 
Later, we discuss the origin of the low gas fraction values in  \S \ref{discuss_section}.

The surface density SFE map (right lower panel of Fig. \ref{maps2}) does not show a clear pattern, some regions clearly show low values of SFE (log(SFE) < -8.8 yr$^{-1}$), while the high values do not correspond to a particular part of the galaxy. The integrated property for this galaxy is log(SFE) = -8.65 yr$^{-1}$ which directly corresponds to a depletion time of log(t$_{\rm dep}$) = 8.65 yr ($\sim$450 Myr). 

\begin{figure*}
    \centering
    \includegraphics[trim={0 0cm 0 0cm},clip,width=0.49\linewidth]{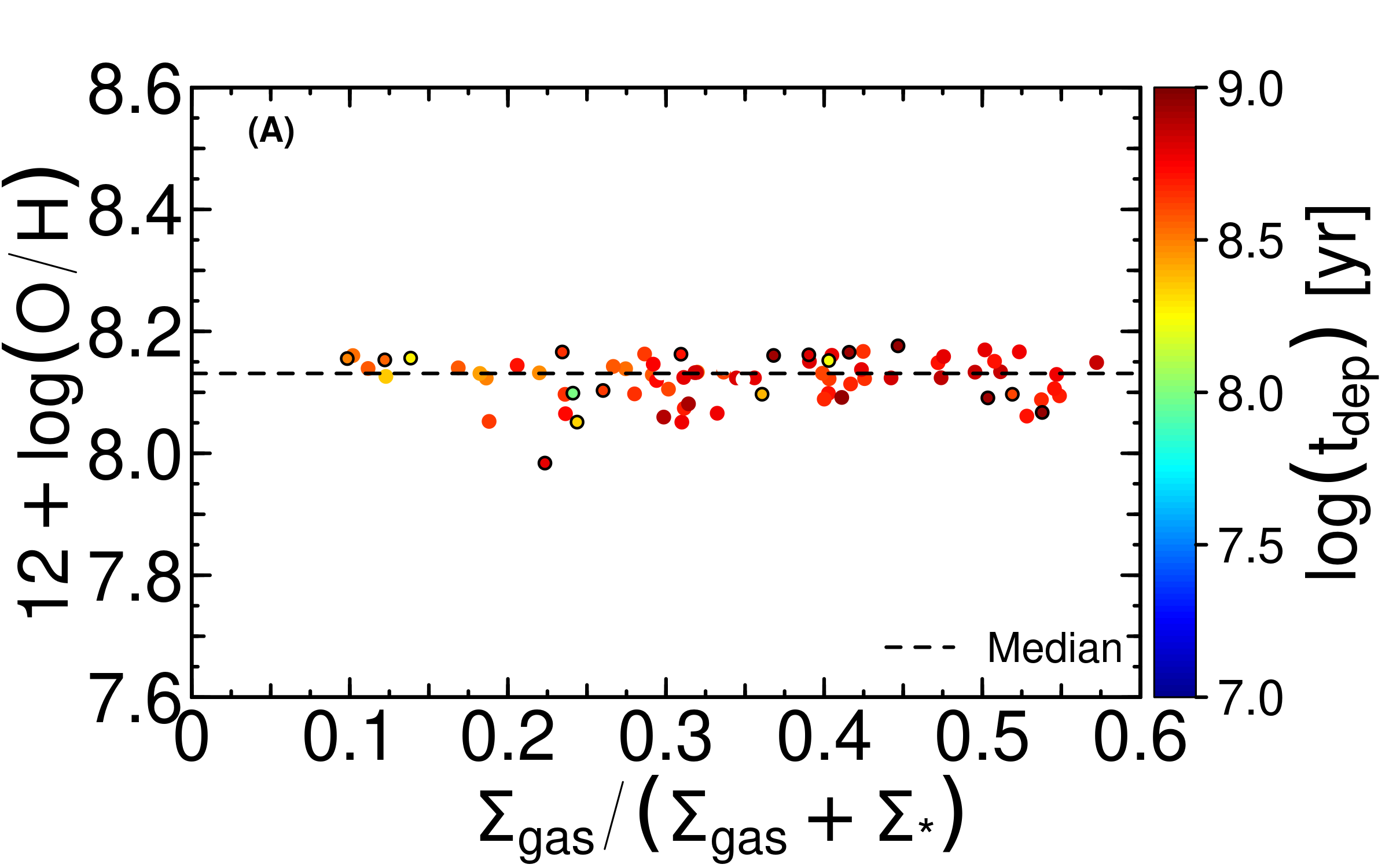}\hspace{0.15cm}
    \includegraphics[trim={0 0cm 0 0cm},clip,width=0.49\linewidth]{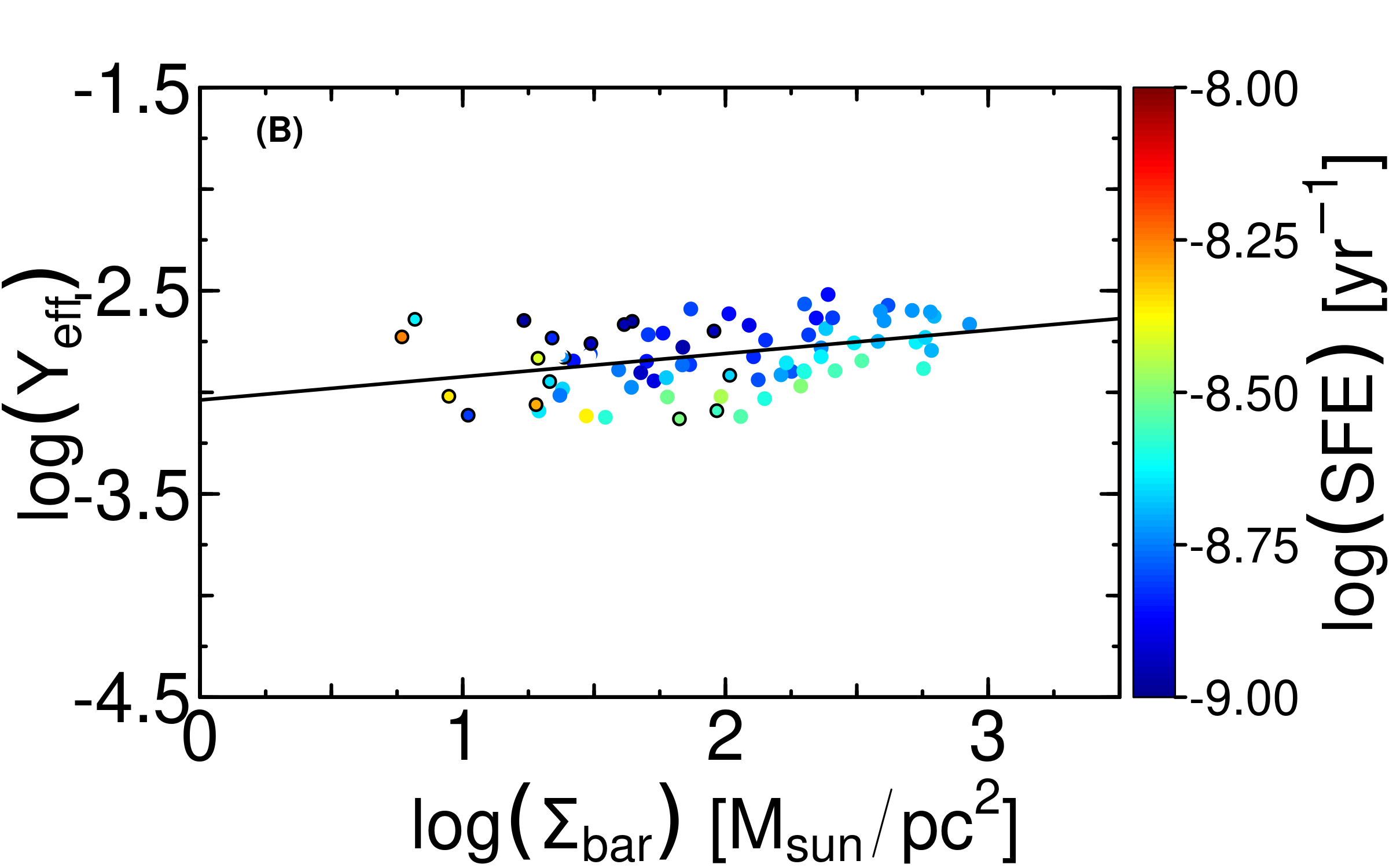}
    
    \includegraphics[trim={0 0cm 0 0cm},clip,width=0.49\linewidth]{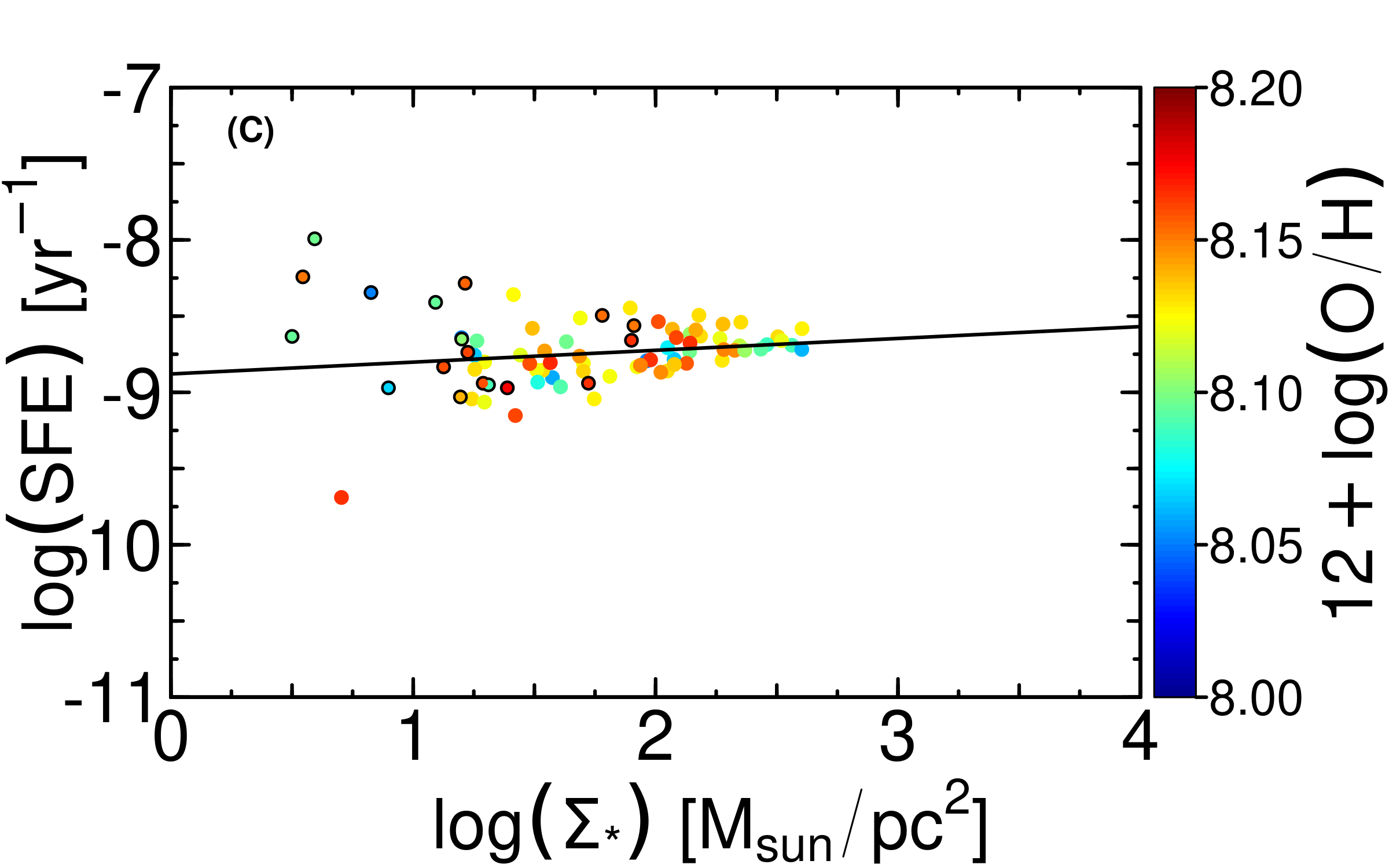}\hspace{0.15cm}
    \includegraphics[trim={0 0cm 0 0cm},clip,width=0.49\linewidth]{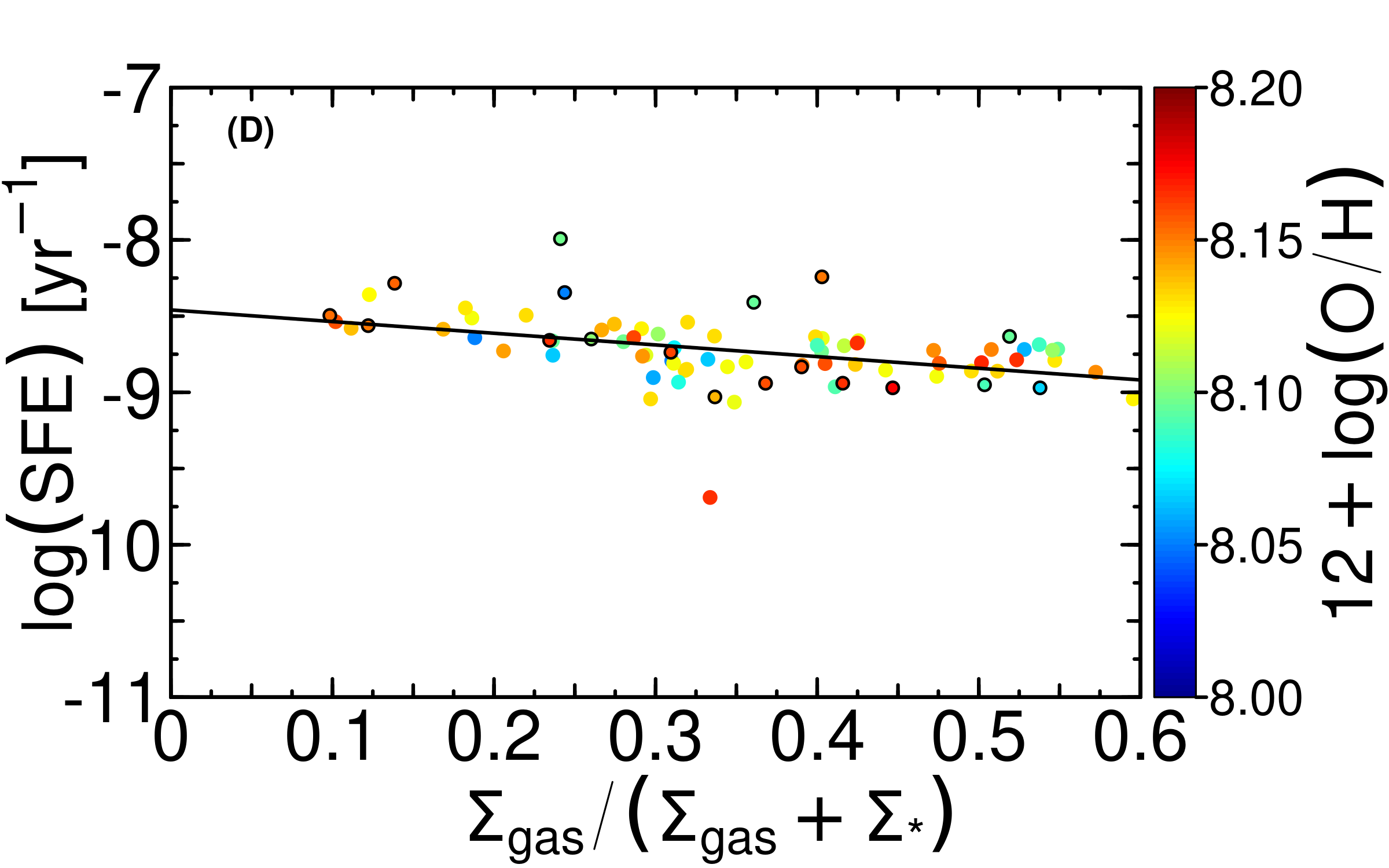}\hspace{0.15cm}
    \includegraphics[trim={0 0cm 0 0cm},clip,width=0.49\linewidth]{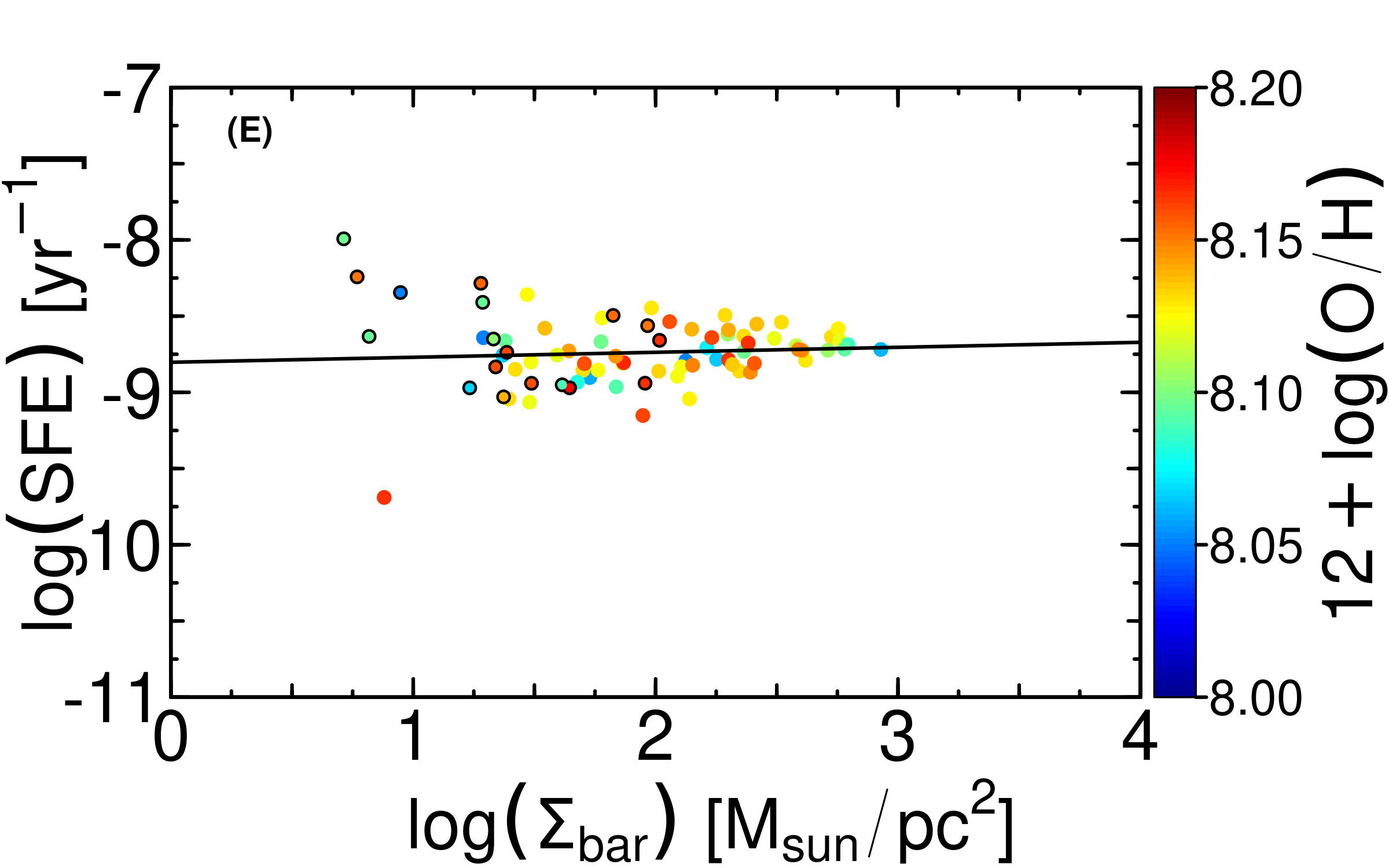}   
    \caption{The $\upmu$-Z (panel A), the \sMbar - \Yeff\ (panel B), the \sMs - SFE (panel C), the $\upmu$ - SFE (panel D) and the \sMbar - SFE (panel E) SRs. The fits to the observed data are shown as black lines.
     The circles with black contours correspond to the spaxels in the tail of NGC 1569. The data is color coded by the properties shown in the colorbar.}\label{SR3_plots}
\end{figure*}

The relation $\upmu$-Z (panel A of Fig. \ref{SR3_plots}) is flat, similar to the \sMZ~relation. As  mentioned earlier, all the data are concentrated towards $\upmu$ < 0.5 which implies a stellar dominated regime across the whole galaxy. The points with black contours (tail of NGC 1569) do not show a special location in the plot with respect to the rest of the data. A high percentage of the data correspond to values of log(t$_{\rm dep}$) > 8.5 yr.

For the relation \sMbar-\Yeff\ (panel B of Fig. \ref{SR3_plots}), we recover a correlation between both properties. The points corresponding to the tail of the galaxy do not show evidence of a particular location. 

The plots of Fig. \ref{SR3_plots} show the relation between \sMs (panel C), $\upmu$ (panel D) and \sMbar\ (panel E) with the SFE. There is no special dependence with the metallicity for the three plots, indeed, since the metallicity range is of only 0.2 dex, there is little dependence with metallicity. For the relations \sMs - SFE and \sMbar - SFE, we appreciate an important feature. In general, if we do not consider the points of the tail (points with black contours), we have a mostly flat relation. However, when we  consider only the points with black contours, our data suggest a possible anticorrelation between both properties. These two relations are the only ones in which the points corresponding to the tail have a clear difference with respect to the others. It is clear how such points have a limit up to log(\sMs) < 2.0 and log(\sMbar) < 2.25. Finally, we see a negative correlation between $\upmu$ and SFE. The regions of the tail are distributed throughout the whole range of $\upmu$ with no special concentration.  


The coefficient values of the fittings of all the SRs mentioned in this section are shown in Table \ref{Fittings}.

\begin{table*}
\centering
\begin{tabular}{ccccccccc}
\hline
\multicolumn{9}{c}{Fitting coefficients} \\ \hline
 & \sMSFR & \sMgSFR & \sMbar - \Yeff & \sMs - SFE & $\upmu$ - SFE & \sMbar - SFE  & \sMHISFR & \sMmolSFR \\ \hline
m & 1.21 & 0.96 & 0.11 & 0.08 & -0.77 & 0.03 & 1 & 0.58  \\ \hline
y$_0$ & -10.64 & -2.68 & -3.04 & -8.88 & -8.46 & -8.8 & -2.55 & -1.62  \\ \hline
$\upsigma_{\rm SD}$ & 0.64 & 0.64 & 0.17 & 0.22 & 0.22 & 0.22 & 0.64 & 0.42  \\ \hline
x$_{\rm RMS}$ & 1.44 & 1.44 & 2.82 & 8.74 & 8.74 & 8.74 & 1.44 & 0.92  \\ \hline
x$_{\rm RMSE}$ & 0.25 & 0.22 & 0.16 & 0.22 & 0.2 & 0.22 & 0.58 & 0.13  \\ \hline
$\uprho$ & 0.92 & 0.94 & 0.36 & 0.08 & -0.46 & 0 & 0.39 & 0.95  \\ \hline
p-value &  $<$ 2.2$\times10^{-16}$ & $<$ 2.2$\times10^{-16}$ & 9.9$\times10^{-4}$ & 0.48 & 1.7$\times10^{-5}$ & 0.99 & 3.3$\times10^{-4}$ &  $<$ 2.2$\times10^{-16}$ \\ \hline
\end{tabular}
\caption{The slopes (m), zero points (y$_0$), standard deviations ($\upsigma_{\rm SD}$), root mean squared (x$_{\rm RMS}$), root mean squared error (x$_{\rm RMSE}$) and Pearson correlation coefficient ($\uprho$) of all fittings reported in this work. The last row corresponds to the p-value of the Pearson correlation.}
\label{Fittings}
\end{table*}

Finally, it is worth to mention that the systematic oﬀsets that can be observed depend on the method used for metallicity estimates \citep[{\it{e.g.}}, ][]{Zurita2021,Groves2023}. The temperature inhomogeneities within the gas can explain the differences between strong line methods \citep{MendezD2023} and thus the proper corrections can be taken into consideration \citep{PenaG2012a,PenaG2012b}. Our work relies in the S-calibrator that takes into account the ionization parameter, the nitrogen-to-oxygen ratio and produces abundances compatible with those computed by the direct T$_{\rm e}$ method. Since our metallicities are mostly constant across all the galaxy with a very low dispersion ({$\rm \sigma \sim$} 0.05), we do not expect changes in the slopes of our SRs.

\section{Discussion}\label{discuss_section}

\subsection{Global properties of NGC 1569}\label{GP_discuss_section}

NGC 1569 is an interesting case of study for its high star formation rate \citep{Kennicutt1998}, low metallicity \citep{Kobulnicky1997} and complex star formation history \citep{Angeretti2005}. Additionally, an extended emission has been observed beyond the optical galaxy data as warm and hot ionized and atomic hydrogen gas \citep{Lianou2014}. NGC 1569 is part of a small group of galaxies, that experienced a possible recent interaction with its close companion \citep{Johnson2013}.

\begin{figure*}
    \centering
    \includegraphics[scale=0.4]{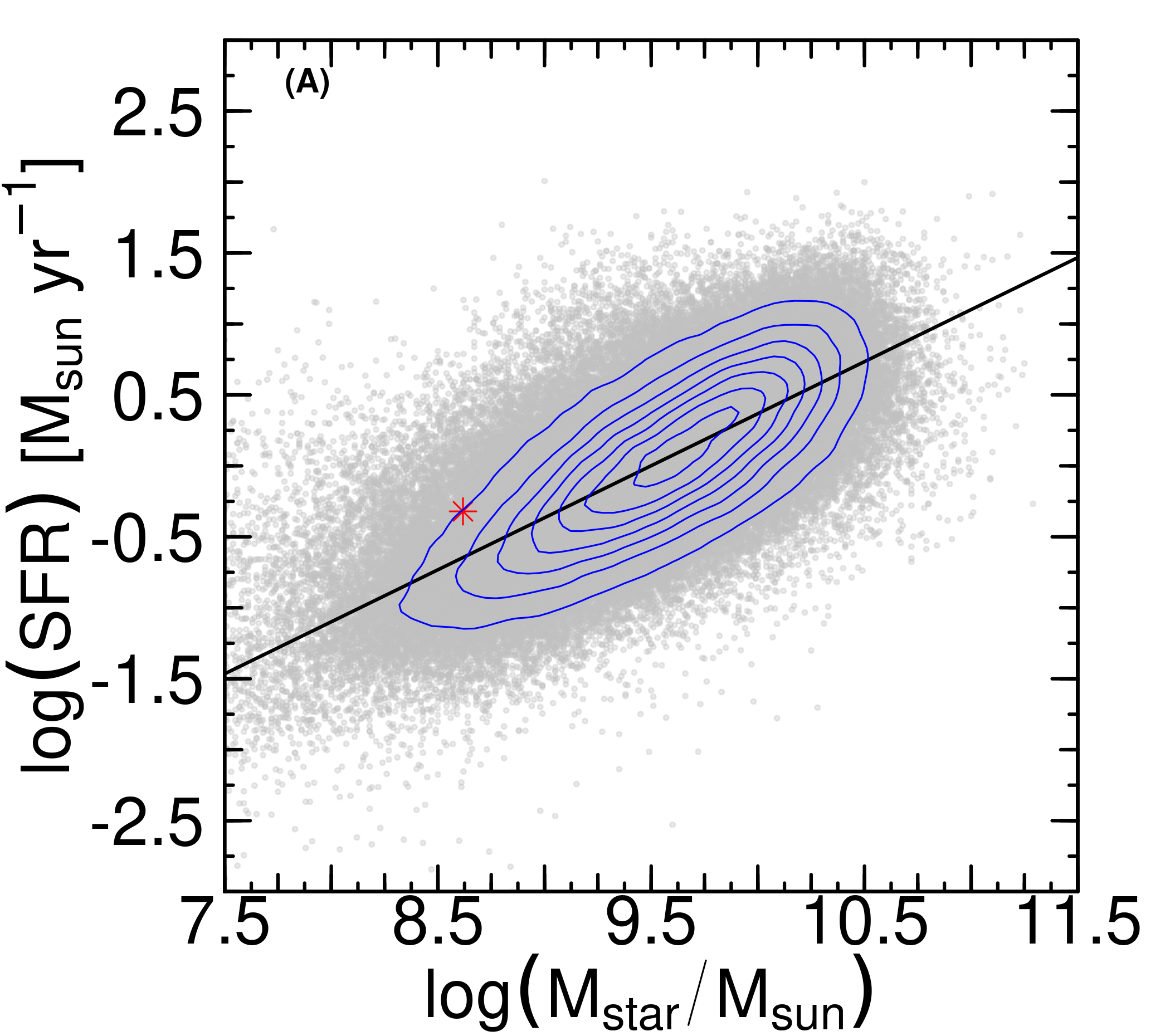}
    \includegraphics[scale=0.4]{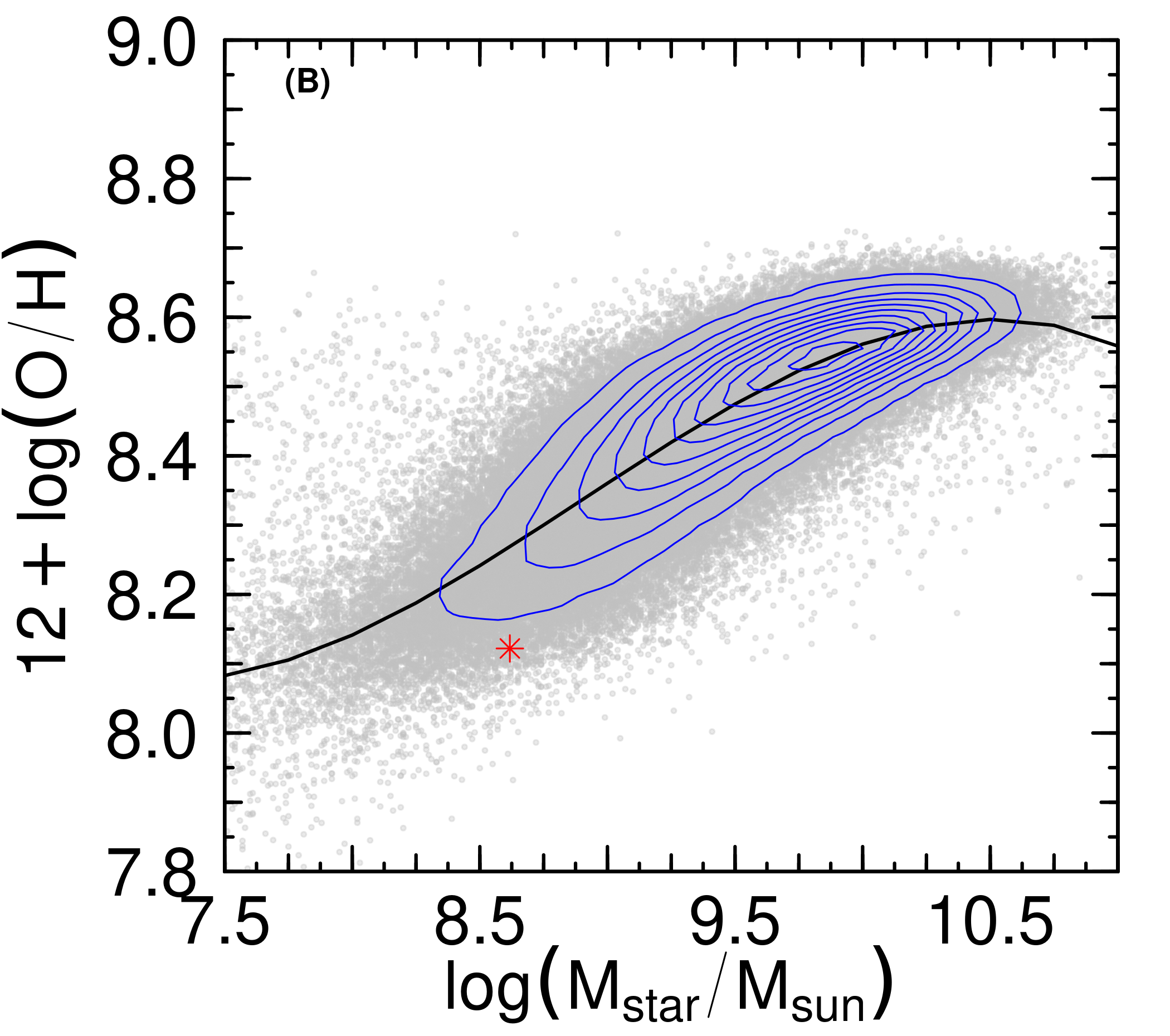}
    \includegraphics[scale=0.4]{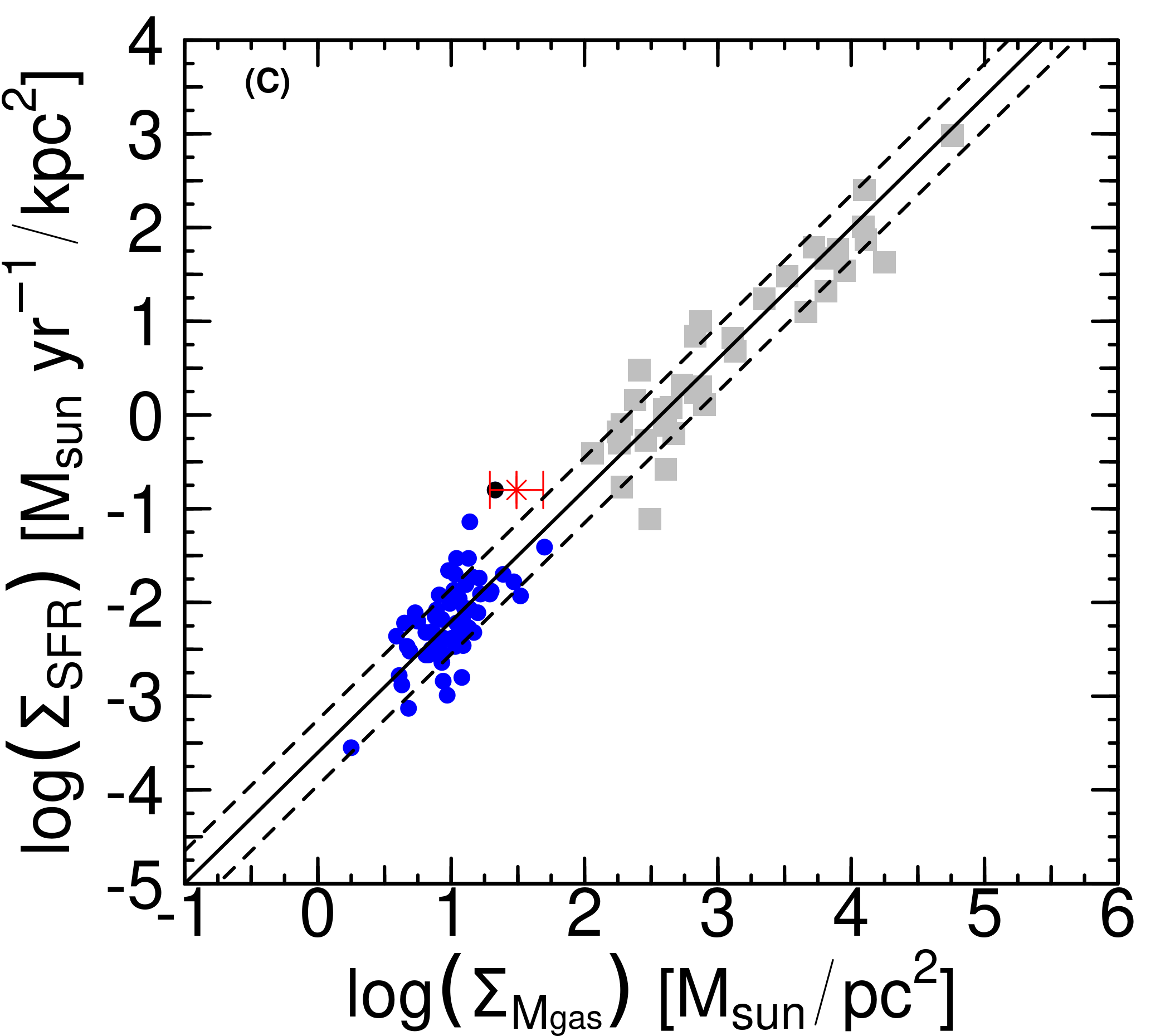}
    \caption{The global \Ms - SFR (panel A), \Ms - Z (panel B) and \sMgas - \sSFR (panel C) SRs using SDSS galaxies. The red asterisk shows our global estimation for NGC 1569. The red asterisk in the panel C corresponds to our estimation of gas mass, preserving the HI mass and CO data reported by \protect\cite{Kennicutt1998}, but using our own \alphaCO (the error bars are plotted following \protect\cite{Kennicutt1998}). In this plot the black dot and the line are the gas mass and the fitting reported by \protect\cite{Kennicutt1998}, respectively (the dashed lines are the one sigma curves). The blue dots are normal spiral galaxies and the gray squares are galaxies considered as starburst. The samples are described in \protect\cite{Kennicutt1998}.}\label{Int_plots}
\end{figure*}

The position of NGC 1569 on some global SRs using SDSS galaxies (red asterisk in the plots of Fig. \ref{Int_plots}) also reveals the importance of analyzing dwarf galaxy systems. Because NGC 1569 has a relatively high SFR, but low metallicity and low stellar mass, this work becomes relevant since it can help to fill in the gap towards the low mass regime. Although NGC 1569 seems to be gas rich, we propose this galaxy should have had even more gas. Later we discuss if this extra quantity of gas was lost via outflows.

By analyzing the classical SRs with the respective position of NGC 1569 (Fig. \ref{Int_plots}), we can test the idea that the presence of pristine inflows of gas enhanced the SFR, while at the same time diluted the gas metallicity; also, the position of NGC 1569 on the KS relation suggests a high SFR with respect to the main sequence of spiral and starburst galaxies. However, when the relation \sMSFR\ is analyzed (see panel A of Fig. \ref{SR1plots}), the role of inflows is not completely clear. Besides, even in the flat \sMZ\ relation (panel B of Fig. \ref{SR1plots}), a high SFR is not seen for the spaxels with lower metallicity. 

SRs such as $\upmu$-Z or \Mb-\Yeff\ help to explore the presence of inflows or outflows by analyzing the changes in effective yields and comparisons with models of galactic chemical evolution. In principle, NGC 1569 does not appear to be comparable with the spiral SDSS galaxies in the $\upmu$-Z or \Mb-\Yeff\ relations \citep{Tremonti2004,LaraLopez2019}, indeed NGC 1569 would  clearly be an outlier in comparison with the those papers. However, \cite{Pilyugin2004} reported different distributions between spiral and irregular galaxies in the $\upmu$-Z diagram. Precisely in the irregular galaxies sample of \cite{Pilyugin2004}, NGC 1569 follows the distribution of such irregular galaxies. The scenario of gas outflows takes high relevance since our global estimation of the total oxygen yield is \Yeff = 0.00185 or log(\Yeff) = -2.73. Empirical estimations of the true oxygen yield ($\rm Y_o$), using oxygen abundances defined with the electronic temperature in the H II regions, result in $\rm Y_o$ = 0.0027 \citep{Pilyugin2004}, $\rm Y_o$ = 0.0032 \citep{Bresolin2004} or $\rm Y_o$ = 0.0030 / 0.035 \citep{Pilyugin2007}. The ratio \Yeff/$\rm Y_o$ provides the estimation for the fraction of the produced oxygen that is rejected from the galaxy through galactic winds. Thus, NGC 1569 would lose around 38\% of the produced oxygen if the $\rm Y_o$ were 0.0030 or would lose around 47\% of the produced oxygen if the $\rm Y_o$ were 0.0035.

Previous works such as \cite{SanchezCruces2022} and references there in, not only reported an extended emission of ionized hydrogen, but also the presence of a high number of supernovae remnants. This latter supports the idea that NGC 1569 is experimenting gas outflows via stellar feedback, in particular high speed galactic winds that are powered by supernova explosions, which is consistent with the conclusions of previous works \citep{Tremonti2004, Pilyugin2004}. As mentioned in \cite{Tremonti2004}, a very high infall of pristine gas in a galaxy would i) reduce its metallicity, ii) enhance its gas fraction and iii) reduce slightly its effective yield. \citet[][]{SanchezA2015} also analyzed how the very extreme low metallicity in galaxies is attributed to infalls of metal poor gas. The presence of metal poor inflows have been also invoked to explain the strong anticorelation in the local SFR-Z${_g}$ relation seen in the metal poor MANGA galaxies \citep{SanchezM2019}. With all these latter facts, the idea of the presence of inflows is not consistent with the observations reported here. We think that outflows are more consistent in this case, since for instance, the gas fraction of NGC 1569 is low and dominated by stellar mass ($\upmu$ = 0.34), as we computed previously. The results of \cite{Dalcanton2007} also support the reduction of the effective yields by outflows -via stellar feedback-.       

We show the position of NGC 1569 in the KS relation (panel C of Fig. \ref{Int_plots}). The plot is taken from \cite{Kennicutt1998}, and shows two distinguishable samples: spiral galaxies (blue dots) and IR galaxies considered as starbursts (squares). It is worth noting that the position of NGC 1569 shows a high \sSFR\ and \sMgas\ with respect to all the spirals. At least in this diagram, NGC 1569 would be in the regime of starburst if it had 1 dex (maybe 1.5 dex) extra of gas mass surface density. With this, we want to propose the idea that such possible extra gas may have been lost via outflows (instead of directly assuming that such gas was converted efficiently into stars). In the next paragraphs, we show how this possible extra gas mass could have been lost by establishing the use of a simple self-regulator model of the SF.

\subsection{Spatially resolved properties of NGC 1569}

Back to our spatially resolved SRs, it is clear in the \sMSFR\ relation (panel A of Fig. \ref{SR1plots}), that the SFRs for the regions of NGC 1569 are on average 1 dex higher compared to other reported surveys ({\it{e.g.}}, MANGA survey); as mentioned previously the fitting of MANGA \citep{Cano2019} in Fig. \ref{SR1plots} (panel A) is for irregular galaxies. As also mentioned, the regime of low mass galaxies has not been analyzed due to observational limitations. Surveys such as MANGA or CALIFA, do not reach low values of stellar mass and metallicity, and hence this difference becomes noticeable when compared with their results. Note that the local \sMZ\ for MANGA galaxies spans from 12+log(O/H) 8.10-8.65 (or 7.95-8.70 once they are converted to our metallicity scale) and NGC 1569 has a metallicity 12+log(O/H) = 8.12. Besides, our \sMZ\ does not have the classic polynomial shape of such relationship, which makes the study of low metallicity galaxies relevant, because it could be evidence of a extremely homogeneous star formation history across the galaxy.

The spatially resolved KS relation for this galaxy (panel C of Fig. \ref{SR1plots}), shows a remarkable similarity to the \sMSFR\ relation, something not usually observed, since the KS is usually tighter. We compare our data with the fitting reported in \cite{Bigiel2008} which is a mean of all the slopes and zero points computed in that work (see panel C of Fig. \ref{SR1plots}). In comparison with \cite{Bigiel2008}, it becomes relevant the large difference in the slopes between the fitting in that work and ours; this behaviour could be attributed to the hypothetical extra gas mass that was lost by stellar feedback. 

From a spatially resolved point of view, some works in the literature have attempted to study the SR $\upmu$-Z (see panel A of Fig. \ref{SR3_plots}). This relation is mainly important because it is a via to compare observations with models of galactic chemical evolution. \cite{Barrera2018} show a study for MANGA galaxies which are contrasted with a couple of analytical models (gas regulator and leaky-box); at the end, the best fitted model is the gas regulator which is also supported by the presence of outflows driven by stellar feedback. \cite{Barrera2018} also highlight the concept of escape velocity that is of particular relevance for low mass galaxies; indeed, such galaxies have a weak gravitational potential that results in a low escape velocities that at the end facilitates the metal loss via stellar feedback. At this point, the case of NGC 1569 is one more time relevant because \cite{Barrera2018} do not have a direct estimation of gas mass and the lower metallicity limit of 12+log(O/H) = 8.10 (7.95 in our metallicity scale) of the study is not low enough.

The work of \cite{Vilchez2019} is relevant since they study spatially resolved oxygen yields in two spiral galaxies, M101 and NGC 628. The novelty of their work is that they establish a reference range of a typical empirical estimation for the oxygen yields under a closed boxed model; then, they compared the values of oxygen yields of such spiral galaxies as a function of the galactocentric radius, finding that at large galactocentric radius the oxygen yields values of M101 are deviated from the given reference range in the beginning. At the end, the authors conclude that there could be gas flows in the outer parts of this galaxy.   


Another important point to take into consideration is that our analysis relies on the methodology of \cite{Calzetti2018}, who reconstructed the SED of NGC 4449 to estimate the global dust masses properties using different analytical models and taking the pertinent assumptions. Such global properties are then used to compute the spatially resolved dust masses and, more importantly, the gas masses using the so called factor D/H dust-to-hydrogen ratio -which is similar to the DGR-. This opens a new way of analysis when either the CO is limited or there is a lack of observations in CO (\MHm) or atomic hydrogen (\MHI), which is particularly common for dwarf galaxies. First, the \MHm\ implies the use of a $\rm \upalpha_{CO}$ which is different galaxy to galaxy; the case of study NGC 1569 helps to fill the values of $\rm \alpha_{CO}$ or X$\rm_{CO}$ particularly in the low mass and metallicity regime for dwarf galaxies, as mentioned in \cite{Bolatto2013}. Second, since the DGR or D/H strongly depends on metallicity \citep[especially also in the low metallicity regime,][]{Remy2014}, NGC 1569 also helps to reduce the large scatter in the observed DGR precisely at such metallicity. Other studies such as \cite{Relano2018} or \cite{Vilchez2019} show the variation of the DGR as a function of the galaxy radius and metallicity, revealing that using a common DGR for all the galaxy must be done carefully.

\subsection{The interplay between the atomic gas, molecular gas, and SFR surface densities}

\begin{figure*}
    \includegraphics[trim={0 0cm 0 0cm},clip,width=0.49\linewidth]{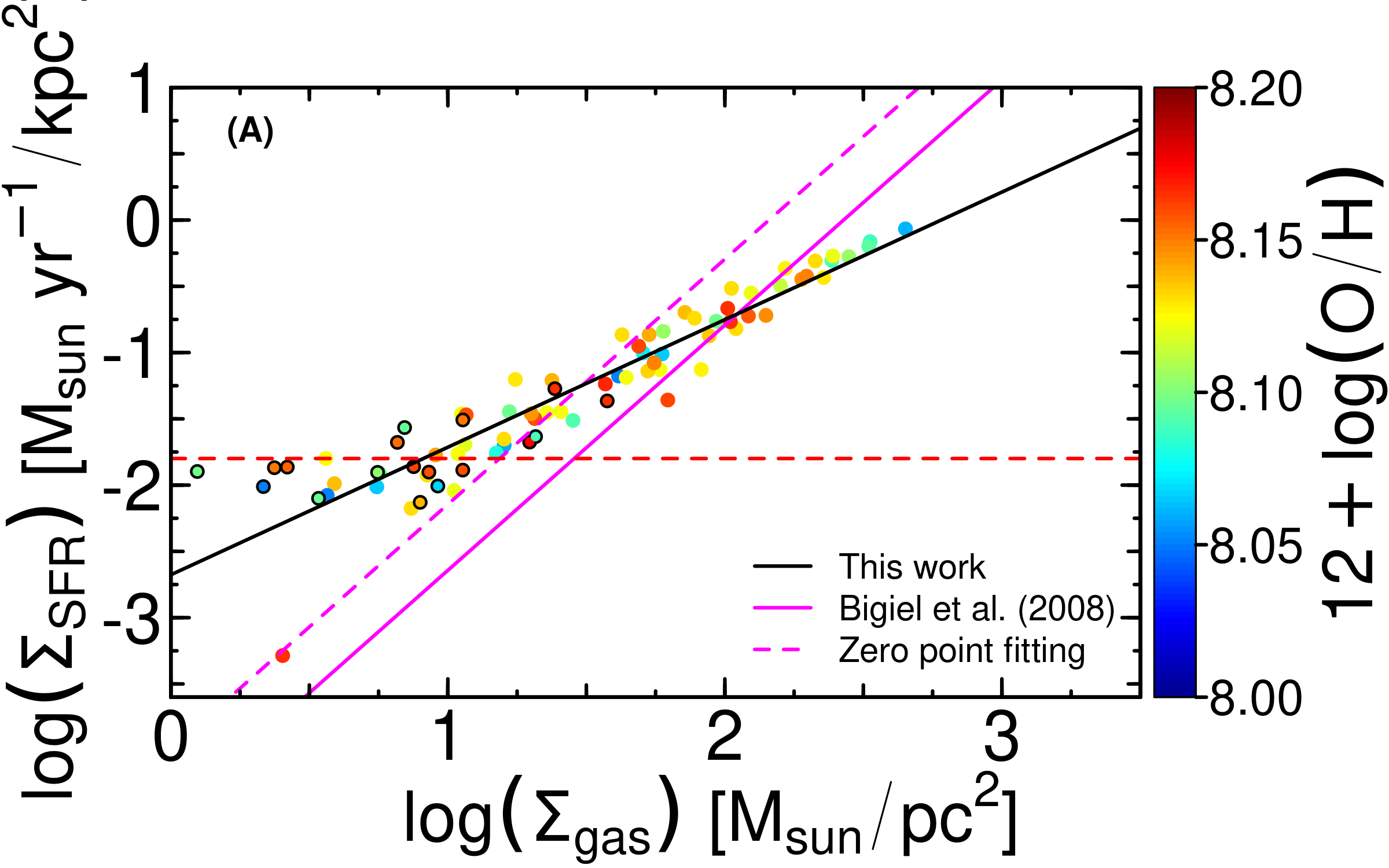}\hspace{0.15cm}
    \includegraphics[trim={0 0cm 0 0cm},clip,width=0.49\linewidth]{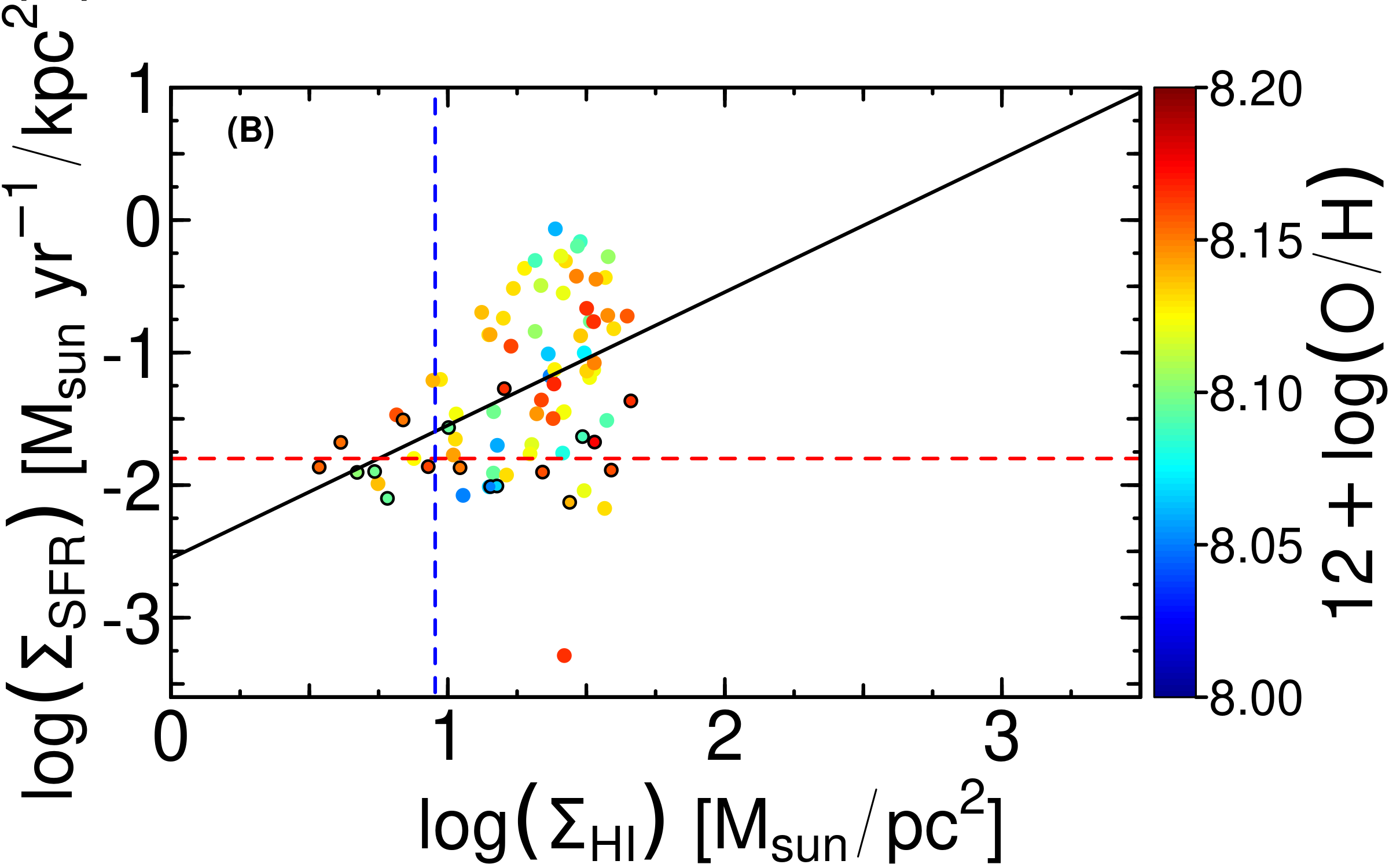}
    \includegraphics[trim={0 0cm 0 0cm},clip,width=0.49\linewidth]{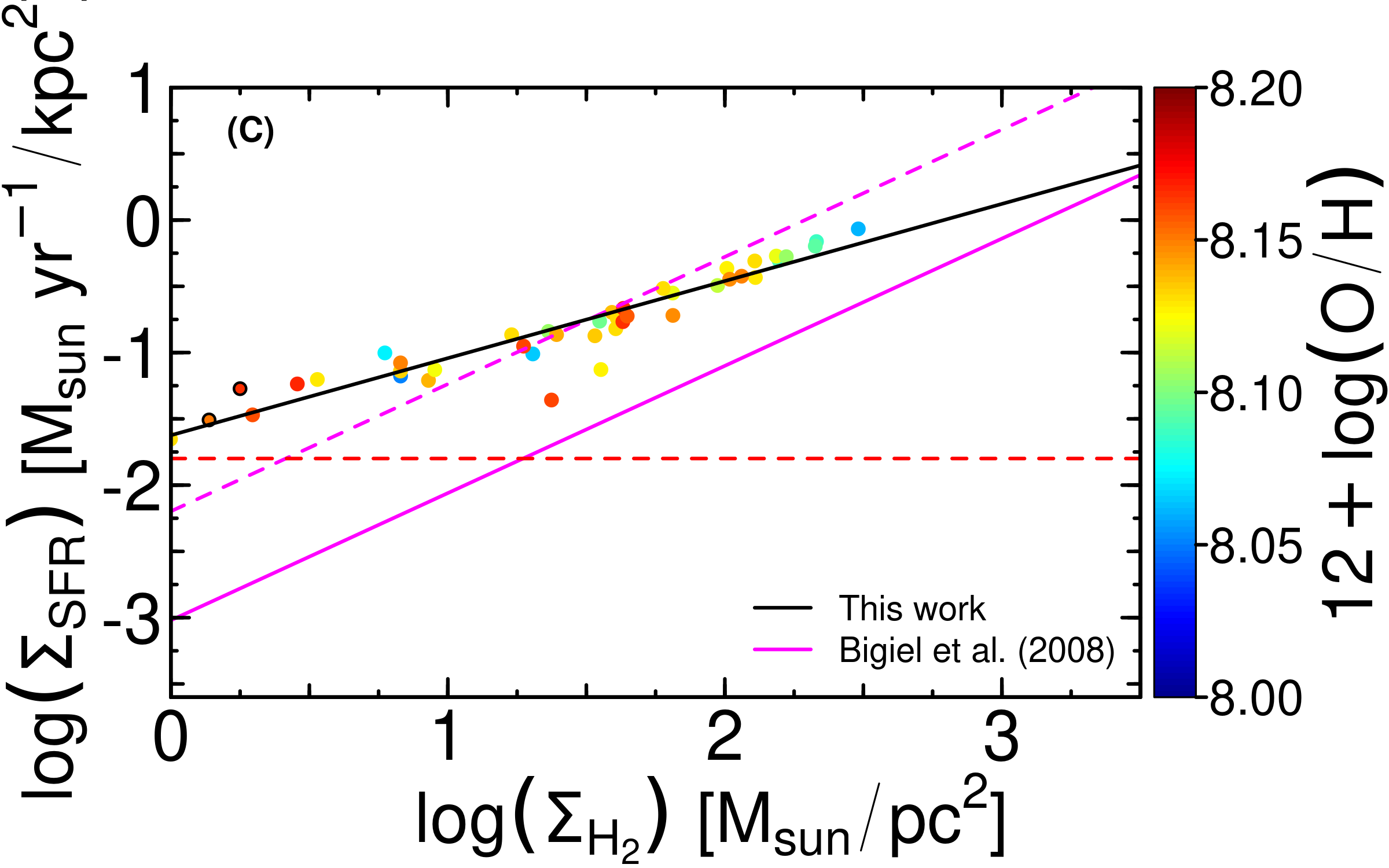} 
\caption{The \sMgSFR\ (panel A), the \sMHISFR\ (panel B) and the \sMmolSFR\ (panel C) SRs. The fits to the data are shown in black solid lines. The purple dashed lines correspond to the fits of \protect\cite{Bigiel2008} for their respective relation. The horizontal dashed red line is the  limit in which there is a valid SFR. The circles with black contours correspond to the spaxels in the tail of NGC 1569. The data is color coded by the property shown in the colorbar of each panel.}
\label{comparisonKS}
\end{figure*}

Figure \ref{comparisonKS} shows a comparison between 3 SRs: in the panel A, the \sMgSFR\ (KS relation), in the panel B the \sMHISFR\ , and in the panel C the \sMmolSFR (molecular KS relation). The \sHI\, is computed directly from the THINGS data. We follow \cite{Calzetti2018} to get an estimation of \sHmol using eq. \ref{eq_masses}. Indeed, once we have a \sMgas\ estimation, and since $\upmu_{\rm{gal}}$ and \sHI\ are known, a value of \sHmol can be computed. There are fewer spaxels since we remove the negative values.

Previous studies pointed out that some relations can be more fundamental than others. The origin of this fact is still a matter of debate since there are works that support different scenarios. For example, \cite{Kennicutt1998} found a stronger correlation between the \sMHISFR\ relation, instead of the \sMmolSFR\ relation, while \cite{Wong2002} found a stronger correlation between the \sMmolSFR\ instead of the \sMHISFR\ relation. However, studies such as \cite{Schuster2007} and \cite{Crosthwaite2007} found that the \sMgSFR\ relation had a tighter correlation than the \sMmolSFR\ relation.    

For the particular case of NGC 1569, our \sMmolSFR\ (panel C of Fig. \ref{comparisonKS}) relation is only very marginally tighter ($\uprho$ = 0.94 and $\uprho$ = 0.95) than our \sMgSFR\ relation (panel A of Fig. \ref{comparisonKS}). For our \sMHISFR\ relation (panel B of Fig. \ref{comparisonKS}), we can not conclude that a correlation exists due to the high dispersion of the data.  

\citet{Leroy2008} analyzed the results of several galaxy surveys in combination with several theoretical models that were called Star Forming Laws, {\it{i.e.}}, different scenarios in which star formation occurs. The work of \cite{Leroy2008} indicates, for example, a clear difference between dwarf and spiral galaxies. Dwarf galaxies form stars at their average rate, while spiral galaxies form stars at about half of their average rate. This work supports the idea that star formation could depend only on the presence of molecular Hydrogen. 


\cite{Bigiel2008} show that the range of the possible power law values for the KS relation is N = 1.1-2.7, while the range of the possible power law values for the molecular KS relation is N = 0.8-1.1. The molecular KS relation reported by \citet{Bigiel2008} is also tighter than the KS relation.  

It is noteworthy that our KS and molecular KS slopes (power law index) are different in comparison with \cite{Bigiel2008} and 
the molecular KS relation for 80 nearby galaxies from \citet{Sun2023}, in which they obtained the coefficient N taking different assumptions and finding a linear molecular KS relation. 
However, \cite{Shetty2014} (and references there in) discuss the origin of a sub-linear $\Sigma_{\rm H_{mol}}-$ $ \Sigma_{\rm SFR}$\ relation with observational support. They explain that a possible origin of such relation could be due to the presence of an important amount of diffuse molecular gas, which is not forming stars. \cite{Casasola2015} also found sub-linear $\Sigma_{\rm H_{mol}}-$ $ \Sigma_{\rm SFR}$\ relations, although for nearby active galactic nuclei. Confronting with our results, our molecular KS relation suggests a deficit of molecular gas in comparison with \cite{Bigiel2008}. Thus, the origin of the slope in our relation is not due to a diffuse molecular gas but a possible gas expulsion by stellar feedback (see below).

Since the slopes in the KS relations are associated to the Star Formation Efficiency (SFE), this concept becomes relevant at the moment of interpreting the physical conditions of a galaxy. \cite{Leroy2008} showed that the SFE is more useful than the \sSFR\
alone to identify where the conditions propitiate the star formation. \citet{Ellison2020} suggested that the \sMmolSFR\ relation could be primarily driven by changes in the SFE, and secondary with a weaker dependence on the gas fraction. 

In conclusion, the SRs for NGC 1569 displayed in Fig. \ref{comparisonKS} are driven by the \sHmol\ supported by the idea that the stars are formed directly from molecular clouds, and thus it seems logic that \sHmol\ and \sSFR\ are immediately more related than \sHI or the total gas mass (\sMgas) with the \sSFR. This idea is also supported by \cite{Leroy2008} who mentioned that under certain SF Laws, indirect evidence for abundant \MHm\ in the central parts of dwarf galaxies can be estimated. 


\subsection{Possible origin of inflows}

In this section, we compare our data with the general trend found by \cite{Bigiel2008} and propose the possible presence of inflows in NGC 1569. Since this result is not statistically significant, we do not mention it as a part of our main results. However, a methodology will be implemented in future analysis. 

Our KS relation (panel A of Fig. \ref{comparisonKS} and also panel C of Fig. \ref{SR1plots}) has a very different slope with respect to \cite{Bigiel2008}. Following the magenta line \citep[fitting of][]{Bigiel2008}, we divide this plot into three zones. The first one with log(\sMgas) < 1.5 is called the zone of outflows (see next subsection), the second one with 1.5 < log(\sMgas) < 2 is called the normal zone, and the third one with log(\sMgas) > 2 is called the inflow zone.

Now, we focus on the inflow zone and particularly on the points that lie to the right of the magenta fit. We define the offset $\upDelta\Sigma_{\rm gas}$ as the difference between $\Sigma_{\rm gas,obs}$ (the observed \sMgas\ value) and $\Sigma_{\rm gas,KS~B08}$ (the \citet{Bigiel2008} fit value at \sMgas). In other words, $\upDelta\Sigma_{\rm gas}$ is the distance between each data point and the  \cite{Bigiel2008} fit in the \sMgas\ axis.

\begin{figure}
\includegraphics[trim={0cm 0cm 0cm 0cm},clip,width=1.1\linewidth]{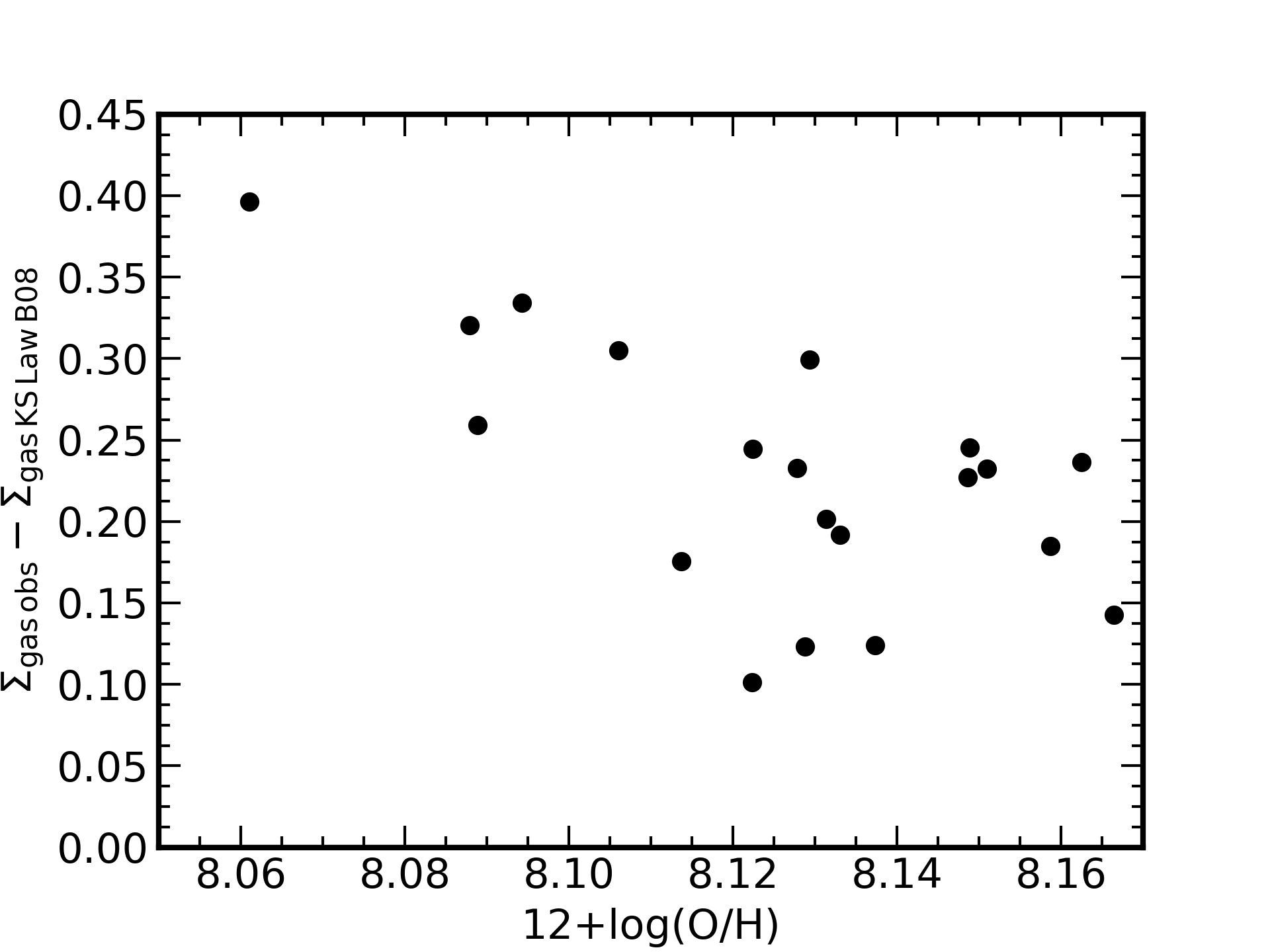}
\caption{The offset $\upDelta\Sigma_{\rm gas}$ vs. metallicity. Such offset is defined as the difference between $\Sigma_{\rm gas,obs}$ (the observed \sMgas value) and $\Sigma_{\rm gas,KS~B08}$.}
\label{gassexcess}
\end{figure}

The offset $\upDelta\Sigma_{\rm gas}$ is plotted versus metallicity in Fig. \ref{gassexcess}, where we show that the larger the gas excess, the lower is the metallicity. This is probably because the points that lie to the right of \cite{Bigiel2008} fit could have a slight gas excess due to the presence of inflows, supported by the idea that pristine gas dilutes the metallicity. These points correspond only to the center part of the galaxy. 

The \sMHISFR\ relation show evidence of an excess of neutral gas, since the HI gas surface density values are larger than in other galaxies, where \sHI\ is not larger than $9\rm{M_{\odot}~pc^{-2}}$ \citep{Bigiel2008}. In Fig. \ref{comparisonKS} (panel B), we show that the majority of \sHI\ values are larger than $10\rm{M_{\odot}~pc^{-2}}$. These large \sHI\ values are still present using higher spatial resolution elements of 700 pc (see Appendix \ref{HI_700}). 

Therefore, the data suggests that the possible excess of gas seen in the KS relation for log(\sMgas) > 2 (in the central part of the galaxy) could be due to an inflow of neutral gas.

We consider that this type of analysis is important in order to examine the interplay between inflows and outflows in the same galaxy. Although our metallicity variation (0.1 dex) due to the presence of inflows is not statistically significant and it is within the errors, the method allows to test changes in metallicity related to inflows. We also tested this methodology using different metallicity calibrators \citep[O3N2, N2,][]{Pettini2004,Marino2013} and found that the shape of the $\upDelta\Sigma_{\rm gas}$ versus metallicity relationship (see Fig. \ref{gassexcess}) is preserved for those calibrations. We will implement this idea in future analysis with a larger sample and appropriate fits for comparison.


\subsection{Possible origin of outflows: a self regulated feedback model}


As mentioned previously, another interesting fact of NGC 1569 is its low gas fraction given its metallicity. The global gas fraction that we report is $\upmu$ = 0.34, which in general means that the total mass of the galaxy is dominated by the stellar mass. Common values of gas fractions ($\upmu$) can also be found in \cite{Pilyugin2004,LaraLopez2019} for galaxies of different types. For spiral galaxies, the mean global gas fraction values are in the range $\upmu$ = 0.50 - 0.85 with a mean 12+log(O/H) in the range 8.4 - 8.7. A particular case is seen for the sample of irregular galaxies in Table 7 of \cite{Pilyugin2004}, for which they report $\upmu$ = 0.20 - 0.83 with a 12+log(O/H) between 7.22 - 8.35 dex. When we compare galaxies with almost the same M$\rm_B$ than NGC 1569 (M$\rm_B$ $\sim$ -15.7) we find galaxies with $\upmu$ from 0.22 to 0.66. The hypothetical 1 - 1.5 dex of extra gas mass that we previously mentioned, could drastically enhance the value of gas fraction.

We propose two scenarios to explain the low gas fraction in NGC 1569. The first one is related to the possible interaction of NGC 1569 with other galaxy. As  mentioned in \cite{Johnson2013}, NGC 1569 is in a system with other 3 galaxies with a possible recent interaction with the nearest companion UGCA 92. \cite{Geha2006} mention that the galaxy internal processes reduce the gas fraction of galaxies, but the presence of a massive or luminous galaxy within 0.5 Mpc could totally remove the gas from a dwarf galaxy. Precisely, one of the companion galaxies of NGC 1569 is a relatively luminous galaxy \citep[IC 342,][]{Johnson2013}, which could imply that the interaction with IC 342 is more relevant than with UGCA 92. 

The removal of gas by an external massive companion should affect also to the neutral gas. However, we report an excess of neutral gas, so the scenario where removal of gas by massive companion occurs is not supported by our analysis. We propose an alternative scenario. A simple model of stellar feedback can explain the deficit of gas.  While stellar feedback is due to different phenomena ({\it{e.g.}}, stellar radiation, stellar winds, supernova explosions), occurring from stellar to galactic scales, 
the spatially resolved star formation self regulator model can be used to parametrize such a complex process. This simple model has been used in previous studies \citep[{\it{e.g.}}, ][]{2017MNRAS.468.4494Z,Barrera2018,2020MNRAS.499.1172Z} and can be described by:

\begin{equation}
\dot{\Sigma}_{\rm{out}}=\upeta \cdot \Sigma_{\rm{SFR}},
\end{equation}

where $\rm \upeta$ is the so-called  mass loading factor, $\dot{\Sigma}_{\rm{out}}$ the gas  outflow rate surface density due to stellar feedback and \sSFR\ is the SFR surface density. Therefore, the mass loading factor, $\upeta$, is the mass surface density outflow rate per unit of SFR surface density provoked by stellar feedback.

Such expression can be written as a function of the gas outflow surface density, $\Delta\Sigma_{\rm{out}}$: 

\begin{equation}
\Delta\Sigma_{\rm{out}}=\upeta \cdot \Delta t \cdot \Sigma_{\rm{SFR}}
\end{equation}

Cosmological and hydrodynamical simulations predict a relation between $\upeta$ and total stellar mass \citep{2015MNRAS.454.2691M,2017MNRAS.465.1682H} or  local gas surface density \citep{,2017ApJ...841..101L}. 
Observationally, the value of $\upeta$ is uncertain, but recent works have made progress to quantify it  as a function of galactocentric radius \citep{2019Natur.569..519K}, stellar mass surface density and total stellar mass  \citep{2020MNRAS.493.3081R,2020MNRAS.499.1172Z}, or the local escape velocity \citep{Barrera2018}. Since we have estimations of $\Sigma_*$, we can estimate $\upeta$ using  the relation between $\Sigma_*$ and $\upeta$ from \citet{2020MNRAS.499.1172Z}:

\begin{equation}
    \log(\upeta + 1) = (-0.32 \pm 0.03) \log(\Sigma_*) + (3.2 \pm 0.3),
    \label{eq_eta}
\end{equation}
which is in agreement with models and theory \citep{2015MNRAS.454.2691M,2017ApJ...841..101L,2017MNRAS.465.1682H}.

The median value of the stellar mass surface density is $\log\Sigma_*=7.9\ \rm{M_{\odot} kpc^{-2}}$, then $\upeta$ = 3.7 according to Eq. \ref{eq_eta}. We assume a value of $\Delta \rm t=4.3\ \rm{Myr}$ in concordance with the characteristic H$\upalpha$ time scale \citep{2020MNRAS.498..235H} and \sSFR $=0.15$\Msun yr$^{-1}$ kpc$^{-2}$ \citep{Kennicutt1998}. We estimate a gas outflow of $\Delta\Sigma_{\rm{out}}$ = 2.4 \Msun pc$^{-2}$.

\begin{figure}
\includegraphics[trim={0 0cm 0 0cm},clip,width=1\linewidth]{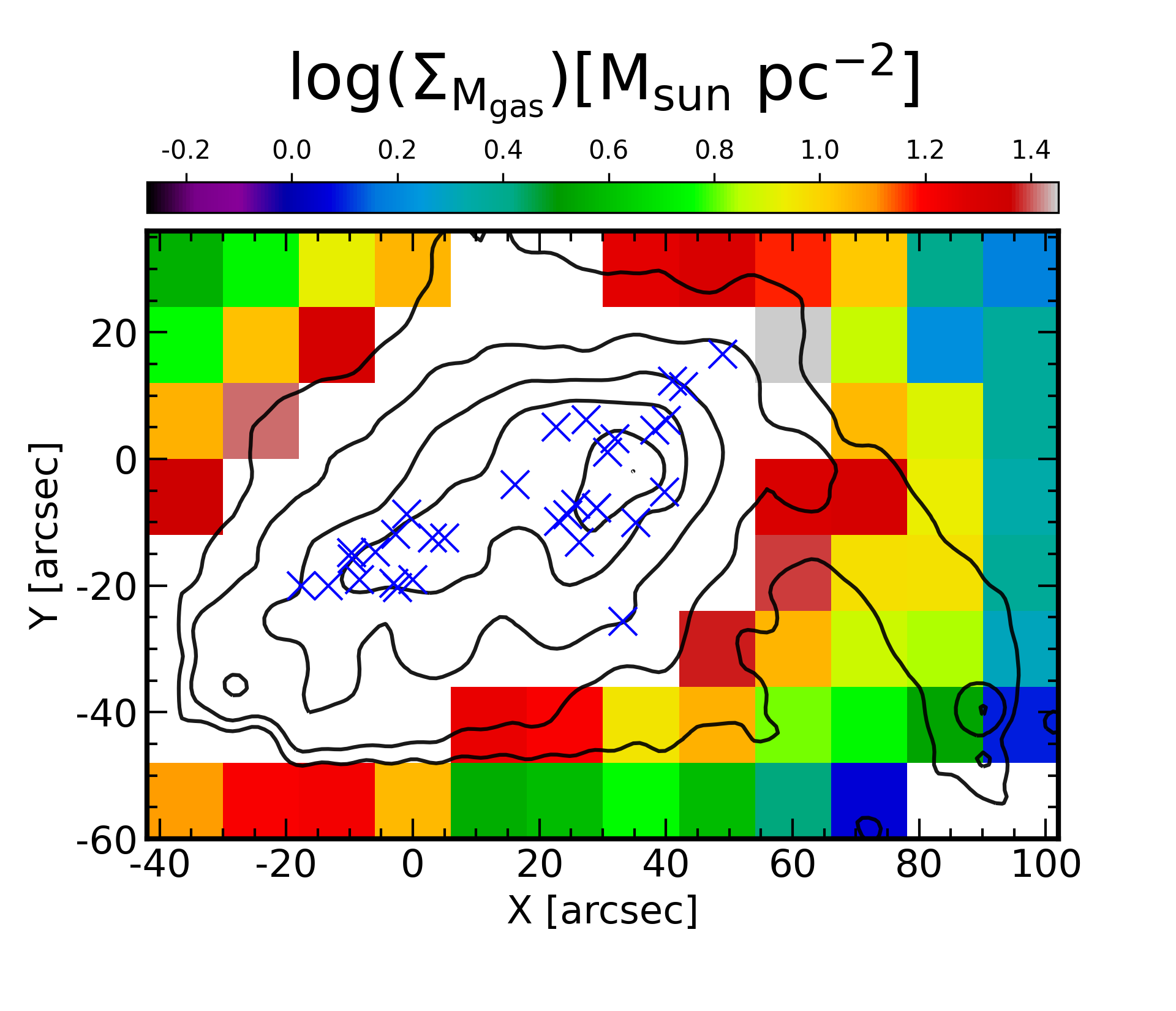}
\caption{The gas mass surface density, \sMgas\ maps for values  log(\sMgas) < 1.5 $\rm{ M_{\odot} pc^{-2}}$, this is the spaxels in the zone of outflows. The black contours is the H$\upalpha$ emission. The blue crosses correspond to supernovae in NGC 1569 (coordinates are taken from \protect\citep{SanchezCruces2022}).}
\label{outflow_dom}
\end{figure}

We define $\Delta \Sigma_{\rm{gas\thinspace}}$ as the difference between the global gas surface densities, that of the global KS relation, and that observed in NGC1569 (our corrected estimation), which are shown as a black dashed line, and as a red symbol in Fig. \ref{Int_plots} (panel C), respectively. We found $\Delta \Sigma_{\rm{gas\thinspace}}=3\ \rm{M_{\odot} pc^{-2}}$. Therefore, stellar feedback due to the recent star formation can explain the gas deficit that we propose.
We plot the spaxels in the zone of outflows, where log(\sMgas) < 1.5, in Fig.\ref{outflow_dom}. We show that these spaxels are in the outer parts of the galaxy, where filaments associated with outflows are usually seen \citep{Johnson2012}. 

Outflows can be also analyzed via X-ray emission, stellar and gas kinematics. For the particular case of NGC 1569, an important quantity of X-ray emission is associated to a diffuse halo and a metal enriched outflows \citep{Heckman1995,Martin2002}. In some particular scenarios, the X-ray emission can be linked to the kinematics of a galaxy. Indeed, for NGC 1569, components of velocity relative to the systemic velocity (v$_{\rm sys}$) reveal the presence of superbubbles or expanding shells of ionized gas \citep{Heckman1995,Martin1998}. Particularly, gas kinematics using HI has allowed the detection of an unusual high mean HI velocity dispersion possibly due to outflows \citep{Stil2002}. Other studies \citep[{\it{e.g.}},][]{Johnson2012} have found evidence of an outflow with a potential expanding shell near the supermassive star cluster A of NGC 1569, supported by both, gas kinematics and the high velocity dispersion of star kinematics. Finally, recent studies \citep[{\it{e.g.}},][]{SanchezCruces2022} performed analysis of physical phenomena in NGC 1569 ({\it{e.g.}}, shocks and supernova remnants) using  high spectral resolution.


\section{Conclusions}\label{conclusion_section}

We compute a set of scaling relations (SRs) and analyze $\sim$100 spaxels of the dwarf galaxy NGC 1569 using the Metal-THINGS, THINGS and CARMA Surveys in combination with DustPedia archival data. Emission line fluxes were derived using STARLIGHT and Gaussian line-profile fittings. We estimated spatially resolved physical properties such as star formation rate surface density ($\Sigma_{\rm{SFR}}$), oxygen gas metallicity (Z) and total gas mass surface density ($\Sigma_{\rm{gas}}$), this latter computed by using the braodband spectral energy distribution modelling code -CIGALE- to get the dust mass surface densities that were converted to total gas surface densities adopting a Dust-to-Gas Ratio (DGR). Such DGR was estimated simultaneously with the CO luminosity-to-molecular
gas mass conversion factor ($\upalpha_{\rm{CO}}$) using the method presented in \citet{Leroy2011} and \citet{Sandstrom2013}. With the $\Sigma_{\rm{gas}}$, it was possible to derive other properties such as the baryonic mass surface density ($\Sigma_{\rm{bar}}$), the local gas fraction ($\upmu$), the local effective oxygen yields (\Yeff), the local star formation efficiency (SFE) and the local depletion time (t$\rm _{dep}$).

For all the mentioned properties, we get spaxel maps with 12" of resolution ($\sim$180 pc). By comparing such properties with each other, we estimate different local SRs.

Our study can be divided into two main sections. In the first one, we analyze the classic \sMSFR, \sMZ\ and \sMgSFR\ relations. We recover the known correlations ($\uprho$ > 0.92) and scatterings (up to 0.25 dex) except for the \sMZ. 

In the second one, we derive the relations $\upmu$ - Z, \sMbar - \Yeff\ and \sMs, $\upmu$, \sMbar\ - SFE. For all SRs (except for the $\upmu$ - Z), we have low correlations ($\uprho$ < 0.36) and a relative low dispersion (up to 0.22 dex). 

We discuss the global and local  properties of NGC 1569, since the global ones reveal the presence of inflows, but the local ones the presence of outflows. Thus, we propose two methodologies to explore both scenarios. Our local estimation for the oxygen yields and local gas fractions reveal a deficiency of gas mass possibly ejected by outflows. For this, one of our methodologies is based on a self regulated feedback model, which show that the stellar feedback plays a stronger role to propitiate outflows.

Finally, given our multiwavelength data, we compute the atomic and molecular Kennicutt-Schmidt (KS) relation in order to discuss the slopes and the SFRs of each one, in comparison with our \sMgSFR.  

Our main results are summarised as follows: 
\begin{enumerate}

   \item We recover the known classic shapes of the \sMSFR\ , \sMgSFR\ and \sMmolSFR\ , all with a low scatter up to 0.25 dex, and high correlation of $\uprho$ > 0.92.
   \item The SFRs in NGC 1569 are $\sim$1.26 dex higher than the SFRs in MANGA galaxies. The slope of our \sMSFR\ is $\sim$1.6 times steeper than the slope in MANGA galaxies.   
   \item Our fittings of the \sMgSFR\ (m = 0.96) and the \sMmolSFR\ (m = 0.58) are flatter than the reported in previous works \citep[{\it{e.g.}},][]{Bigiel2008}.
   \item The shape of our \sMZ\ is not similar to the global one, rather we obtain a flat relation mainly due to the constant metallicity 12+log(O/H) $\sim$ 8.12. This, in combination with the flat gradient, differ from what it is observed in spiral galaxies. In fact, flat metallicity gradients are rather common in young low mass galaxies that are still assembling their mass.
   \item We report a log(DGR) = -3.08 and log(\alphaCO) = 1.6 \Msun pc$^{-2}$ for NGC 1569, establishing new estimations for the gas mass log(\sMgas\ ) = 1.49 \Msun pc$^{-2}$ and molecular gas log(\sHmol\ ) = 1.05 \Msun pc$^{-2}$. This is 0.16 and 0.95 dex higher than the reported in \cite{Kennicutt1998}, respectively.
   \item We show in the \sMs - SFE and \sMbar - SFE relations that the tail of NGC 1569 could play an important role in establishing the slope of the fittings.
   \item Although the global SRs show a possible evidence of inflows, our local relations do not confirm such scenario. On the contrary, they support the idea of the presence of outflows because NGC 1569 has a lower oxygen yield (log(\Yeff) = -2.83) and a lower gas fraction ($\upmu$ = 0.34) compared with galaxies of similar masses. As pointed out in \cite{Dalcanton2007}, outflows can be the mechanisms that really reduce the oxygen yield. Also, as mentioned in \cite{Tremonti2004}, an inflow could produce an enhancement in the gas fraction but it is not enough to explain low values in the oxygen yields.
   \item The position of NGC 1569 in the global KS relation, the shape of the local relation, and the values of $\upmu$ and \Yeff\ could reveal a deficiency of gas mass -possibly ejected by outflows-. We show how the simple spatially resolved star formation self-regulator model can explain the absence of gas mass that is lost by stellar feedback. We conclude that the stellar feedback plays a stronger role to propitiate outflows and thus to explain the low gas fraction of NGC 1569.
\end{enumerate}

The origin of SRs go down to the star formation process itself, and its wide range of star formation efficiencies for galaxies. However, the analysis have never been very precise because star formation spans a wide range of scales, from cluster-forming cores to molecular clouds to the whole interstellar medium. Moreover, the combination with other physical parameters, for instance, dust distribution as a tracer of gas mass, is important in understanding the evolution and formation of stars from galaxy to galaxy, or the conversion of gas into stars, that control not only the formation history of stars within galaxies, but also their chemical enrichment. 

We highlight the importance of dwarf galaxies in the cosmological context. Indeed, it is thought that dwarf galaxies have similar properties to galaxies at high redshift; hence, the analysis of the local ones could uncover insights of evolution and  physical processes of the first galaxies. With the advent of the James Webb Space Telescope (JWST), a new era of IFU observations of dwarf galaxies is possible, not only for local galaxies but also for high redshift ones. 

In a future study we will contrast our results with simulated low mass galaxies to explore the role of inflows, outflows and the possible changes in global SRs.





\section*{Acknowledgments}

This research was supported by the International Space Science Institute (ISSI) in Bern, through ISSI International Team project No. 505 (The Metal-THINGS Survey of Nearby Galaxies). LEG and MV gratefully acknowledge to the Consejo Nacional de Ciencia y Tecnología del Estado de Puebla (CONCYTEP) for their generous financial support for telescope observations during this project. MALL acknowledges support from the Spanish grant PID2021-123417OB-I00, and the Ramón y Cajal program funded by the Spanish Government (RYC2020-029354-I).
MR acknowledge financial support from the CONACYT project CF-86-367. LSP acknowledges support from the Research Council of Lithuania (LMTLT) grant P-LU-PAR-23-28. MEDR acknowledge support from PICT-2021-GRF-TI-00290 of ANPCyT (Argentina). SPO acknowledges support from the Comunidad de Madrid Atracción de Talento program via grant 2022-T1/TIC-23797.

\section*{Data Availability}

The data underlying this article will be shared on reasonable request to the corresponding author. All the data cubes, images and spectra will be made available on the Metal-THINGS website (in preparation).



\bibliographystyle{mnras}
\bibliography{Bibliografia.bib} 




\appendix

\section{Dust masses through Hershel-SPIRE bands}\label{Dustmass160250}

\begin{figure*}
\includegraphics[trim={0cm 0cm 0cm 0cm},clip,width=0.8\linewidth]{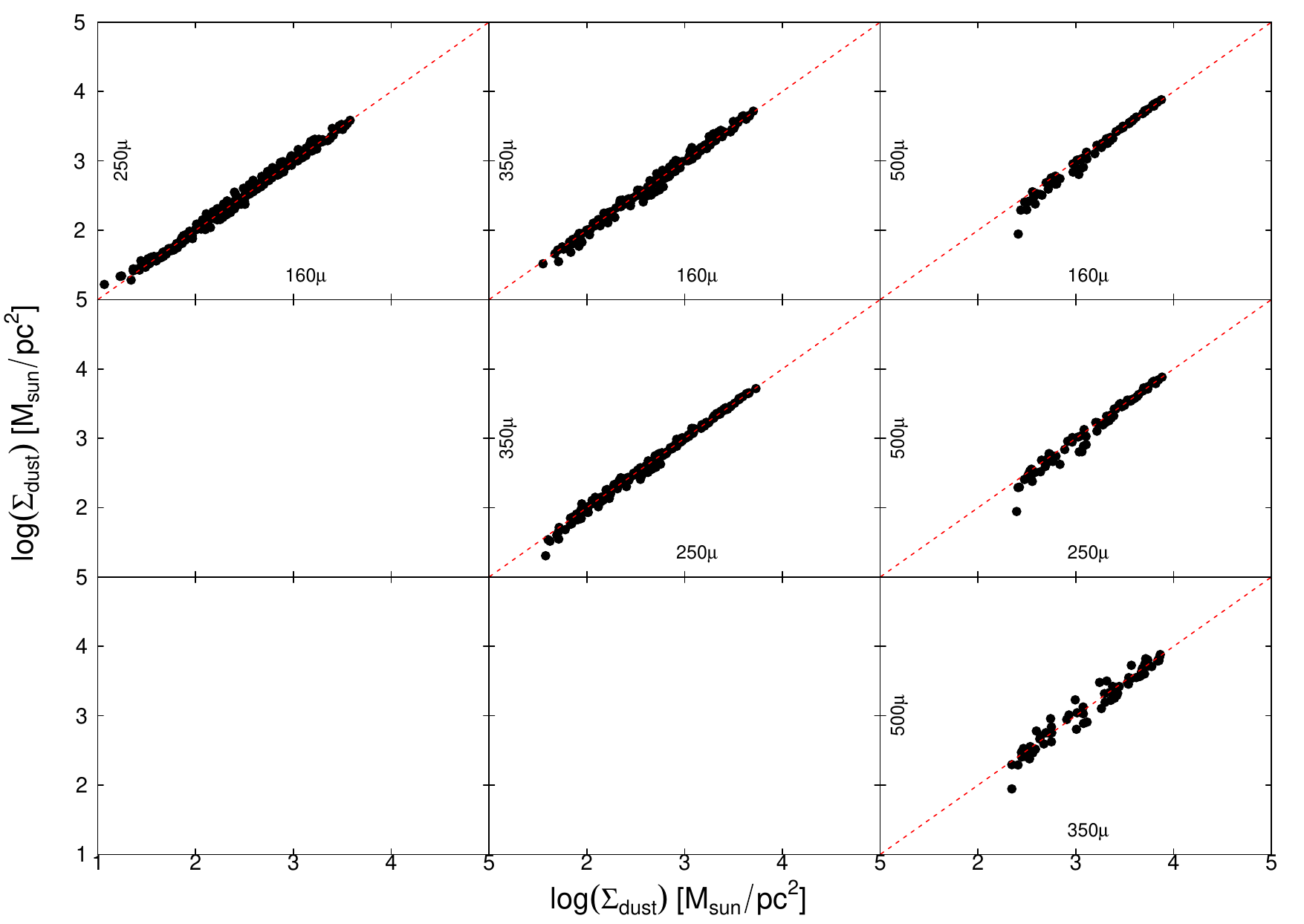}
\caption{Comparison between different dust mass estimates using different Hershel-SPIRE bands.}
\label{DustComparison}
\end{figure*}

This section shows the correlation between the different values of dust masses using CIGALE, computed using  different bands of HERSCHEL-PACS/SPIRE. As  shown in Fig. \ref{DustComparison}, the differences between them are not critical, hence we can use the dust masses up to different bands. 
For example, to compute the DGR we use the values of dust masses up to 160 $\upmu$m, because we have more values to have statistics. All our work is based on the reprojection and convolution of all our spaxels maps up to 250 $\upmu$m, because we have enough scale and number of values to plot our SRs.     

\section{CIGALE: SEDs examples}

\begin{figure*}
    \centering
    \includegraphics[trim={0 0cm 1.5cm 0cm},clip,width=0.45\linewidth]{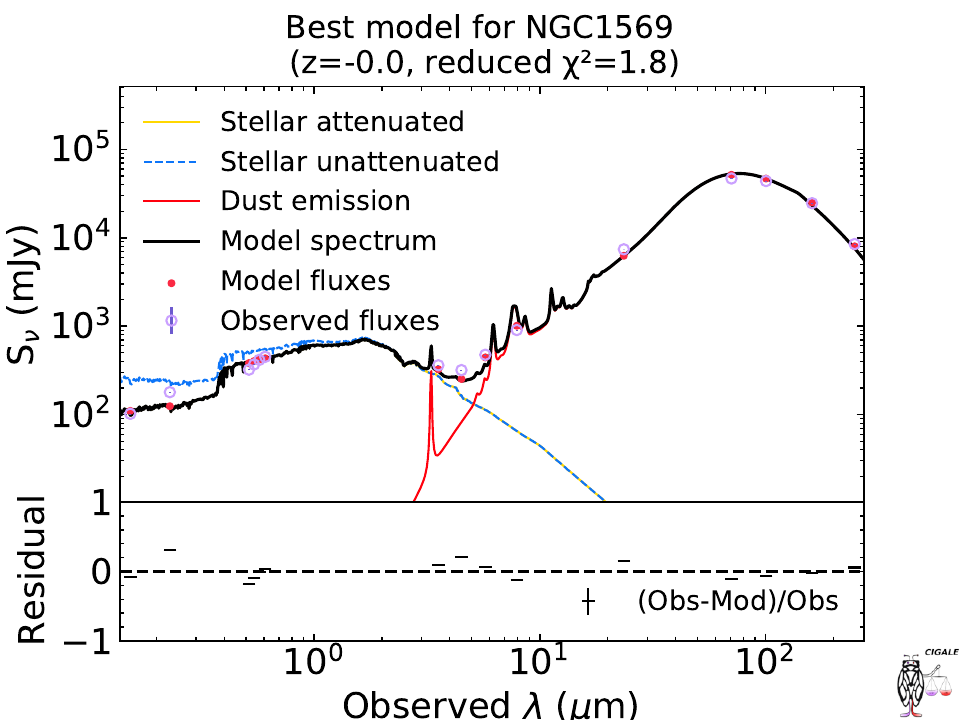}
    \includegraphics[trim={0 0cm 1.5cm 0cm},clip,width=0.45\linewidth]{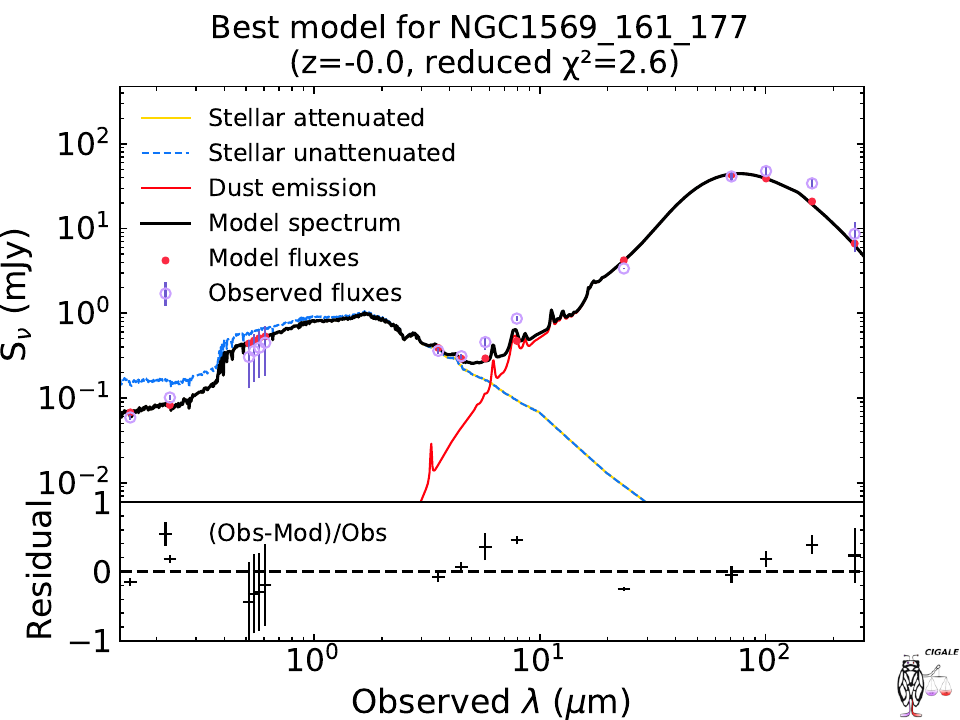} 
    \includegraphics[trim={0 0cm 1.5cm 0cm},clip,width=0.45\linewidth]{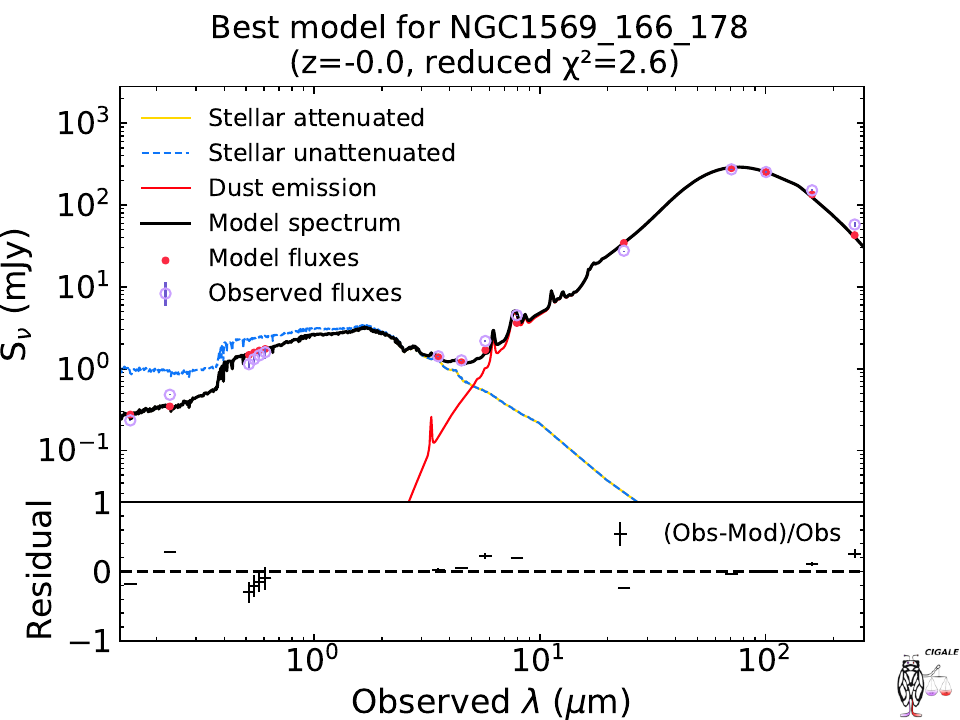} 
    \includegraphics[trim={0 0cm 1.5cm 0cm},clip,width=0.45\linewidth]{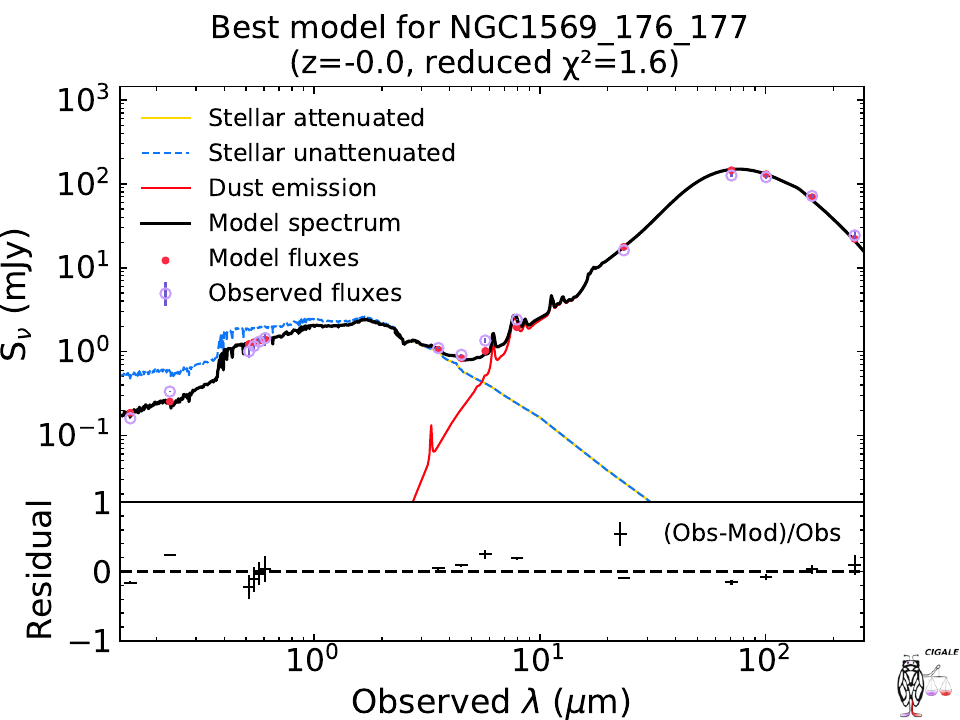}    
    \caption{Different spectra computed by CIGALE. The integrated spectrum for all the galaxy is shown in the upper left panel. The rest of the spectra were selected across a horizontal axis from the center of the galaxy to the edge.}\label{CIGALE}
\end{figure*}

In order to evaluate the quality of our SEDs, which we use to compute the dust masses, we show some SEDs fits. The first example correspond to the spaxel closest to the center of the galaxy (upper right panel of Fig. \ref{CIGALE}). We also show a couple of extra SEDs corresponding to the right edge of the galaxy (bottom panels of Fig. \ref{CIGALE}). The global SED of NGC 1569 is shown in the upper left panel of the same figure.

\section{Gas excess up to 500 microns HERSHEL-SPIRE band}

\begin{figure}
\includegraphics[trim={0cm 0cm 0cm 0cm},clip,width=1\linewidth]{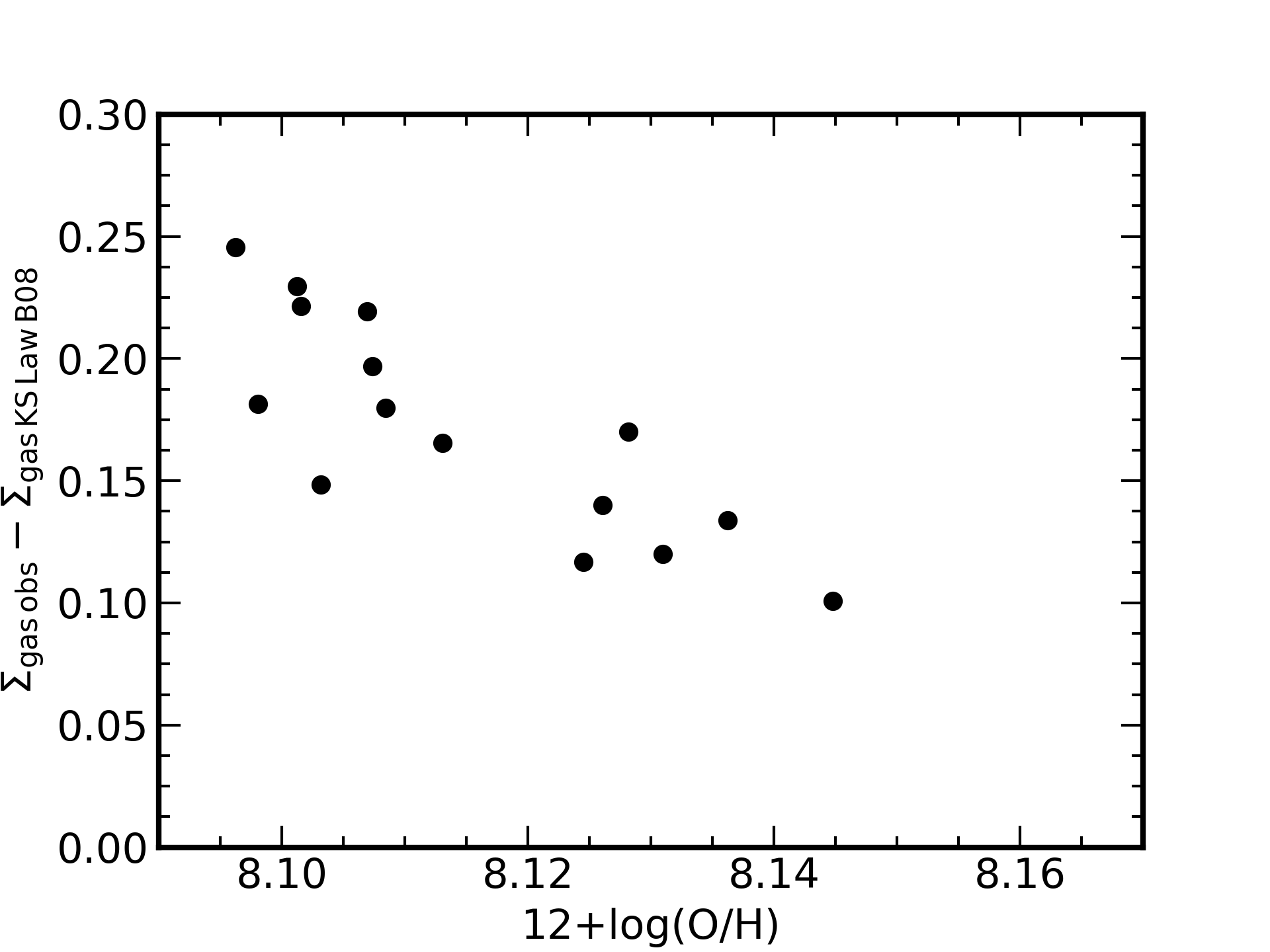}
\caption{The offset $\upDelta~\Sigma_{\rm gas}$ vs. metallicity. Such offset is defined as the difference between $\Sigma_{\rm gas,obs}$ (the observed \sMgas value) and $\Sigma_{\rm gas,KS~B08}$. These  values correspond to the gas/dust mass estimated up to 500 $\upmu$m.} 
\label{Gasexcess2}
\end{figure}

To evaluate if the inverse relation between excess of gas and metallicity shown in Fig. \ref{gassexcess} is preserved for larger spatial scales, we show the same relation at the maximum scale available (that of 500 $\upmu$m). We show in Fig. \ref{Gasexcess2} that the same offset of 0.05 dex is obtained.

\section{The HI KS relation at higher spatial scales (700 pc)}\label{HI_700}

\begin{figure}
\includegraphics[trim={0cm 0cm 0cm 0cm},clip,width=1.1\linewidth]{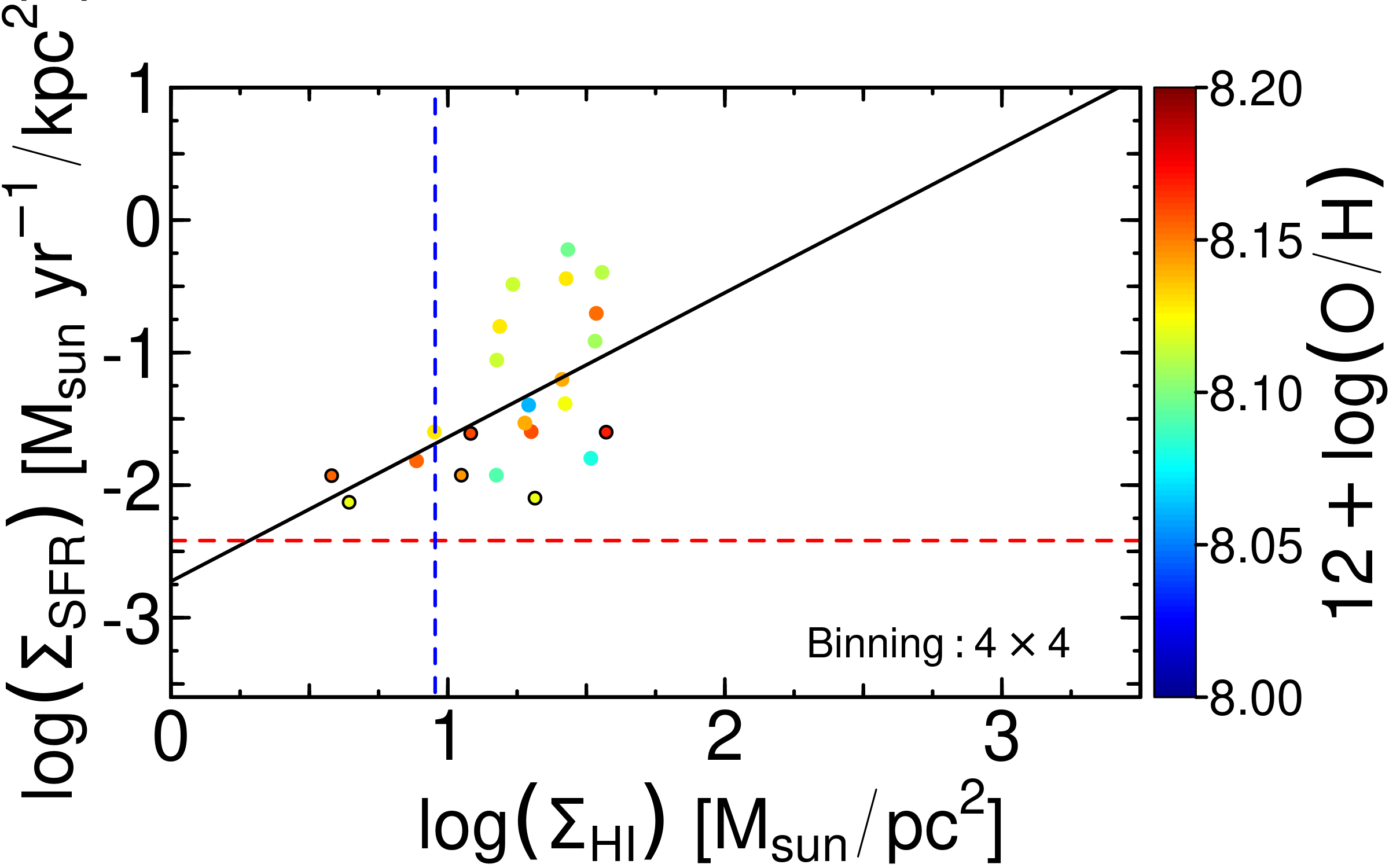}
\caption{The \sMHISFR\ scaling relation up to $\sim$700 pc. The horizontal dashed red line is the  limit in which there is a valid SFR. The circles with black contours correspond to the spaxels in the tail of NGC 1569. The data is color coded by the oxygen abundance. The vertical dashed blue line shows the values larger than $9\rm{M_{\odot}/pc^2}$. }
\label{HI700pc}
\end{figure}

Fig. \ref{HI700pc} shows the \sMHISFR\ relation at a similar scale (750pc) as in \cite{Bigiel2008}. This probe that even at larger scales, our result of HI excess is preserved. 

\section{Coefficients for global scaling relations}

In Table \ref{Fittings_global}, we report the coefficient values of the fits  for the global SRs shown in Fig. \ref{Int_plots}. We use a sample of SDSS galaxies for the \Ms - SFR and \Ms - Z, and the sample of \cite{Kennicutt1998} to plot the $\Sigma_{\rm M_{gas}}-$ $ \Sigma_{\rm SFR}$ relation.

\begin{table}
\centering
\begin{tabular}{cccc}
\hline
\multicolumn{4}{c}{Global fitting coefficients} \\ \hline
 & \Ms - SFR & \Ms - Z & $\Sigma_{\rm M_{gas}}-$ $ \Sigma_{\rm SFR}$ \\ \hline
m & 0.73 & -0.03, 0.83, -7.10  & 1.4  \\ \hline
y$_0$ & -6.97 & 27.97 & 0.00025  \\ \hline
\end{tabular}
\caption{The slopes (m) and zero points (y$_0$) for the global scaling relations reported in this work. For the \Ms - Z relation, a third order polynomial of  was fitted, the values shown in the table correspond to the x$^3$, x$^2$ and x coefficients respectively. The values for the global $\Sigma_{\rm M_{gas}}-$ $ \Sigma_{\rm SFR}$ relation are taken from \protect\cite{Kennicutt1998}.}
\label{Fittings_global}
\end{table}

\section{Gas metallicity recalibration}

\begin{figure}
\includegraphics[trim={0cm 0cm 0cm 0cm},clip,width=1\linewidth]{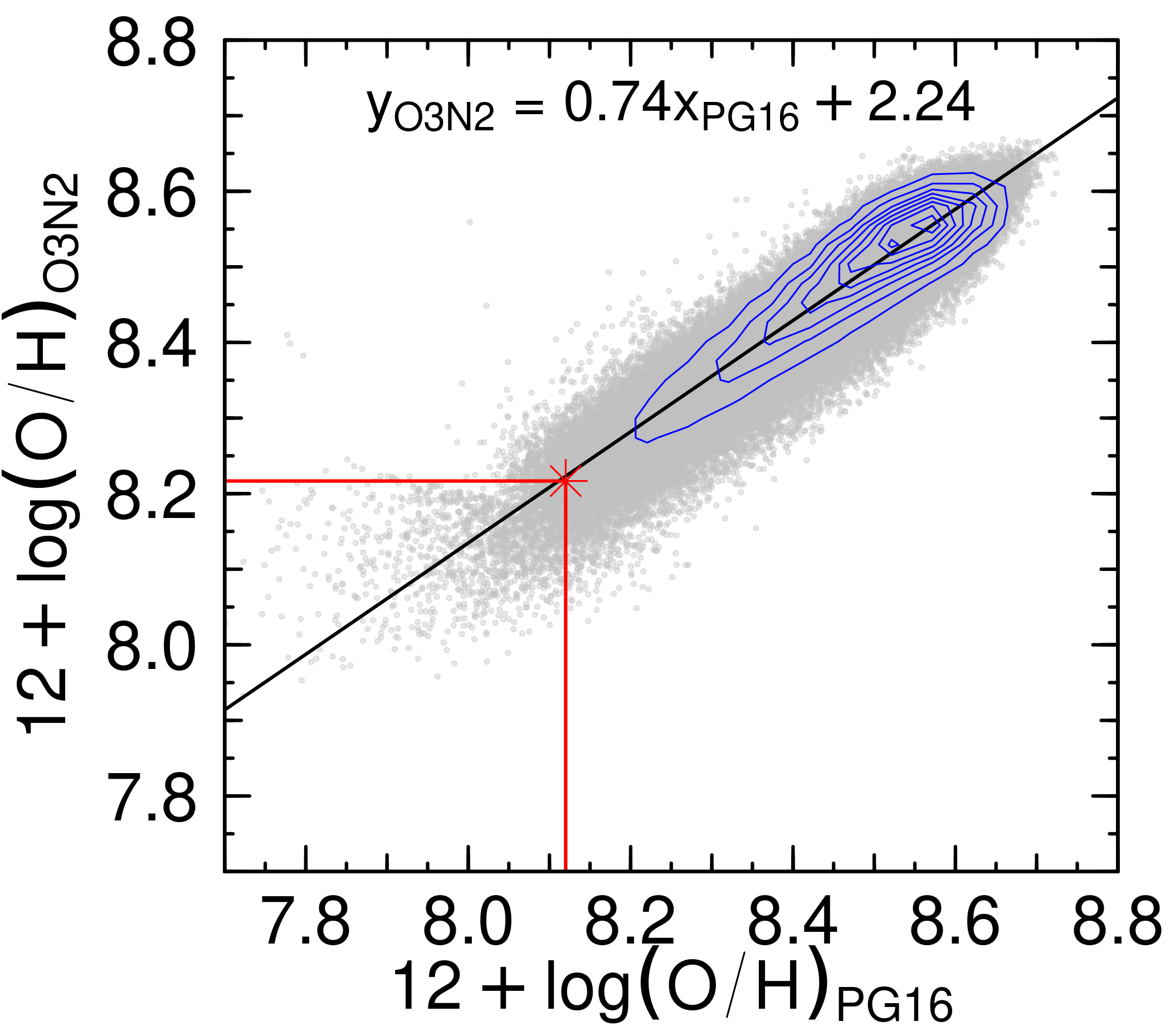}
\caption{The metallicity recalibration between \protect\cite{Pilyugin2016} and \protect\cite{Marino2013}. The red asterisk shows the metallicity of NGC 1569 in both scales.}
\label{HZrecal}
\end{figure}

Since we compare our metallicities with MANGA galaxies, we compute the recalibration of metallicities between \cite{Pilyugin2016} and \cite{Marino2013}, also using SDSS galaxies. In Fig. \ref{HZrecal} is shown such recalibration as well as the value of NGC 1569 in both metallicity scales. 

\section{Gas metallicity estimates for the MANGA galaxies}\label{Amet}

\begin{figure}
\includegraphics[trim={0cm 0cm 0cm 0cm},clip,width=1\linewidth]{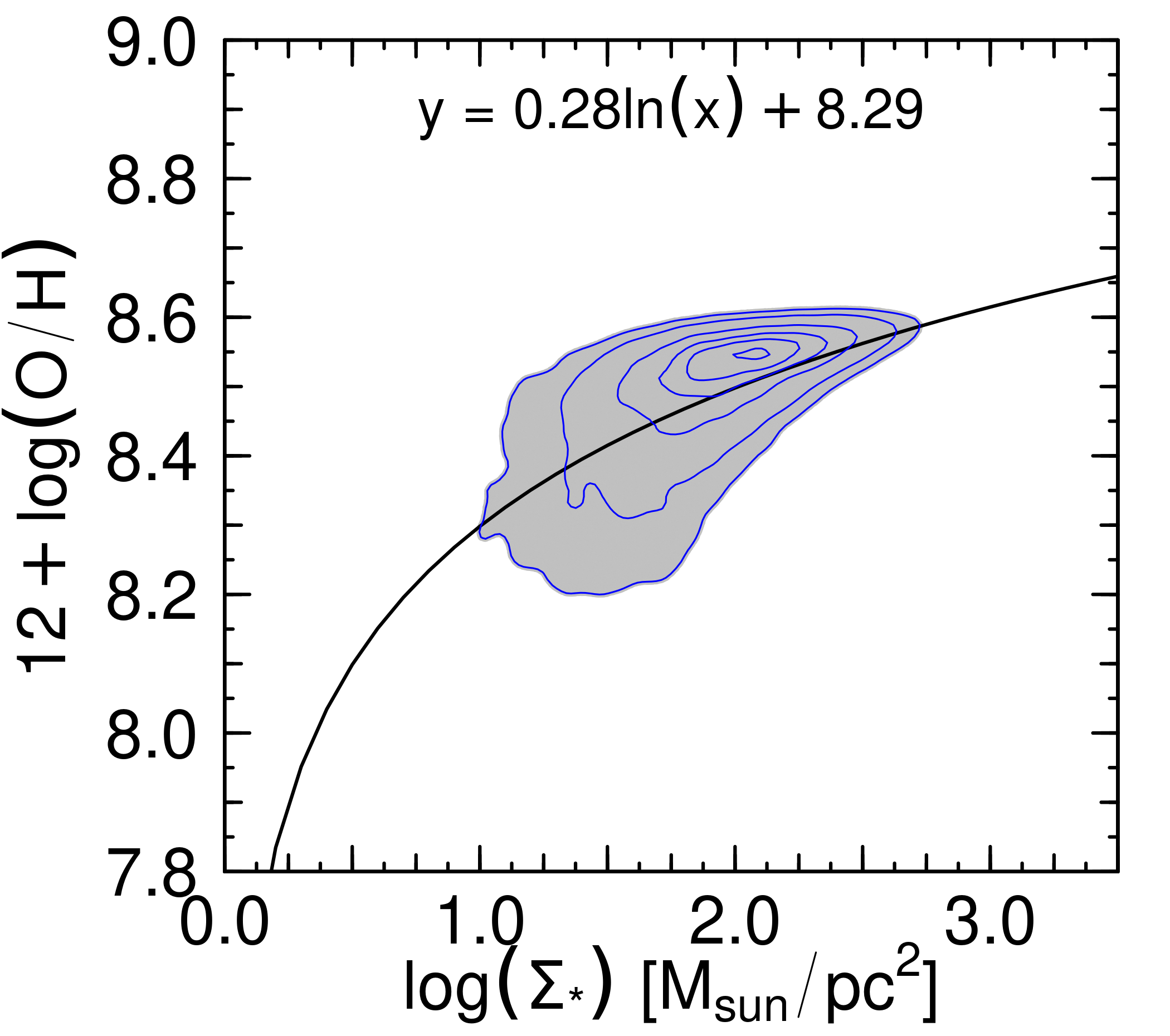}
\caption{The local \sMZ\ for the MANGA galaxies using the S-calibrator \protect\cite{Pilyugin2016}.}
\label{MZmanga}
\end{figure}

In order to have a fair comparison of our \sMZ\ with other works, we use the MANGA DR15 data from \cite{Zinchenko2021} who also derive the oxygen abundance (O/H) using the S-calibrator \citep{Pilyugin2016}. We take the SF spaxels enclosed in the 60\% area of the density contour plot (Fig. \ref{MZmanga}). Then, we fit a logarithmic curve with the following parameters: y=0.28ln(x) +8.29. This fitting is  used in panel B of Fig. \ref{SR1plots}.

\bsp	
\label{lastpage}
\end{document}